\documentclass[reprint,amsmath,amssymb,aps,prc,superscriptaddress,floatfix]{revtex4-2}
\usepackage{float,natbib,hyperref,graphicx,dcolumn,bm,color,amsmath,slashed,multirow,amssymb,appendix,enumitem}
 
\usepackage[normalem]{ulem}  
\renewcommand\sout{\bgroup \color{red} \ULdepth=-.5ex \ULset}

\begin{document}


\title{Electromagnetic emission from strongly interacting hadronic and partonic matter created in heavy-ion collisions}


\author{Adrian William Romero Jorge}
\email{jorge@itp.uni-frankfurt.de}
\affiliation{Frankfurt Institute for Advanced Studies (FIAS), Ruth Moufang Str. 1, 60438 Frankfurt, Germany}
 \affiliation{Institut f\"ur Theoretische Physik, Johann Wolfgang Goethe	University, Max-von-Laue-Str. 1, 60438 Frankfurt, Germany}
 \affiliation{Helmholtz Research Academy Hessen for FAIR (HFHF), GSI Helmholtz	Center for Heavy Ion Physics. Campus Frankfurt, 60438 Frankfurt, Germany.}


\author{Taesoo Song} 
\affiliation{GSI Helmholtzzentrum f\"ur Schwerionenforschung GmbH, Planckstraße 1, 64291 Darmstadt, Germany}

\author{Qi Zhou} 
\affiliation{Key Laboratory of Quark \& Lepton Physics (MOE) and Institute of Particle Physics, Central China Normal University, Wuhan 430079, China}

\author{Elena Bratkovskaya}%
 \affiliation{GSI Helmholtzzentrum f\"ur Schwerionenforschung GmbH, Planckstraße 1, 64291 Darmstadt, Germany}
 
\affiliation{Institut f\"ur Theoretische Physik, Johann Wolfgang Goethe	University, Max-von-Laue-Str. 1, 60438 Frankfurt, Germany}
\affiliation{Helmholtz Research Academy Hessen for FAIR (HFHF), GSI Helmholtz	Center for Heavy Ion Physics. Campus Frankfurt, 60438 Frankfurt, Germany.}

\date{\today}


\begin{abstract}
We investigate dilepton production in heavy-ion, proton-proton, and proton-nucleus collisions from low energies of 1 AGeV (SIS) to ultra-relativistic energies (LHC) using the Parton-Hadron-String Dynamics (PHSD) transport approach. PHSD is a microscopic, non-equilibrium approach that integrates hadronic and partonic degrees of freedom, providing a comprehensive description of relativistic heavy-ion collisions—from initial nucleon-nucleon interactions to quark-gluon plasma (QGP) formation, hadronization, and final-state interactions.
Key dilepton sources in PHSD include hadronic decays, bremsstrahlung,  QGP radiation ($q+\bar q \to e^+e^-$, \ $q+\bar q \to g+ e^+e^-$, \ $q+g \to q+ e^+e^-$), primary Drell-Yan production, and semileptonic decays of correlated charm and bottom pairs. PHSD well describes dilepton data from HADES, STAR, and ALICE experiments. We examine in-medium effects, such as the vector meson spectral function broadening, and present the excitation function for the dilepton "excess"  in the invariant mass range $0.4<M_{ee}<0.75$ GeV/c$^2$ and its centrality dependence at various energies.
For the first time we report on the baryon chemical potential $\mu_B$-dependence of the QGP radiation  calculated on a basis of the Dynamical-Quasi-Particle Model (DQPM) in PHSD. The influence of $\mu_B$ on the QGP yield grows at lower collision energies, where $\mu_B$ becomes large, however, its impact on the total dilepton spectra is small due to the lowering of the QGP volume with decreasing energy. The excitation function of QGP dileptons, including the $\mu_B$-dependent quasiparticle masses and widths, is presented versus correlated charm, confirming that the QGP radiation overshines charm contributions at $\sqrt{s} \simeq 25-30$ GeV in central Au+Au collisions, providing access to thermal QGP dileptons at BES RHIC and FAIR. 
Additionally, we find that the contribution from partonic processes to dilepton spectra in p+p collisions at intermediate masses - although being small - increases with energy, suggesting that after subtracting correlated charm and bottom contributions the QGP radiation in high multiplicity p+p collisions might also be experimentally accessible at LHC energies.
\end{abstract}

\maketitle

\section{Introduction}\label{Intro}

Present experiments at heavy-ion colliders have successfully recreated conditions similar to microseconds after the Big Bang. During this early phase the universe transitioned from a quark-gluon plasma (QGP) — a state of quarks, antiquarks, and gluons — to  color-neutral hadronic matter, where partonic degrees of freedom become confined within interacting hadrons and resonances. Understanding the nature of confinement and the dynamics of this phase transition remains a fundamental challenge in modern physics. Initial theories assumed the QGP to be a weakly interacting system of massless partons that could be describable by perturbative Quantum Chromodynamics (pQCD). However, experimental results have revealed that the medium created in ultra-relativistic heavy-ion collisions interacts more strongly than anticipated previously,  showing properties of an almost perfect liquid \cite{Hirano:2005wx,Shuryak:2008eq}. This conclusion is supported by the observation of strong radial expansion and the scaling of elliptic flow with the number of constituent quarks  at Relativistic Heavy Ion Collider (RHIC) energies \cite{Molnar:2003ff,STAR:2003wqp}. 

Electromagnetic probes, such as dileptons, can provide additional information about the properties of the strongly interacting hadronic and partonic matter created in heavy-ion collisions since they are undisturbed by the strong final state interaction.
The dileptons are emitted during the whole evolution of the expanding system from many different sources -  from hadron decays to QGP thermal radiation - such that the consistent description of dilepton production requires an application of dynamical models.

A first generation of relativistic heavy-ion collision experiments, conducted during the 1990s, observed an enhancement in dilepton production at low invariant mass in heavy systems compared to predictions from conventional hadronic cocktails \cite{Agakishiev:1995xb,Mazzoni:1994rb}. This enhancement was later attributed to the in-medium modification of the $\rho$ meson spectral function. At that time, two primary scenarios were proposed to explain these findings: a reduction in the $\rho$ meson mass, as suggested by the Brown-Rho scaling hypothesis \cite{Brown:1991kk} and the prediction of the Hatsuda-Lee sum rule \cite{Hatsuda:1991ez}, or a collisional broadening of its spectral function as predicted by many-body hadronic models \cite{Rapp:1997fs,Friman:1997tc,Peters:1997va,Rapp:1999qu,Lutz:2001mi}.
Although these experiments clearly demonstrated the necessity of incorporating in-medium effects, due to the large experimental error bars they could not conclusively determine whether the additional dilepton strength at lower invariant masses arose from a dropping vector meson mass or the broadening of its spectral function. A significant step towards resolving this ambiguity came with higher resolution measurements by the NA60 Collaboration \cite{Arnaldi:2006jq}, whose data strongly supported the broadening scenario. This conclusion was further supported by CERES data \cite{Adamova:2006nu} and subsequently reinforced by experiments at RHIC conducted by the PHENIX \cite{PHENIX:2009gyd} and STAR \cite{STAR:2023wta} Collaborations as well as at LHC by the ALICE Collaboration \cite{ALICE:2018ael}.
Theoretical calculations could provide a reasonable description of dilepton data at intermediate invariant masses only by incorporation of in-medium effects (cf. reviews \cite{Linnyk:2015rco,Bleicher:2022kcu,Gale:2025ome} and references therein).

In-medium modifications of vector meson properties have been investigated in proton-nucleus (p+A) and nucleus-nucleus (A+A) collisions at low energies, specifically around 1 AGeV. Initial measurements conducted by the DLS Collaboration at the BEVALAC in Berkeley during the 1990s revealed an unexpectedly high dilepton yield in C+C and Ca+Ca collisions within the invariant mass range of 0.2 to 0.5 GeV/c$^2$ \cite{Matis:1994tg,Wilson:1997sr}. This observation, known as the "DLS puzzle," indicated that the dilepton production was approximately five times larger than predictions from various transport models that incorporated conventional dilepton sources such as bremsstrahlung, $\pi^0$, $\eta$, $\omega$, and $\Delta$ Dalitz decays, as well as direct decays of vector mesons ($\rho$, $\omega$, $\phi$) \cite{Wolf:1993pja,Bratkovskaya:1996bv,Xiong:1990bg}.
Despite attempts to resolve this discrepancy by integrating different scenarios for in-medium modifications of vector meson properties — such as dropping mass or collisional broadening of the $\rho$ and $\omega$ spectral functions — the mismatch between experimental data and theoretical models persisted \cite{Ernst:1997yy,Bratkovskaya:1997mp,Fuchs:2005zga,Bratkovskaya:1998pr}. 
The persistent discrepancy motivated the development of the HADES (High Acceptance DiElectron Spectrometer) detector at GSI in Germany \cite{HADES:2007nok,HADES:2008wxk,HADES:2009mrh,HADES:2009cui, HADES:2011nqx}. High-statistics measurements performed by the HADES Collaboration confirmed the elevated dilepton yields observed by DLS in C+C collisions at 1.0 AGeV \cite{HADES:2007nok,HADES:2008wxk}.

Advances in theoretical transport approaches and effective models for elementary nucleon-nucleon (NN) reactions have contributed to resolving the DLS puzzle. In  Ref. \cite{Bratkovskaya:2007jk} it has been proposed that incorporating in the Hadron-String-Dynamics (HSD) transport model  the enhanced proton-neutron (pn) and proton-proton (pp) bremsstrahlung contributions consistent with updated One-Boson-Exchange (OBE) model calculations  \cite{Kaptari:2005qz}, can account for the previously unexplained dilepton yields. Specifically, the HSD model, when adjusted to include enhanced bremsstrahlung cross sections and collisional broadening of vector-meson spectral functions, successfully reproduced the HADES and DLS experimental data for C+C and Ca+Ca collisions at energies of 1 and 2 AGeV \cite{Bratkovskaya:2007jk}. Similar results have been achieved by other independent transport models, such as the Isospin Quantum Molecular Dynamics (IQMD) and the Rossendorf BUU approaches \cite{Thomere:2007cj,Barz:2009yz}.

Although theoretical predictions indicate that the properties of vector mesons undergo significant modifications even at relatively low collision energies of 1–2 AGeV, an experimental observation of these changes remains challenging. At these energy scales the production rates of $\rho^0$ and $\omega$ mesons are notably low, while the mass region of interest, $M_{ee} > 0.4$ GeV/c$^2$, is dominated by substantial contributions from other hadronic dilepton sources. These include $\Delta$-Dalitz decays and $pN$ bremsstrahlung processes, which create a considerable background. 

The FAIR SIS100 accelerator at GSI Darmstadt will play a pivotal role in advancing the study of heavy-ion collisions by enabling detailed measurements of dileptons with high statistics,  very high net-baryon densities and moderate temperatures in nucleus–nucleus collisions at beam energies up to 35 AGeV. This will allow for detailed studies of Au + Au collisions up to 11 AGeV beam energy and p + A collisions up to 30 GeV. SIS100 will achieve high rates of $10^5$ to $10^7$ Au-Au collisions per second \cite{Bhaduri:2022cql,Hohne:2014zia}. 

Furthermore, dileptons are particularly suited for studying heavy-ion collisions because their invariant mass provides an additional scale beyond that offered by photons, allowing for a partial disentanglement of the production mechanisms originating from the early, potentially partonic phase, and those arising during the later hadronic phase \cite{Gale:1987ki,Kvasnikova:2001zm,Dusling:2006yv,Renk:2006qr,Rapp:2013nxa,Linnyk:2015rco}. Over decades of experimental and theoretical efforts it has been established that dileptons with invariant masses below approximately 1.2 GeV/c$^2$ predominantly result from hadronic decays, offering insights into the modification of hadronic properties in the dense and hot medium created during collisions \cite{Rapp:2013nxa,Linnyk:2015rco,Song:2018xca}. In contrast, dileptons in the intermediate mass range, 1.2  $< M_{ee} <$ 3 GeV/c$^2$, are expected to provide crucial information on thermal radiation from the quark-gluon plasma, which includes contributions from processes such as $q+\bar{q} \to e^+e^-$, $q+\bar{q} \to g+\gamma^*$, and $g+q(\bar{q}) \to q(\bar{q})+e^+e^-$. Additionally, this mass range is sensitive to correlated semileptonic decays of open charm pairs ($c\bar{c} \to D\bar{D}$), produced early in the collision.

At high-energy collisions, such as those at RHIC and LHC, the dilepton signal from the QGP in the intermediate mass regime is overshadowed by a significant background from $D\bar{D}$ decays \cite{Manninen:2010yf, Linnyk:2012pu, Linnyk:2015rco,Song:2018xca}. However, at lower bombarding energies, the production of charm pairs decreases significantly, potentially opening a window where partonic sources dominate the intermediate mass range. 
The dilepton study with the Parton-Hadron-String Dynamics model in Ref. \cite{Song:2018xca} showed that the QGP over-shines the correlated charm at $\sqrt{s_{NN}}  \leq 30$ GeV. These calculations included  a fully microscopic analysis of the charm dynamics and the angular correlations of charm-anticharm pairs - including the rescattering of $c, \bar c$ quarks in the QGP and $D, \bar{D}$ hadronic rescattering as well as a modeling of dilepton emission from the partonic processes in the strongly interacting QGP (sQGP) at different temperatures $T$ and baryon chemical potential $\mu_B$.

We stress that the dynamical transport approaches are particularly suited for simulating 
dilepton emission from the heavy-ion collisions, as they track the evolution from the initial nucleon-nucleon collisions to the final hadronic freeze-out. The Parton-Hadron-String Dynamics  model 
\cite{Cassing:2008sv,Cassing:2008nn,Cassing:2009vt,Bratkovskaya:2011wp,Konchakovski:2011qa,Moreau:2019vhw}
offers a fully microscopic description of strongly interaction matter, in both hadronic and partonic phases, based on off-shell Kadanoff-Baym dynamics.  The PHSD models the whole time evolution of heavy-ion collisions - from primary NN scattering to the formation of a QGP, described in terms of strongly-interacting quasiparticles - quarks and gluons with $(T,\mu_B)$-dependent  masses and widths defined by the Dynamical Quasi-Particle Model (DQPM) \cite{Cassing:2007yg, Cassing:2007nb, Moreau:2019vhw}, to the  partonic scatterings, hadronization, and hadronic rescatterings.  This allows to capture the various sources of dileptons, including hadronic decays, bremsstrahlung, thermal QGP radiation, primary Drell-Yan (DY) and correlated heavy-flavor decays, in a single framework \cite{Linnyk:2015rco,Song:2018xca}.

In this study, we employ the PHSD to investigate the production of dileptons from SIS energies ($\sqrt{s_{NN}} \approx 2\text{--}3\,\mathrm{GeV}$) to LHC energies ($\sqrt{s_{NN}} \approx 5\text{--}13\,\mathrm{TeV}$).
We focus on  the in-medium effects on dilepton production from vector mesons  at low invariant masses and the interplay between dilepton emission from the QGP and correlated heavy-flavor decays at intermediate masses. For the first time within the PHSD we study the influence of the  baryon chemical potential $\mu_B$ (additionally to the  temperature $T$) on the dilepton yield from the sQGP and estimate the relative contribution of primary Drell-Yan to the dilepton spectra in the  wide energy range.
Our analysis thus provides a comprehensive overview of dilepton emission over a wide phase space, revealing how the evolution from hadronic to partonic degrees of freedom can be probed by  electromagnetic observables.

This paper is organized as follows: 
In Sec.\ref{sec1}, we present the Parton-Hadron-String Dynamics (PHSD) transport approach, including its off-shell formulation and the Dynamical Quasi-Particle Model for describing the quark-gluon plasma properties. Next, in Sec.\ref{sec2}, we outline the major dilepton production mechanisms in PHSD, covering hadronic decays, and partonic channels. 
In Sec. \ref{sec3} we discuss the time evolution of the fireball created in heavy-ion collisions at different energies. Sec. \ref{sec3d} details the influence of the  $(T,\mu_B)$-dependent EoS on dilepton production.
In Sec.~\ref{sec7}, we compare our results to experimental data in the wide range of collision energies, from SIS18 up to RHIC and LHC. In  Sec.\ref{sec5} we show the dilepton excess in the low-mass region and its connection to in-medium effects and thermal QGP radiation.   Sec~\ref{sec6} focuses on the excitation function of the dielectron yield, emphasizing the interplay between thermal QGP emission, open charm decays, and Drell-Yan contributions. 
In Sec.\ref{sec4}, we analyze the transverse mass spectra of intermediate-mass dileptons.   Finally, we summarize our main conclusions in Section \ref{conclusions}. The details of the evaluation of primary Drell-Yan are presented in Appendix \ref{appendix:DY}.

\section{The Parton-Hadron-String Dynamics  Transport Approach}\label{sec1}

The Parton-Hadron-String Dynamics  is a non-equilibrium microscopic transport approach that combines both hadronic and partonic degrees of freedom \cite{Cassing:2008sv, Cassing:2008nn, Cassing:2009vt, Bratkovskaya:2011wp, Konchakovski:2011qa, Linnyk:2015rco, Moreau:2019vhw}. PHSD provides a comprehensive description of the complete evolution of relativistic heavy-ion collisions, beginning with initial out-of-equilibrium nucleon-nucleon ($NN$) collisions, progressing through the formation of the quark-gluon plasma, and encompassing partonic interactions, hadronization and final-state interactions of the resulting hadrons. The dynamical evolution of the interacting system is modeled by solving the Cassing-Juchem generalized off-shell transport equations in the test-particle representation \cite{Cassing:1999wx, Cassing:1999mh}, based on a first-order gradient expansion of the Kadanoff-Baym equations \cite{KadanoffBaym}, which are applicable for strongly interacting degrees of freedom \cite{Juchem:2004cs, Cassing:2008nn}
(see also a book for the details \cite{Cassing:2021fkc}).

The hadronic sector of PHSD builds upon early developments of the HSD transport approach \cite{Ehehalt:1996uq, Cassing:1999es}, including the baryon octet and decouplet, ${0}^{-}$ and ${1}^{-}$ meson nonets, and higher resonances. Multi-particle production in elementary baryon-baryon ($BB$) reactions above $\sqrt{s_{BB}^{th}} = 2.65$ GeV, meson-baryon ($mB$) collisions above $\sqrt{s_{mB}^{th}} = 2.35$ GeV, and meson-meson ($mm$) reactions above $\sqrt{s_{mm}^{th}} = 1.3$ GeV is described by the Lund string model \cite{NilssonAlmqvist:1986rx}. These are realized through event generators FRITIOF 7.02 \cite{NilssonAlmqvist:1986rx, Andersson:1992iq} and PYTHIA 6.4 \cite{Sjostrand:2006za}, which are "tuned" for a more accurate description of low and intermediate energy reactions - cf. \cite{Kireyeu:2020wou}.

The "PHSD tune" of the Lund string model includes also a modification of string fragmentation and the properties of produced hadrons in the hot and dense medium created in heavy-ion collisions, in particular,   an implementation of chiral symmetry restoration via the Schwinger mechanism for string decay in the dense medium \cite{Cassing:2015owa, Palmese:2016rtq}; accounting for the initial state Cronin effect for $k_T$ broadening in the medium and implementation of the in-medium properties of hadrons in the string fragmentation by incorporation of the in-medium spectral functions for mesonic and baryonic resonances with momentum, density and temperature dependent widths (instead of non-relativistic spectral functions with constant width) \cite{Bratkovskaya:2007jk,  Song:2020clw}. 

The strings decay into leading hadrons and 'pre-hadrons,' which are newly produced "unformed" mesons and baryons ("remnants" of the string color flux tube breaking in the Lund picture) within a formation time $t_F=\tau_F\gamma$, where $\gamma = 1/\sqrt{1 - v^2/c^2}$ and $v$ is the velocity in the initial $NN$ center-of-mass frame and $\tau_F=0.8$ fm$/c$. While the leading hadrons (related to the leading (anti-)quark $q, \bar q $ or diquarks $qq$ can interact immediately with  reduced hadronic cross sections \cite{Cassing:1999es}, the  pre-hadrons are not allowed to interact for  $t<t_F$ if they are in "cold" matter with local energy density $\varepsilon < \varepsilon_C$, i.e. below the critical energy density $\varepsilon_C \sim 0.4$~GeV/fm$^3$ fixed according to the lattice quantum chromodynamic (lQCD)  calculations \cite{Borsanyi:2022qlh}. Below this critical energy density, 'pre-hadrons' evolve into asymptotic hadronic states after the formation time $t_F$ and interact via hadronic cross-sections.

The transition between hadronic and partonic degrees of freedom occurs if $\varepsilon \simeq \varepsilon_C$: then the "pre-hadrons" are melting to "thermal" partons - quarks/antiquarks and gluons - which properties and interactions are defined along with the DQPM model.
 
The partonic interactions during the QGP phase are described  within the effective {\bf Dynamical Quasi-particle Model (DQPM)}, which is formulated in the two-particle irreducible (2PI) propagator  representation \cite{Cassing:2007yg, Cassing:2007nb, Moreau:2019vhw, Soloveva:2019xph}.   The DQPM describes non-perturbative strongly-interacting QCD matter in accordance with the lattice QCD equation-of-state (EoS) at finite temperature $T$ and baryon chemical potential $\mu_{\mathrm{B}}$ \cite{Aoki:2009sc, Cheng:2007jq}.
In the DQPM the partonic degrees of freedom are  off-shell quasiparticles with  properties  defined by complex self-energies: the real part of the self-energies are attributed to a dynamically-generated mass, and the imaginary part contains information about the interaction rates in the system, which results in spectral functions with non-vanishing widths. The functional form of $(T, \mu_{\mathrm{B}})$-dependence of the dynamical parton self-energies is chosen using a hard-thermal-loop inspired parametrization, with one parameter, which is fixed by comparing the DQPM calculations of the entropy density (or pressure or energy density) with lattice QCD results at $\mu_{\mathrm{B}}=0$,
we recall that the strong coupling $g^2$ can be directly obtained from the parametrization of lQCD data for the entropy density at zero baryon chemical potential.  All thermodynamics quantities at non-zero $\mu_B$ are predictions of the DQPM. 

The thermodynamic properties of the sQGP in terms of transport coefficients, such as ratios of shear viscosity $\eta/s$ and bulk viscosity  $\zeta/s$  to entropy density $s$, ratio of the electric conductivity to temperature $\sigma_0/T$, as well as the scaled baryon diffusion coefficient  $\kappa_B/T^2$ and the full diffusion coefficient matrix for nuclear matter, have been evaluated  in Refs. 
\cite{Moreau:2019vhw,Soloveva:2019xph,Fotakis:2021diq,Soloveva:2020ozg,Fotakis:2021diq,Soloveva:2021quj,Grishmanovskii:2023gog} by explicitly computing parton interaction rates based on DQPM couplings and propagators and are found to provide a good description of available lQCD data at finite $T$ and $\mu_B$.
Respectively, the DQPM framework provides a consistent description of the QCD thermodynamic and the QGP properties in terms of transport coefficients in the whole $(T,\mu_B)$ plane as well as a non-equilibrium description - via Kadanoff-Baym equations - of strongly interacting degrees of freedom propagated in a self-generated scalar mean-field potential \cite{Cassing:2009vt}.  

On the partonic side the following processes are included in the PHSD:  elastic $2 \leftrightarrow 2$  scatterings  $qq \leftrightarrow qq$, $\bar{q} \bar{q} \leftrightarrow \bar{q}\bar{q}$, $gg \leftrightarrow gg$ as well  as  inelastic $2 \leftrightarrow 1$ processes $gg \leftrightarrow g$, $q\bar{q} \leftrightarrow g$  described by a relativistic Breit–Wigner cross section. The elastic partonic cross sections are evaluated by computing the leading order diagrams for scattering of off-shell massive quasiparticles using “dressed” propagators,  $(T,\mu_B)$ dependent strong coupling and their self-energies, exploiting 'detailed-balance'. For the detailed calculations of partonic cross sections we refer the reader to Refs. \cite{Moreau:2019vhw, Grishmanovskii:2023gog} where the $(\sqrt{s}, T, \mu_B)-$ dependence and angular distributions have been evaluated explicitly. 

As the fireball expands, the probability of parton hadronization increases near the phase boundary between hadronic and partonic phases, which is a crossover according to the lQCD data  \cite{Borsanyi:2022qlh}.  The resulting hadronic system is further described by off-shell HSD dynamics, which optionally include self-energies for hadronic degrees of freedom \cite{Cassing:2003vz, Bratkovskaya:2007jk,Song:2020clw}.

PHSD has been employed successfully in studies of p+p, p+A, and A+A collisions from SIS18 to LHC energies, reproducing many hadronic, dilepton, and photon observables \cite{Cassing:2008sv, Cassing:2008nn, Cassing:2009vt, Bratkovskaya:2011wp, Konchakovski:2011qa, Linnyk:2015rco, Moreau:2019vhw, Song:2018xca,Moreau:2021clr}.
We note that the present study is realized with the advanced version of the PHSD v6.1 (for internal version numeration) which incorporates advantages of $(T,\mu_B)$- dependent EoS in the DQPM for the QGP dynamics as in PHSD v5.0 \cite{Moreau:2019vhw} with the latest developments in strangeness production \cite{Song:2020clw} and  in charm/bottom dynamics \cite{Song:2023zma} as well as improved dilepton production from the QGP as will be discussed in the next section.

\section{Dilepton production  in the PHSD}\label{sec2}

Dileptons ($e^+e^-$ and $\mu^+\mu^-$ pairs) are produced during the whole time evolution in heavy-ion collisions via hadronic and partonic processes.  In the PHSD the following processes are accounted for:

\textbf{I. Hadronic sources of dileptons} \\
At low invariant masses ($M_{ee} < 1$ GeV/$c^2$), the dileptons are produced i) via Dalitz decays of resonances, which are 3-body decays involving the emission of a dilepton pair along with an additional particle. It includes the decays of mesons and baryons: $\pi^0, \eta, \eta^\prime \to \gamma e^+ e^-$, $\omega \to \pi^0 e^+ e^-$, $a_1 \to \pi e^+ e^-$, $\Delta \to N e^+ e^-$; ii) by  direct decays, which are 2-body decays where vector mesons  directly decay into a dilepton pair: $\rho^0, \omega, \phi \to e^+ e^-$. iii) Additional contributions arise from hadronic bremsstrahlung processes:  $NN \to NN e^+ e^-, \ \pi N \to \pi N e^+ e^- $.

In the PHSD the dynamics of the vector meson ($V= \rho, \omega, \phi, a_1$) is realized including an in-medium  modification of their spectral functions. We recall that the influence of in-medium effects on the vector mesons has been extensively studied within the PHSD approach in the past (cf. Refs. \cite{Bratkovskaya:2007jk,  Bratkovskaya:2013vx, Linnyk:2015rco, Song:2018xca}) and it has been shown that the collisional broadening scenario for the in-medium  vector-meson spectral functions is consistent with  experimental dilepton data from SPS to LHC energies in line with the findings by other groups (cf. the review \cite{Rapp:2013nxa}). Accordingly,  in the present study we adopt the collisional broadening scenario for the vector-meson spectral functions as the  'default' scenario.

We recall that in the PHSD the spectral functions of vector mesons are modeled via a Breit-Wigner spectral function: 
\begin{align}
A_{V}(M, \rho) =  C \frac{2}{\pi} 
\frac{M^2 \Gamma^{\ast}_{V}(M, \rho)}{[M^2 - M_V^{\ast 2}(\rho)]^2 + M^2 \Gamma^{\ast 2}_{V}(M, \rho)}, 
\label{eq:spec_func}
\end{align}
where $C$ is a normalization constant determined from the condition 
\begin{align}
\int_{M_{\mathrm{min}}}^{M_{\mathrm{max}}} A_{\phi}(M, \rho) \ dM = 1,  
\label{eq:normalization}
\end{align}
in which $M_{\mathrm{min}} = 2 m_{e}$ ($m_e$ is the mass of electron) while  $M_{\mathrm{max}}$ is set to 4 GeV.
$M_V^{\ast 2}(\rho)$ is the pole mass which can be modified in-medium as for the dropping mass scenario 
\cite{Brown:1991kk,Hatsuda:1991ez} or taken as the free pole mass as in case of the collisional broadening scenario which is considered here as the default case.

The density dependence of the total width $\Gamma^{\ast}_{V}(M, \rho)$ of a vector meson $V$  is given as 
\begin{align}
\Gamma^{\ast}_{V}(M, \rho) &= \Gamma_{V}(M) + \alpha^V_{coll} \frac{\rho}{\rho_0},
\label{eq:width_density}
\end{align}
where $\Gamma_{V}(M)$ is the vacuum decay width of the vector meson $V$  \cite{Bratkovskaya:2007jk} and $\alpha^V_{coll}$ is a “broadening coefficient”.  In order to explore uncertainties in the modeling of the collisional broadening scenario 
within the linear density approximation (\ref{eq:width_density})  we will vary $\alpha^V_{coll}$:\\
1) for a modest broadening as in Ref. \cite{Song:2018xca} - which will be the default scenario we use - $\alpha^\rho_{coll} = 70$ MeV for $\rho$-mesons,  $\alpha^\omega_{coll} = 40$ MeV for $\omega$-mesons and $\alpha^\phi_{coll} = 25$ MeV  for $\phi$- mesons;  \\
2) for a  stronger broadening - as incorporated in the early PHSD  dilepton studies at SIS energies \cite{Bratkovskaya:2007jk,Bratkovskaya:2013vx} - we use  $\alpha^\rho_{coll} = 150$ MeV,  $\alpha^\omega_{coll} = 70$ MeV, $\alpha^\phi_{coll} = 40$ MeV, which are also in line with the experimental analysis from Ref.   \cite{Metag:2007hq}.

We note that the consistent incorporation of the in-medium effects requires the  off-shell formulation of transport theory - based on Kadanoff-Baym dynamics - which  is realized in the PHSD for the production, propagation and interactions of the vector mesons.

For intermediate masses ($1 < M_{ee} < 3$ GeV/$c^2$), dileptons arise from semileptonic decays of correlated charm $D+\bar{D}$  and bottom $B+\bar{B}$ pairs. 
For the details of charm/bottom dynamics in the PHSD and calculation of dileptons from the correlated charm and bottom, we refer the reader to the PHSD study in Ref. \cite{Song:2018xca}.

\textbf{II. Partonic sources of dileptons} \\
The partonic processes for the dilepton emission   (up to next-to-leading-order contributions)  include:\\
-- $q\bar{q}$ annihilation:  $q\bar{q} \to e^+ e^-, \ \ q\bar{q} \to g  e^+ e^-$  and\\
-- Compton scattering: $qg \to q e^+ e^-, \ \ \bar{q}g \to \bar{q} e^+ e^- $. \\
Since in the PHSD the QGP partons are off-shell strongly interacting quasiparticles, the cross sections for dilepton production from partonic scattering are evaluated with the DQPM propagators and couplings  and differ from those for the pQCD scattering of massless on-shell partons. The  matrix elements and cross sections for the off-shell partonic processes for dilepton emission are given explicitly in Ref. \cite{Song:2018xca} (Appendix A).

We note that in the PHSD  the partonic dileptons are accounted for by both i) scattering of partons from the thermal QGP as well as ii) from  scattering of leading quarks/antiquarks from the string ends with the partons from the QGP. The latter processes are related to non-equilibrium scattering at early times when the density of strings is still high, however, the formation of the QGP (due to the dissolution of  "pre-hadrons" to off-shell partons) is already happening.

The novel aspects of this study are: \\
-- We investigate how the chemical potential $\mu_B$- dependence of the QGP EoS influences the dilepton production in a wide energy range in addition to the $T$-dependence of the QGP EoS explored previously at high bombarding energies \cite{Linnyk:2015rco,Song:2018xca}; \\
-- We estimate the primary Drell-Yan dileptons from $NN$ binary collisions in the PHSD. 
The primary Drell-Yan processes in A+A collisions are computed by summing up  dileptons from all possible binary collisions as calculated in the PHSD. The details of the calculations for primary Drell-Yan dileptons from NN collisions are given in Appendix A.

{\bf Time integration method for dileptons:}\\
The dilepton yield from hadronic and partonic sources in the PHSD is calculated  using a time integration method, also known as the "shining" method (which was first introduced in studies such as \cite{Heinz:1991fn}). This approach allows for the continuous emission from vector mesons and baryon resonances until absorption or hadronic decay. This method accurately incorporates in-medium dynamics, as detailed in \cite{Bratkovskaya:2007jk, Bratkovskaya:2013vx, Galatyuk:2015pkq, Staudenmaier:2017vtq, Larionov:2020fnu}, such that for vector meson decay it is given by
\begin{equation}
\frac{dN_{e^+e^-}}{dM} = \frac{\Delta N_{e^+e^-}}{\Delta M} = \sum_{j=1}^{N_{\Delta M}} \int_{t_j^{i}}^{t_j^{f}} \frac{dt}{\gamma} \frac{\Gamma_{e^+e^-}(M)}{\Delta M}, 
\end{equation}
where the electromagnetic decay width, $\Gamma_{e^+e^-}(M)$, represents the resonance decay, and $t_i$ and $t_f$ indicate the times when the resonance is produced and disappears (by strong decay or absorption processes), respectively.

\section{Time evolution of the fireball}\label{sec3}

In this section we show the characteristic behavior of the energy density $\varepsilon$, temperature $T$ and baryon chemical potential $\mu_B$ achieved in heavy-ion collisions at various energies as evaluated in the PHSD.
This is essential for the description of the QGP phase, since the properties of the quasiparticles depend on $(T,\mu_B)$, as well as for the modeling of the crossover transition from the hadronic to the partonic phase.

We recall that in the transport approach, the temperature $T$ and baryon chemical potential $\mu_B$ are not directly accessible. However, they can be related to the energy density and baryon density - calculated in mean-field level - on the spacial grid cells by averaging over all parallel ensembles by the equation-of-state.  To extract $T$ and $\mu_B$ from the corresponding equation of state of the partonic system (which is the lQCD EoS in the PHSD), a set of equations must be solved - cf. Ref. \cite{Moreau:2019vhw} for details:
\begin{equation}
\left\{\begin{array}{l}
\varepsilon^{\mathrm{EoS}}\left(T, \mu_B\right)=\varepsilon^{\mathrm{PHSD}} / r(x) \\
n_B^{\mathrm{EoS}}\left(T, \mu_B\right)=n_B^{\mathrm{PHSD}}. \label{eosq}
\end{array}\right.
\end{equation}
These equations connect the left-hand side, which includes energy density and baryon density obtained in the lQCD Taylor expansion method, to the unknown variables $T$ and $\mu_B$. The right-hand side represents the energy density and baryon density calculated within PHSD. In Eq. (\ref{eosq}), the energy density from PHSD ($\varepsilon^{PHSD}$) is normalized by a function that accounts for pressure anisotropy following Ref. \cite{Ryblewski:2012rr}:
\begin{equation}
r(x)=\left\{\begin{array}{l}
\frac{x^{-1 / 3}}{2}\left[1+\frac{x \operatorname{arctanh} \sqrt{1-x}}{\sqrt{1-x}}\right] \text { for } x \leq 1 \\
\frac{x^{-1 / 3}}{2}\left[1+\frac{x \arctan \sqrt{x-1}}{\sqrt{x-1}}\right] \text { for } x \geq 1
\end{array}\right.
\end{equation}
The anisotropy parameter $x$ is approximated based on pressure components, with $x = (P_{\perp}/P_{||})^{3/4}$, where $ P_{\perp}$   and  $P_{\parallel}$ are, respectively, the transverse and longitudinal pressures. The function $r(x)$  approximately accounts for the momentum anisotropy in the considered cell.

\begin{figure*}[h!]
    \centering
\includegraphics[width=0.35\linewidth]{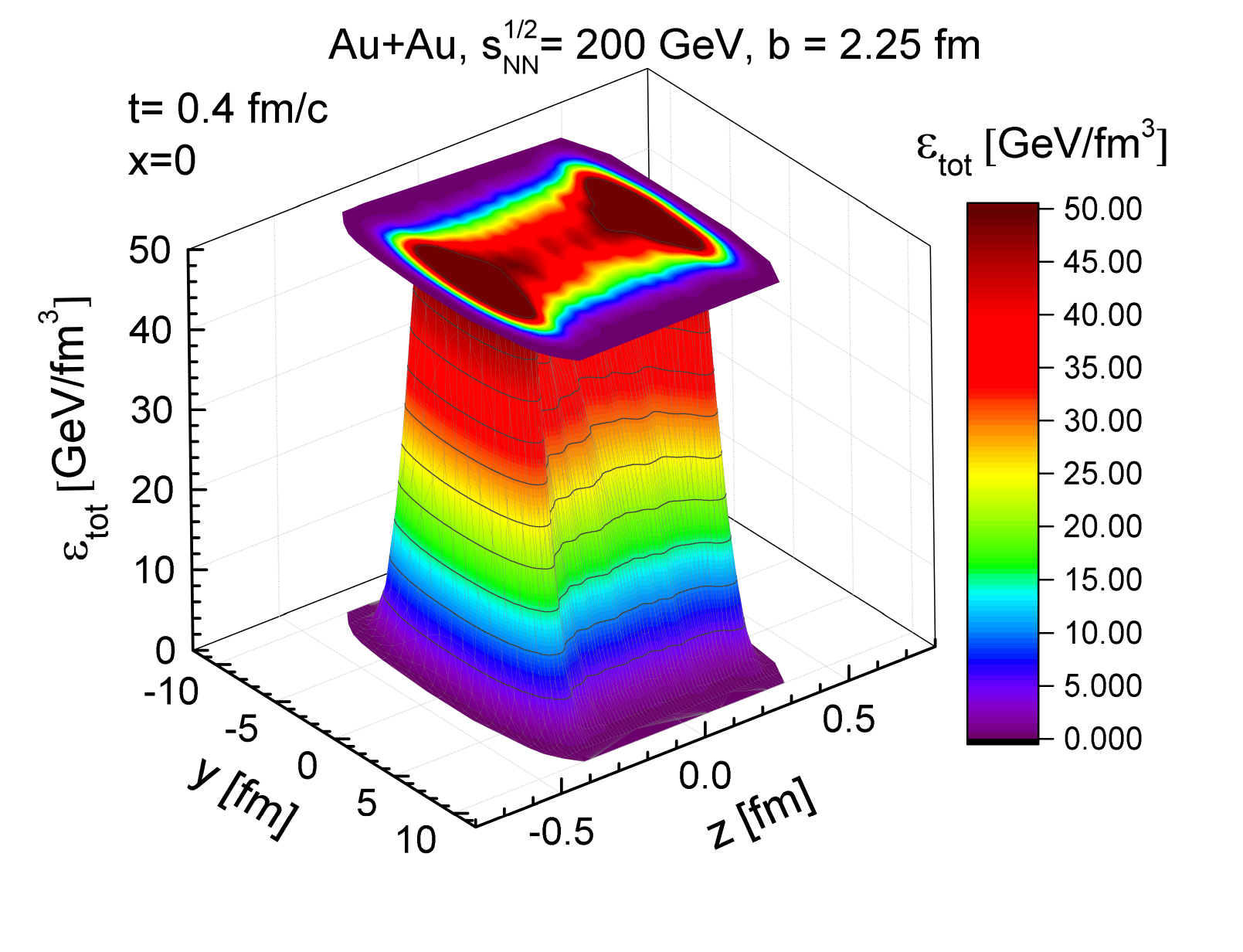}
\includegraphics[width=0.35\linewidth]{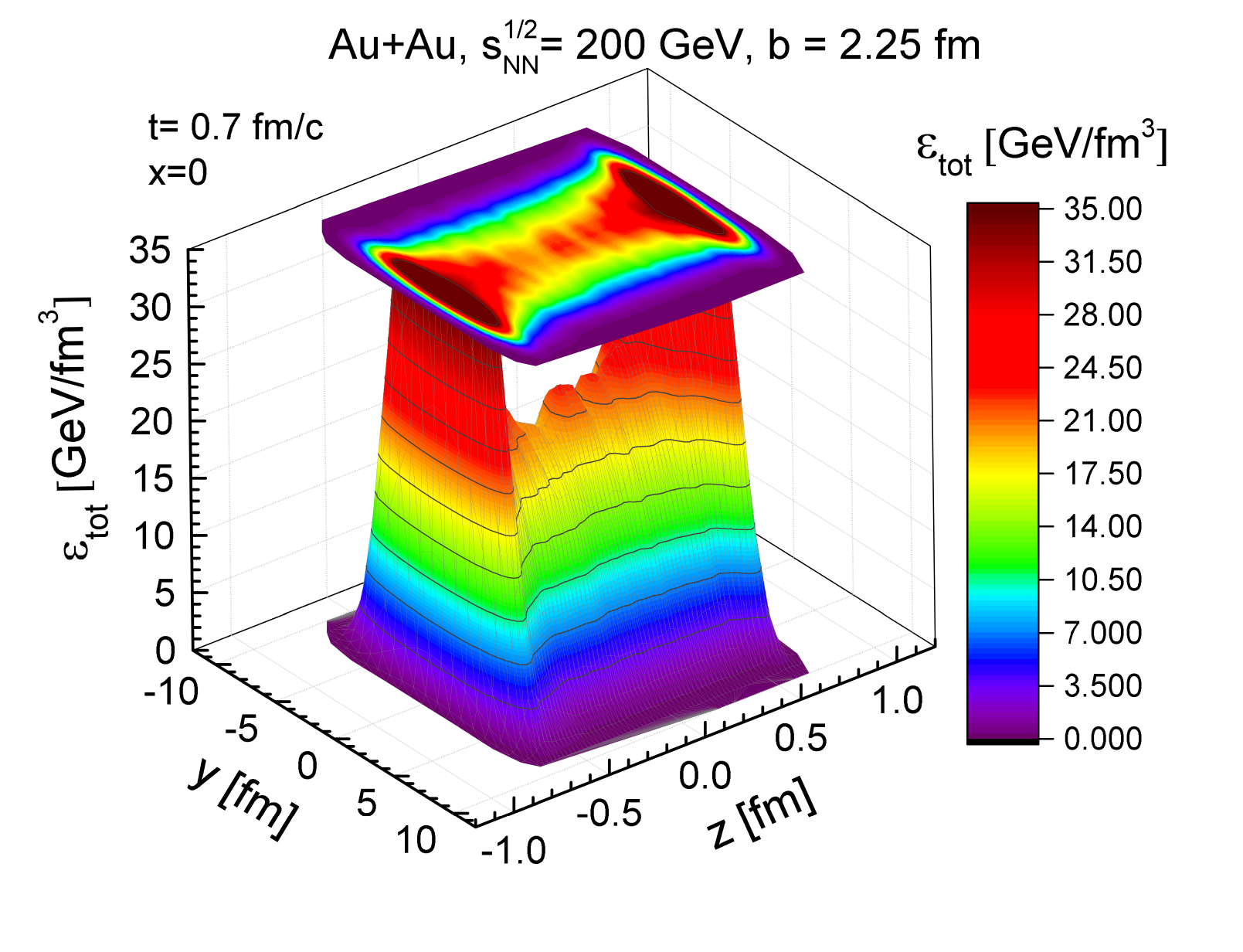}
\includegraphics[width=0.35\linewidth]{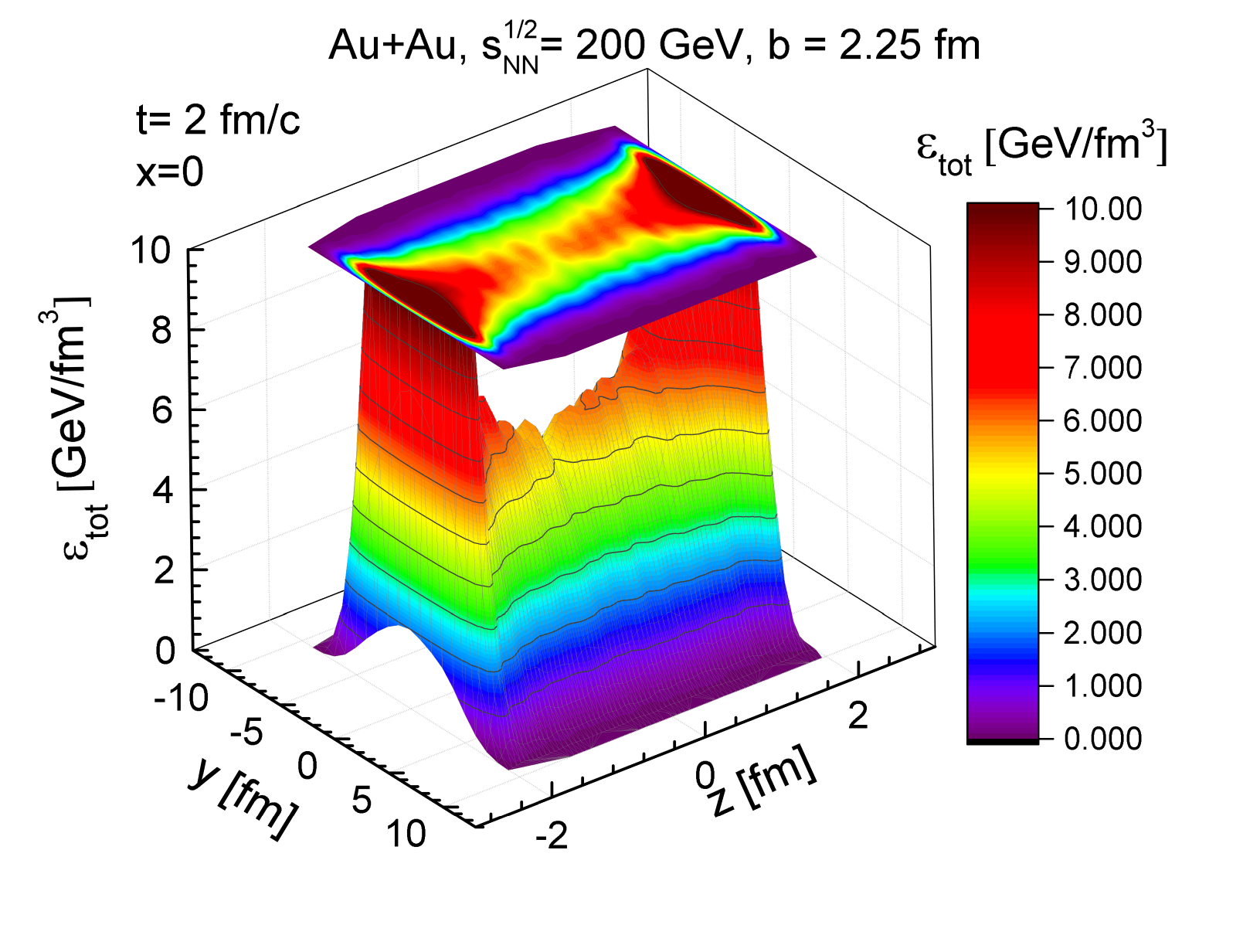}
\includegraphics[width=0.35\linewidth]{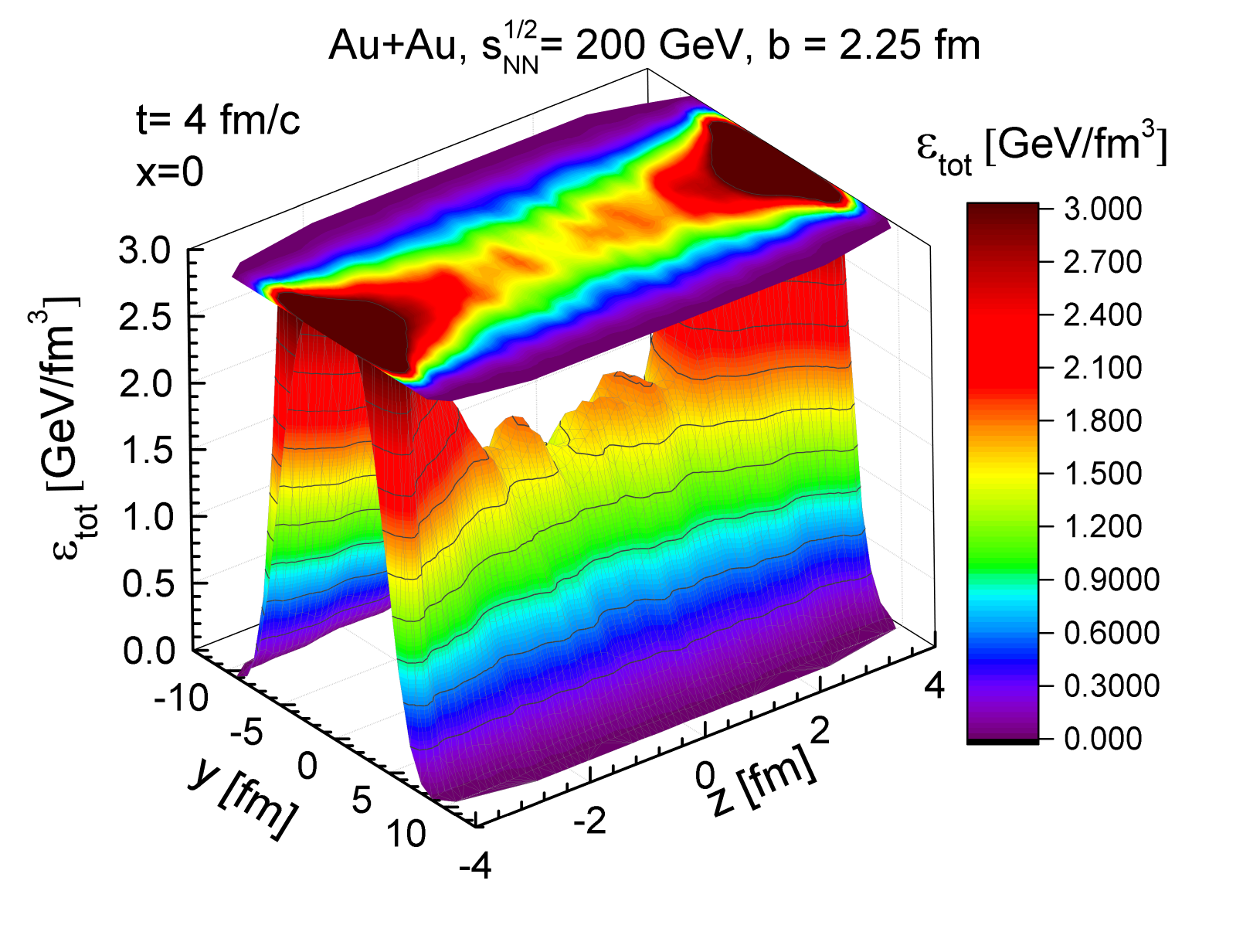}
    \caption{Total energy density $\varepsilon_{tot}$ in the $(y,z)$ plane for $x=0$ for central ($b=2.25$ fm) Au+Au collisions at $\sqrt{s_{NN}}=200 $ GeV  for the times $t=0.4, \ 0.7, \ 2$ and 4 fm/c. Note: the color coding and the scale in $z$-axis are different for better visibility.}
    \label{3D-En-200GeV}
\end{figure*}
\begin{figure*}[h!]
    \centering
\includegraphics[width=0.35\linewidth]{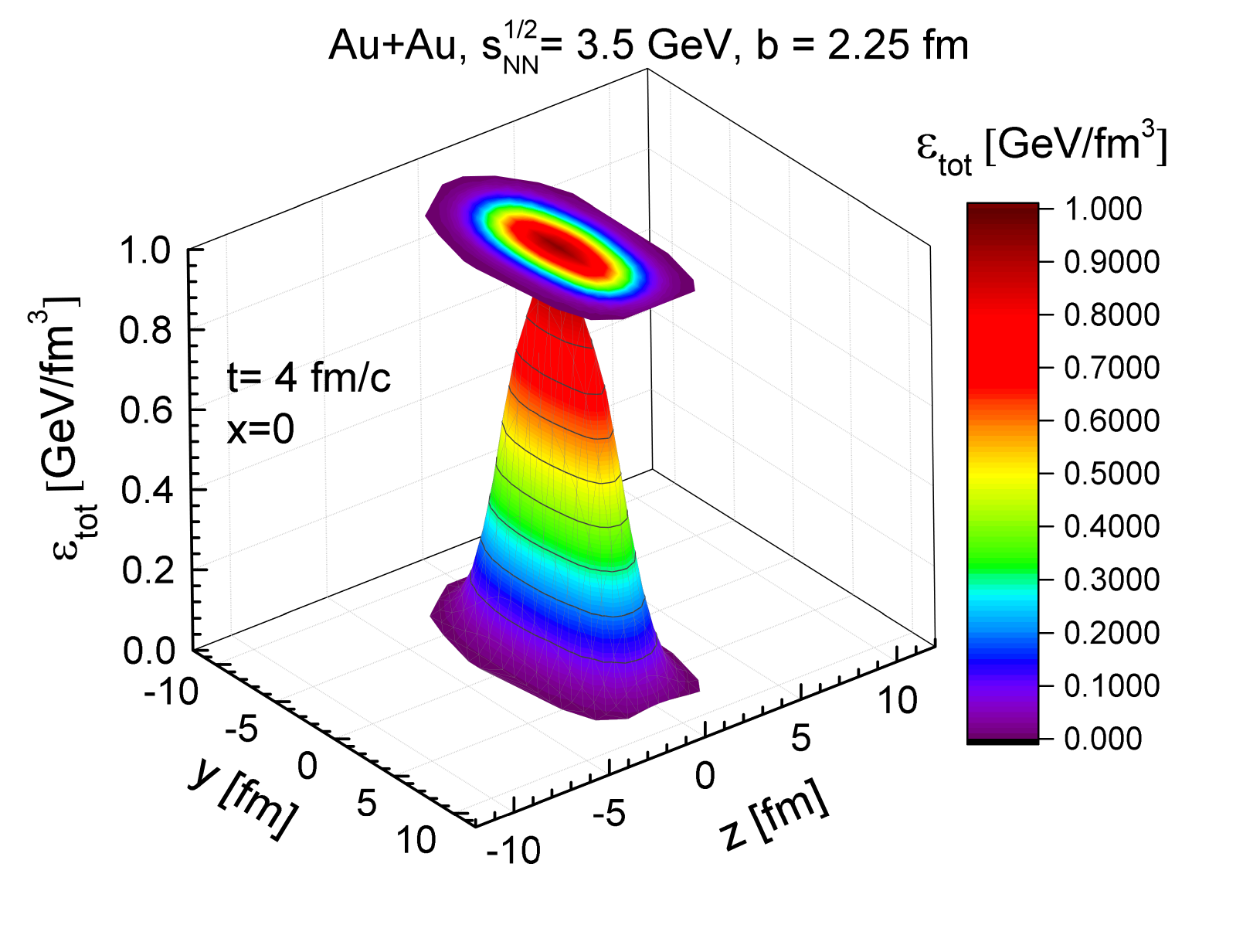}
\includegraphics[width=0.35\linewidth]{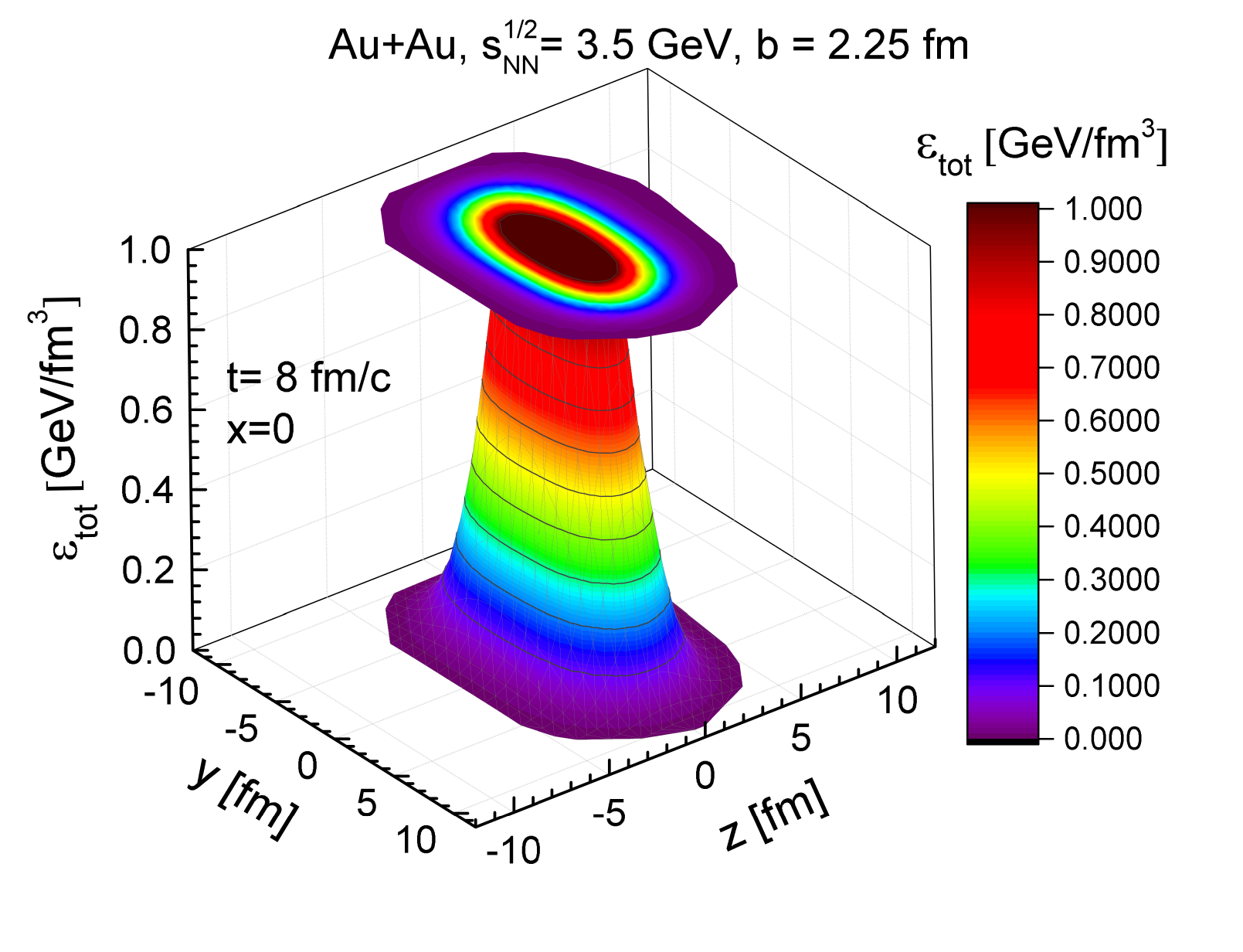}
\includegraphics[width=0.35\linewidth]{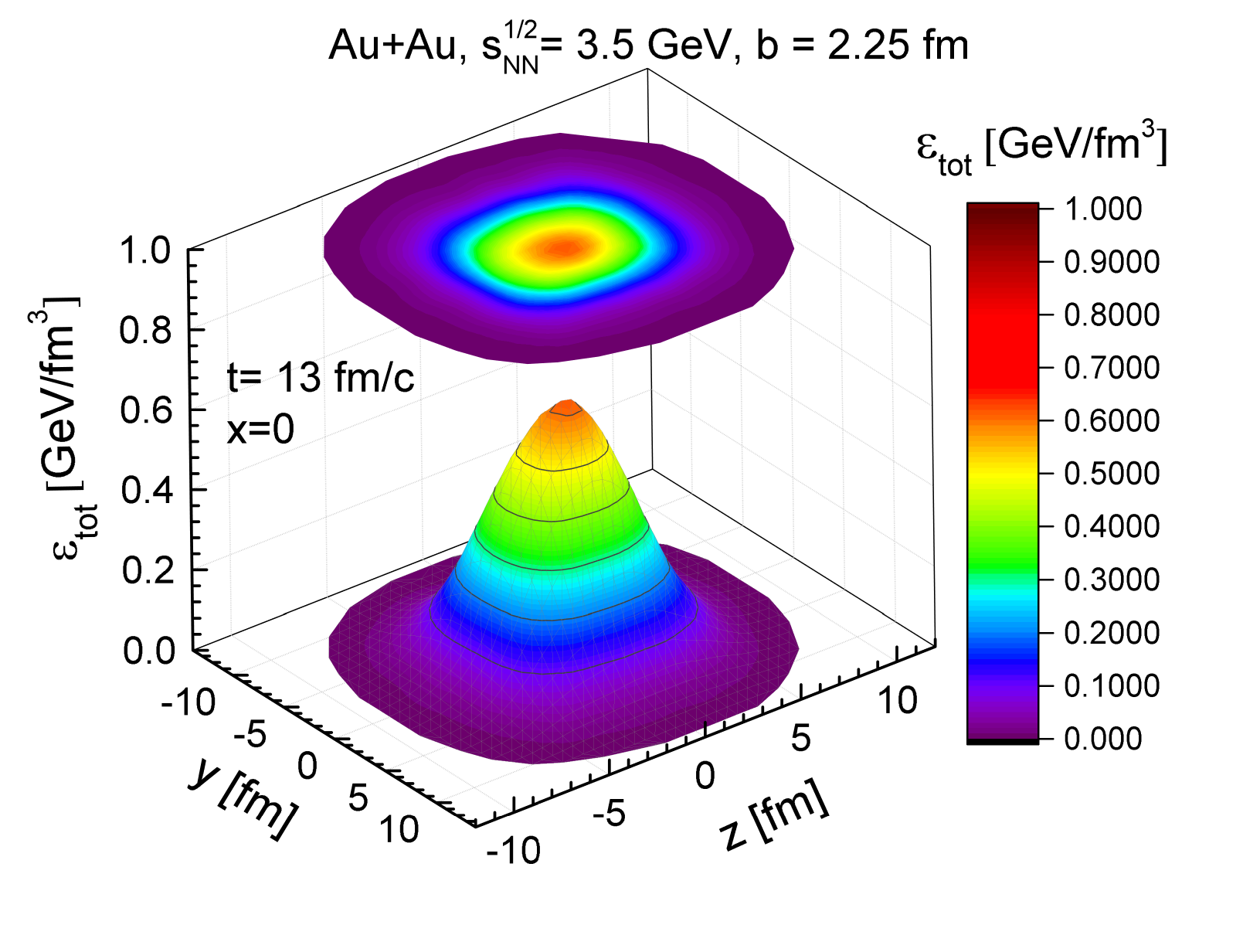}
\includegraphics[width=0.35\linewidth]{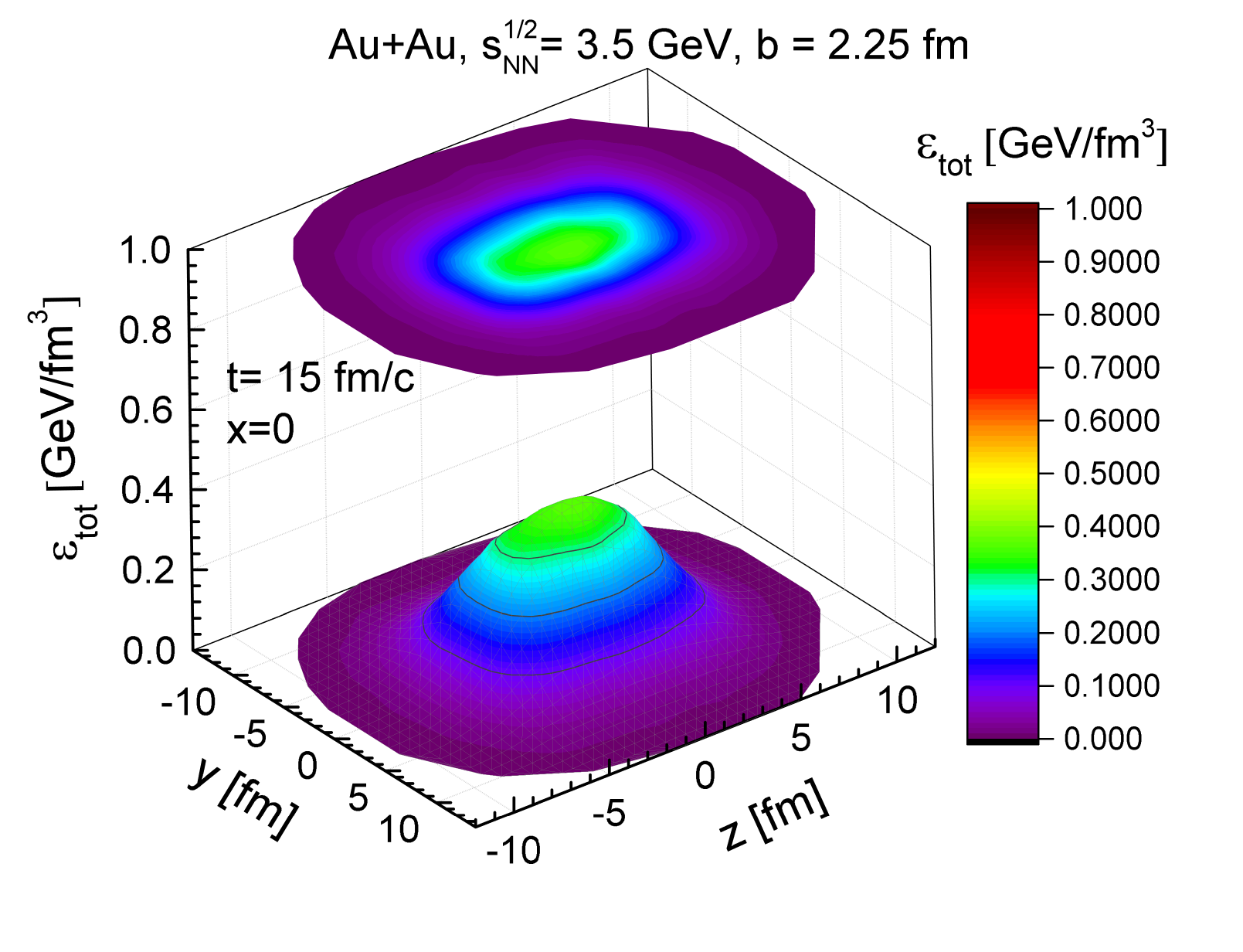}
     \caption{Total energy density $\varepsilon_{tot}$ in the $(y,z)$ plane for $x=0$ for central ($b=2.25$ fm) Au+Au collisions at $\sqrt{s_{NN}}=3.5 $ GeV  for the times $t=4, \ 8, \ 13$ and 15 fm/c.   }
    \label{3D-En-35GeV}
\end{figure*}

Fig. \ref{3D-En-200GeV} shows the  time evolution of the total energy density $\varepsilon_{tot}$ in the $(y,z)$ plane for $x=0$ and contour plots with the projections of $\varepsilon_{tot}$ on $(y,z)$ for central ($b=2.25$ fm) Au+Au collisions at $\sqrt{s_{NN}}=200 $ GeV. The snapshots are taken for the times $t= 0.4, \ 0.7, \ 2$ and 4 fm/c by averaging over 30 events which provides a smooth distribution. 
One can see a very high energy density phase at the early time of 0.4 fm/c, shortly after the overlap of the colliding nuclei, when the matter is dominantly in the non-equilibrium phase and the expansion has just started.  
$\varepsilon_{tot}$ decreases in time, however, rather hot matter remains in the  central region with energy density much above the critical $\varepsilon_C=0.4$ GeV/fm$^3$ \cite{Aoki:2009sc, Cheng:2007jq}.  
The QGP exists  in a large volume at $\sqrt{s_{NN}}=200 $ GeV, the "hot spots" related to the initial participants expand in $z$ and survive for a long time.
Similarly, in Fig. \ref{3D-En-35GeV} we present  $\varepsilon_{tot}$  at $\sqrt{s_{NN}}=3.5 $ GeV  for the times $t=4, \ 10, \ 13$ and 15 fm/c, averaged over 200 events.  Since the bombarding energy is low the overlap time is much longer than at 200 GeV and  $\varepsilon_{tot}$ is smaller, however, above $\varepsilon_{tot} > \varepsilon_C$ in a central volume, which shows that a small fraction of QGP is expected even at such modest energies. 

\begin{figure}[t!]
    \centering
\includegraphics[width=0.9\linewidth]{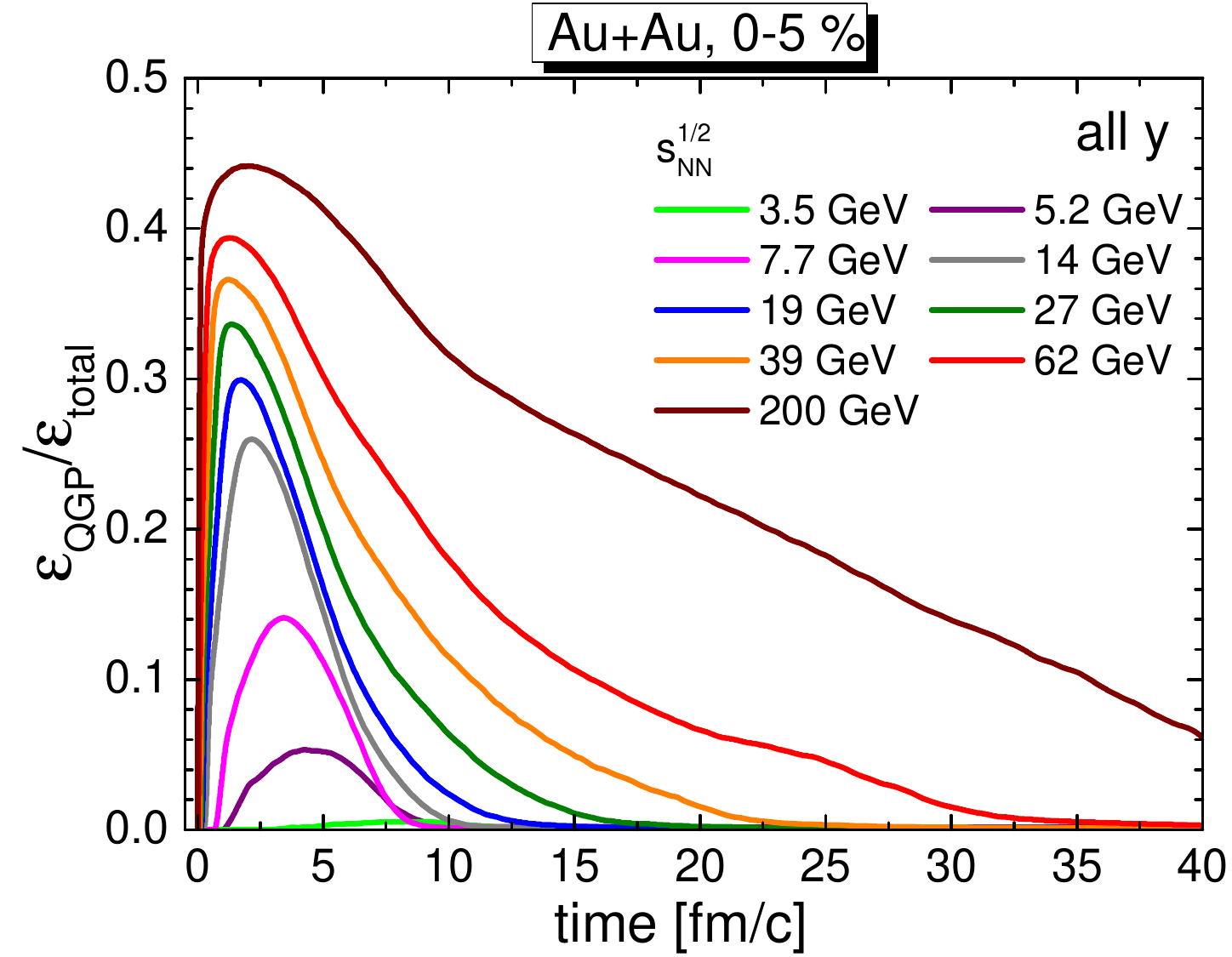}
\includegraphics[width=0.9\linewidth]{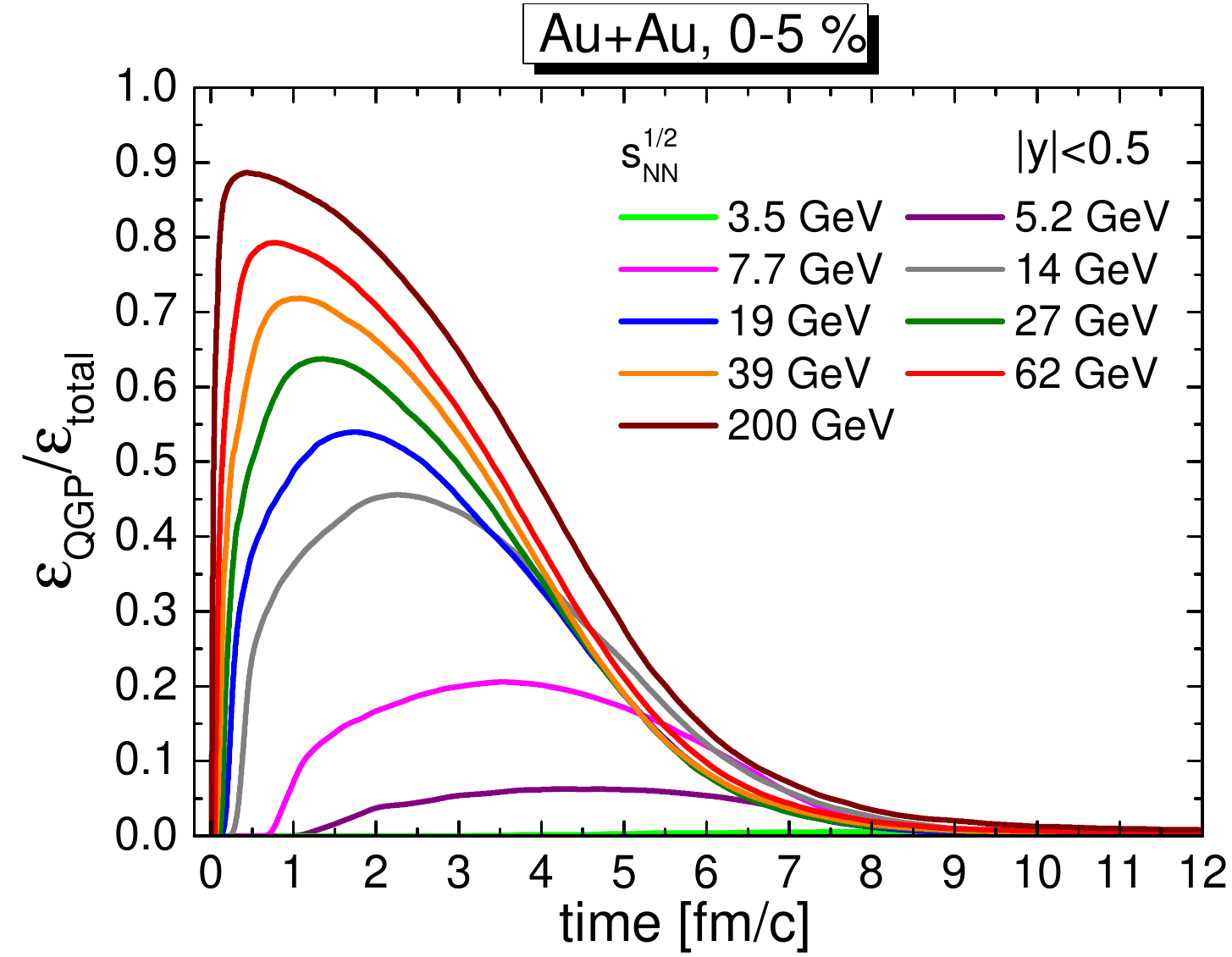}
\caption{ The fraction of energy in the QGP phase as a function of time $t$ for central Au+Au collisions (impact parameter $b = 2.25$ fm) at various collision energies $\sqrt{s_{NN}}$ from 3.5 to 200 GeV. The top panel considers all rapidities, while the bottom panel focuses on the midrapidity range $|y| < 0.5$. 
   }
    \label{Fraction_QGP}
\end{figure}
In order to quantify this observation we show  in Fig. \ref{Fraction_QGP}  the fraction of energy in the QGP phase as a function of time $t$ for central Au+Au collisions (for the impact parameter $b = 2.25$ fm, i.e.  approximately to 0-5\% centrality)  at various collision energies $\sqrt{s_{NN}}$ from 3.5 to 200 GeV. The top panel considers all rapidities, while the bottom panel shows only the midrapidity range $|y| < 0.5$.  
One can see that the QGP dominates the early time evolution of high energy collisions such that up to 90\% of matter at midrapidity and about 50\% for all $y$ (covering the whole system) at $\sqrt{s_{NN}}=200$ GeV is in a partonic phase. The QGP fraction decreases with decreasing collision energy. This is consistent with early PHSD estimates \cite{Cassing:2009vt, Moreau:2021clr} for selected energies. One can see also that the QGP fraction for the case "all $y$" (upper plot) exists for a long time due to the high energy density in the expanding target/projectile participant regions (as seen from Fig. \ref{3D-En-200GeV}). Since in most of experiments only the midrapidity region is measured,  the high density regions at projectile/target rapidity practically do not influence the observables at midrapidity. We note that even for the lowest energy of $\sqrt{s_{NN}}$ from 3.5 GeV (green lines), a very small fraction of QGP ($<1$\% of total energy) exists, too. 

\begin{figure*}[h!]
    \centering
\includegraphics[width=0.35\linewidth]{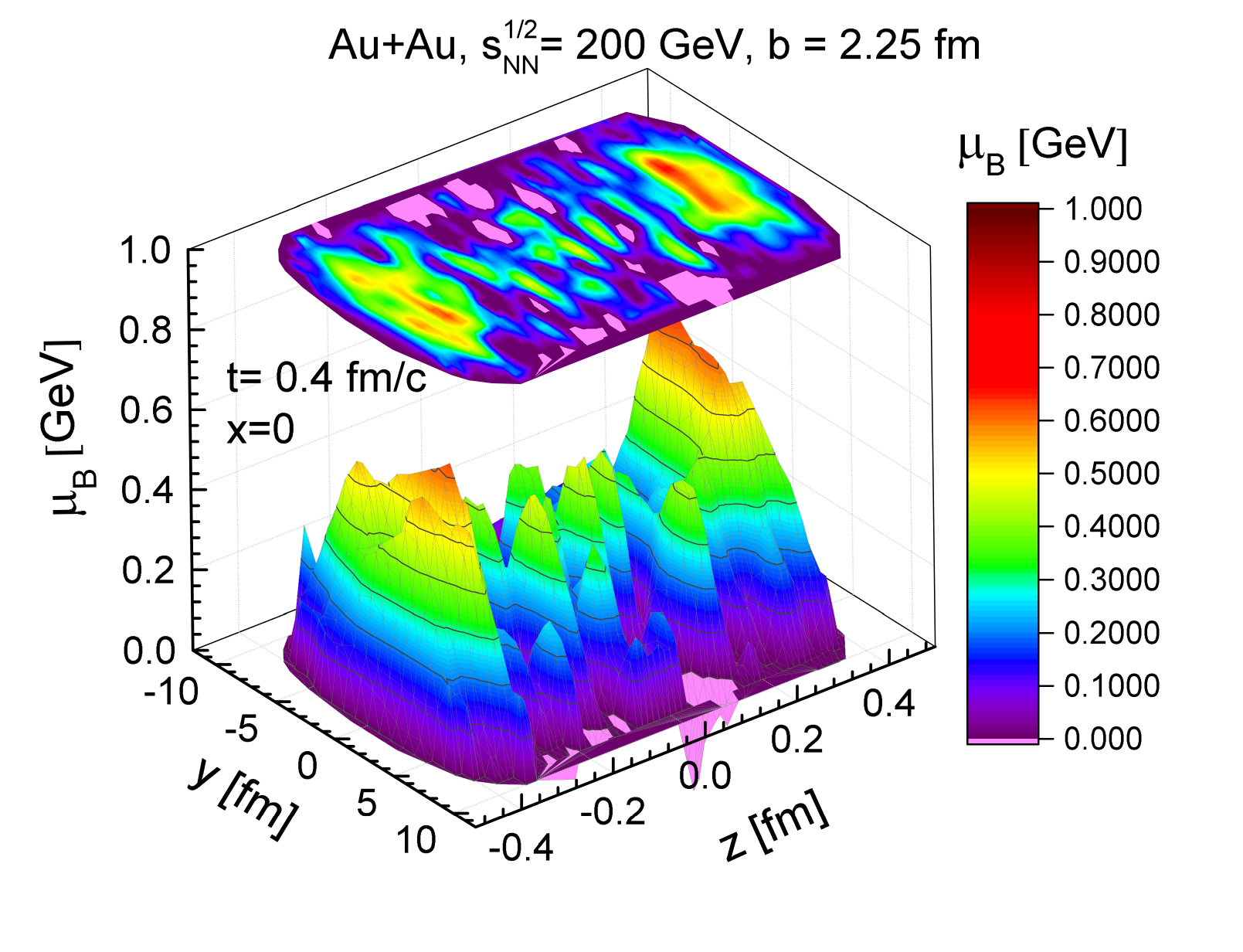}
\includegraphics[width=0.35\linewidth]{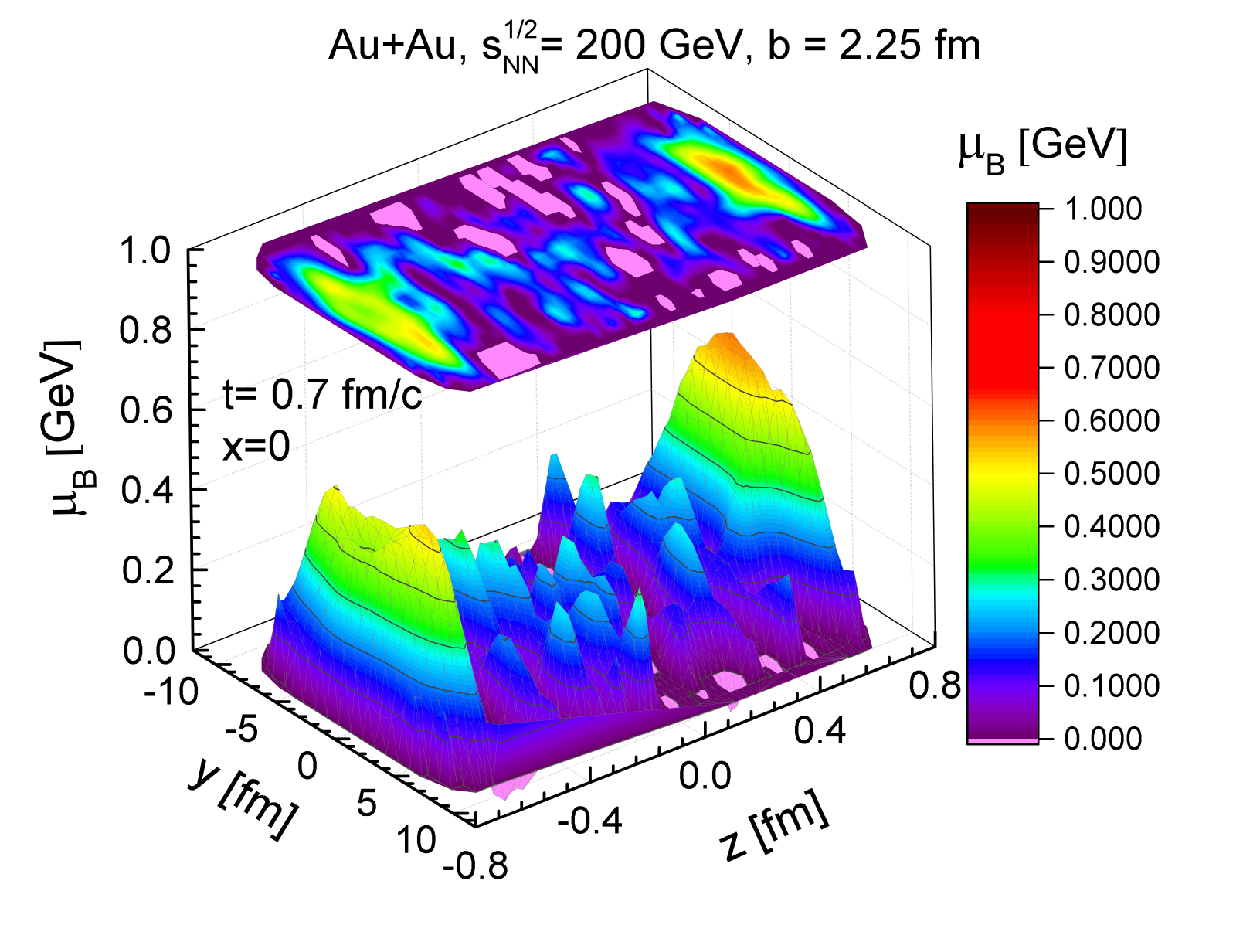}
\includegraphics[width=0.35\linewidth]{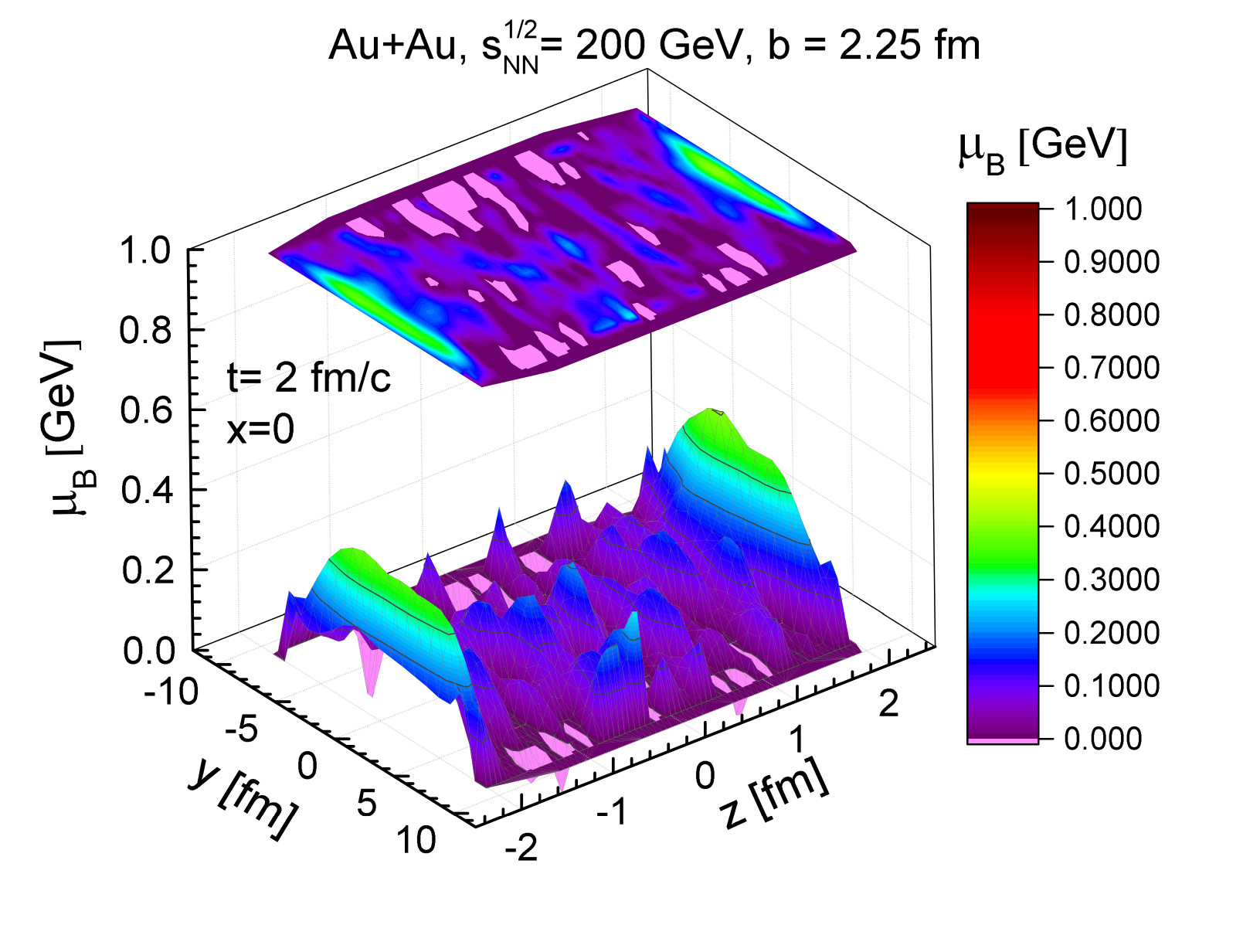}
\includegraphics[width=0.35\linewidth]{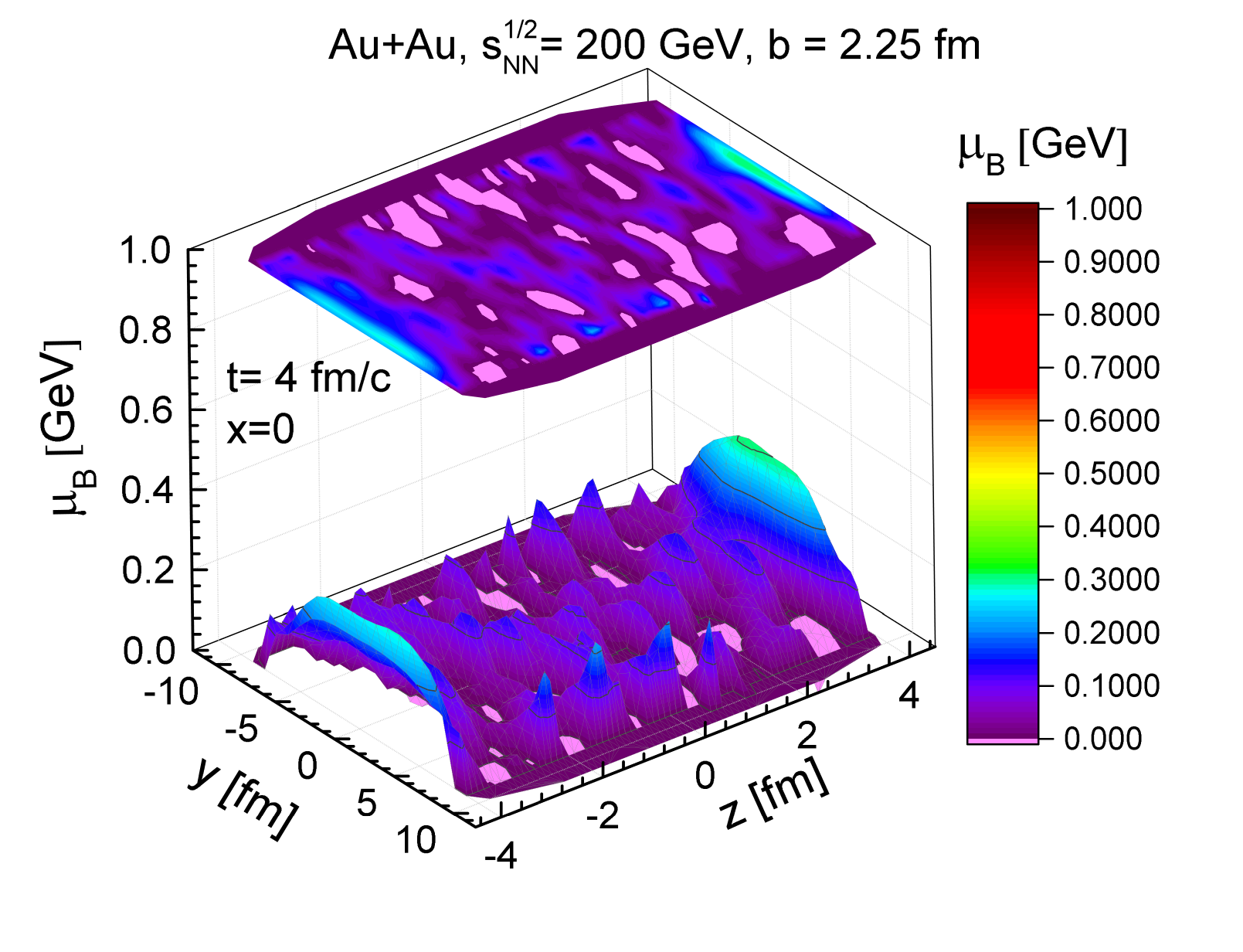}
    \caption{Baryon chemical potential $\mu_B$ in the $(y,z)$ plane for $x=0$ for central ($b=2.25$ fm) Au+Au collisions at $\sqrt{s_{NN}}=200 $ GeV  for times $t=0.4, \ 0.7, \ 2$ and 4 fm/c. The negative  $\mu_B$ is shown by pink color. Note that the scale in the $z$-axis is different for different times.}
    \label{3D-En-200GeV_muB}
\end{figure*}

\begin{figure*}[h!]
    \centering
   \includegraphics[width=0.35\linewidth]{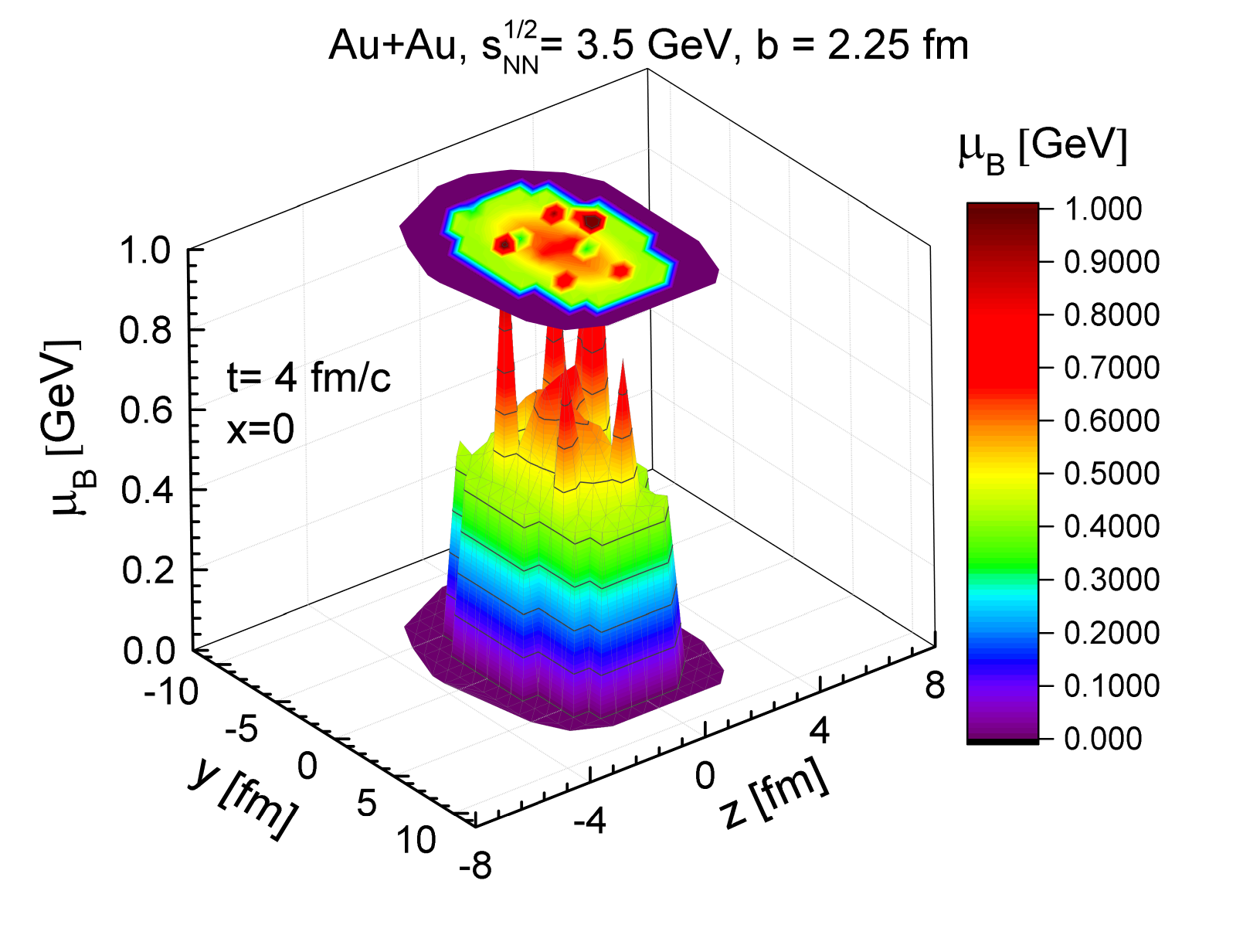}
    \includegraphics[width=0.35\linewidth]{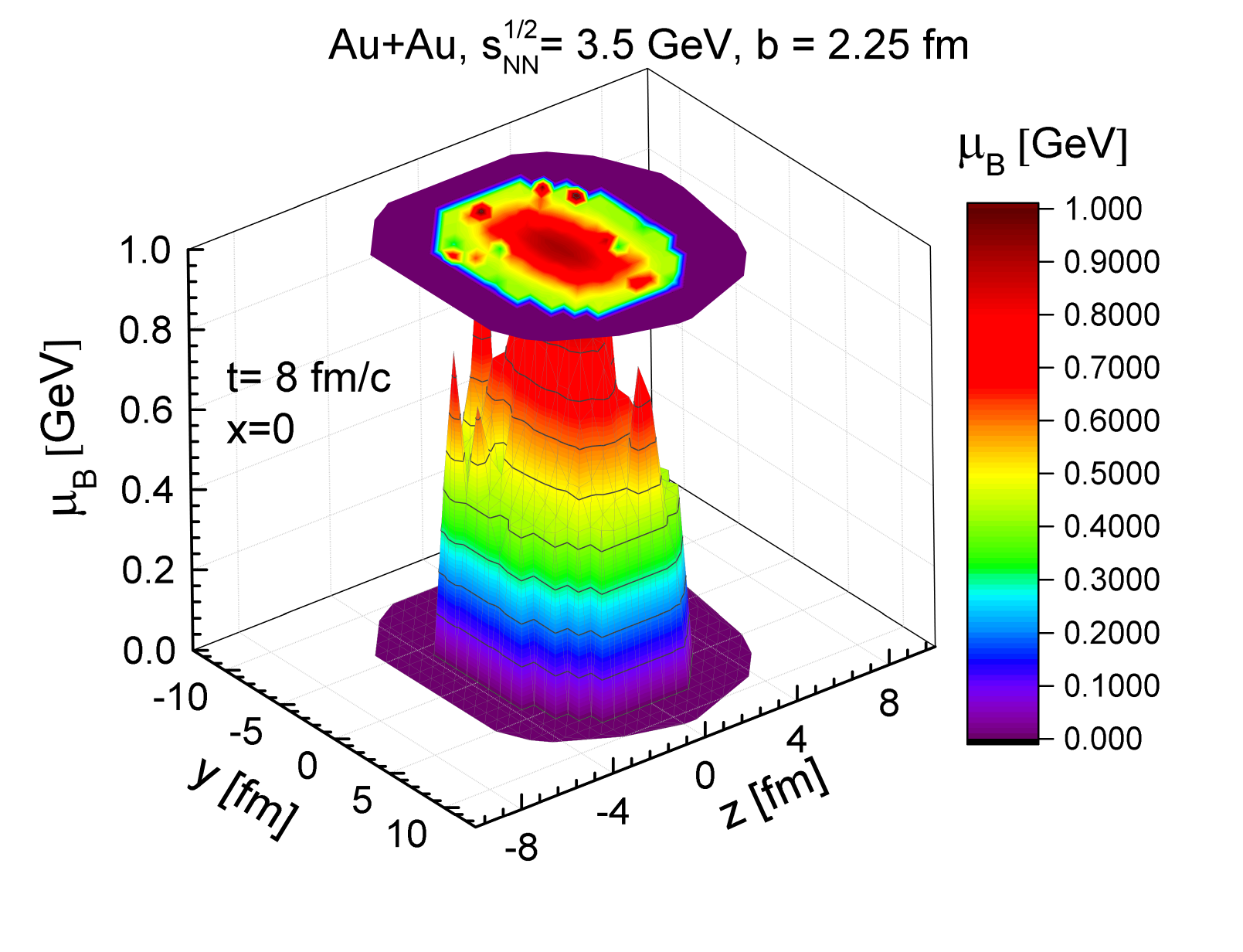}
     \includegraphics[width=0.35\linewidth]{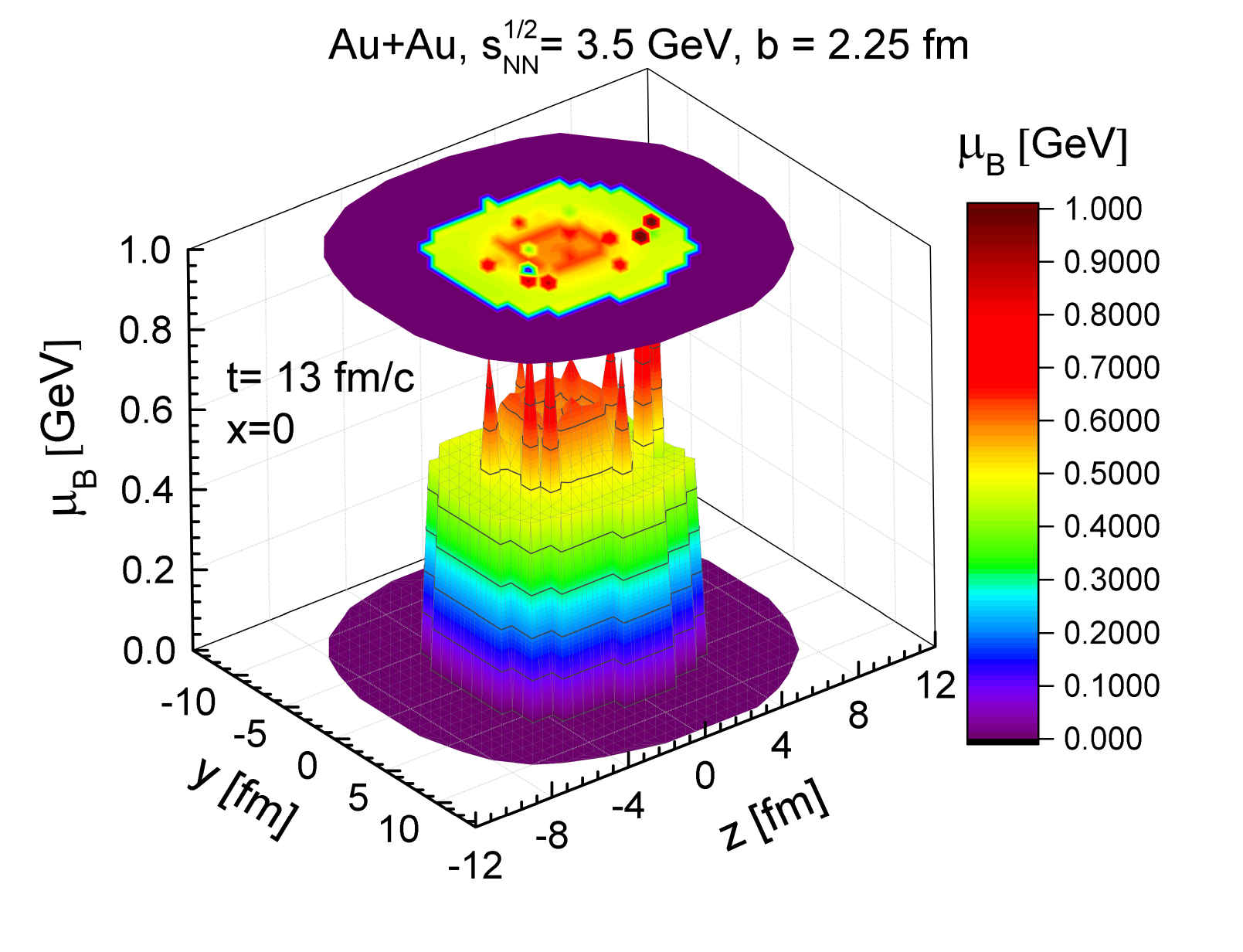}
      \includegraphics[width=0.35\linewidth]{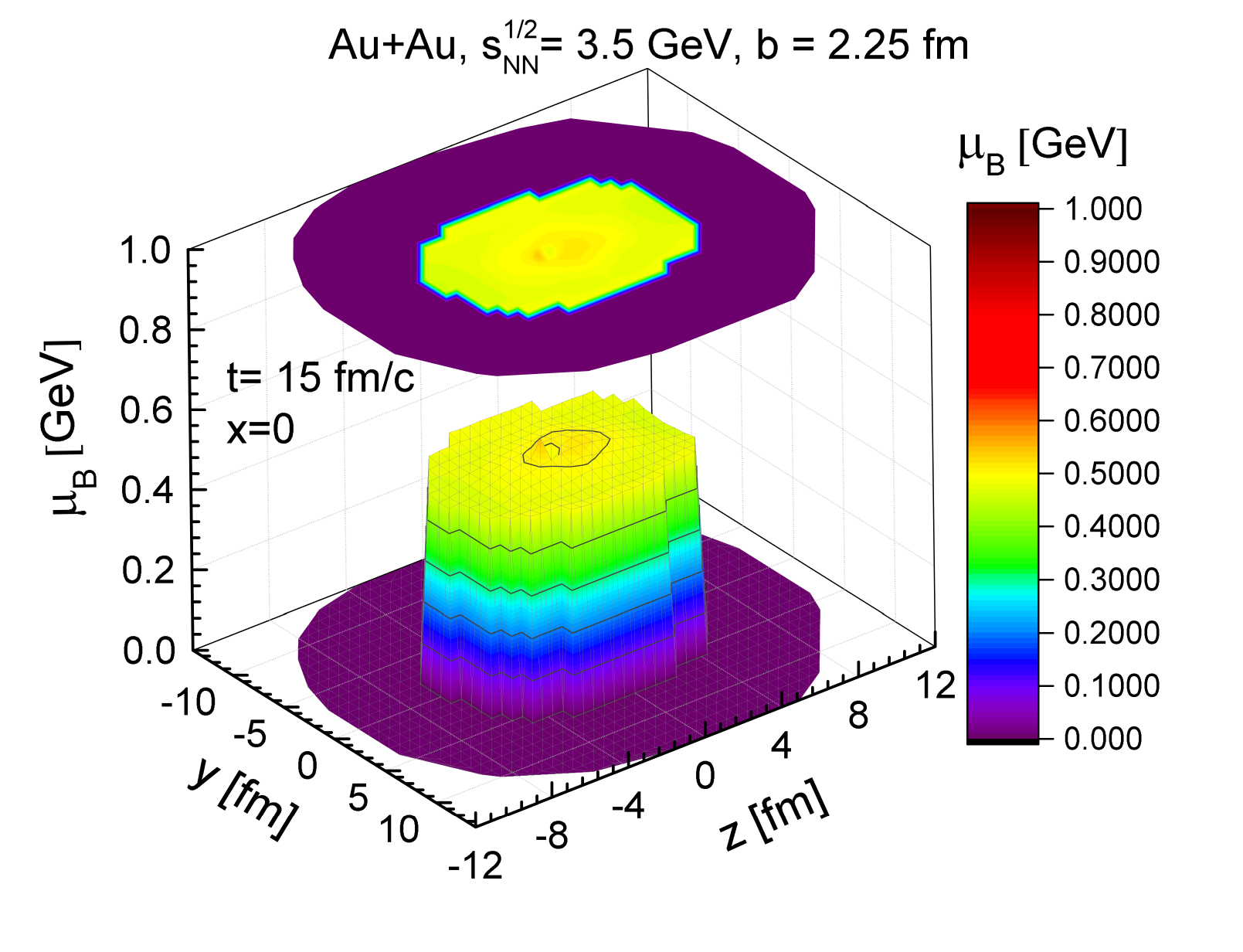}
    \caption{Baryon chemical potential  $\mu_B$ in the $(y,z)$ plane for $x=0$ for central ($b=2.25$ fm) Au+Au collisions at $\sqrt{s_{NN}}=3.5 $ GeV  for the times $t=4, \ 8, \ 13$ and 15 fm/c.
    }
    \label{3D-En-35GeV_muB}
\end{figure*}


In Fig. \ref{3D-En-200GeV_muB} we illustrate the  baryon chemical potential  $\mu_B$ in the  $(y,z)$ plane for $x=0$ for central ($b=2.25$ fm) Au+Au collisions at $\sqrt{s_{NN}}=200 $ GeV  for times $t=0.4, \ 0.7, \ 2$ and 4 fm/c; the resulting distribution is averaged over 30 events.
One can see that in the central region $\mu_B$ decreases fast due to the expansion and it fluctuates around zero. It can became negative in some cells where more antibaryons/antiquarks are located due to fluctuations.

Contrary, by  lowering the bombarding energy to $\sqrt{s_{NN}}=3.5 $ GeV  the baryon chemical potential becomes large and positive (since the energy is not sufficient for $B \bar B$ production) -
see Fig. \ref{3D-En-35GeV_muB}. The time evolution is much slower, such that in the central region the baryon chemical potential reminds large for a long time. This means that in spite that the QGP fraction at low energies is very small, we probe the QGP EoS at finite $\mu_B$ by dileptons emitted from such "hot spots/droplets".

\section{Influence of the  $(T,\mu_B)$-dependent EoS on dilepton production }\label{sec3d}

Here we study how the inclusion of the $\mu_B$- dependence of the QGP equation-of-state affects the dilepton production from partonic sources.
For that purpose we compare 2 scenarios:
\begin{enumerate}
\item  "DQPM $(T, \mu_B=0)$": the QGP thermodynamics is described based on the DQPM with $T$-dependent quasiparticle  properties -- masses and widths, strong coupling, partonic interaction cross sections as well as dilepton production cross section from QGP sources. This scenario corresponds to the lQCD EoS for $\mu_B=0$, which has been explored early for dilepton production  within the PHSD  in Refs. \cite{Linnyk:2015rco,Song:2018xca}, however, here we use the novel DQPM model based on updated lQCD data \cite{Aoki:2009sc, Cheng:2007jq} which indicate $\varepsilon_C \simeq 0.4$ GeV/fm$^3$ instead of 0.5  GeV/fm$^3$ as used in the early PHSD calculations.
\item "DQPM $(T, \mu_B)$": the QGP thermodynamics is based on the DQPM with the $(T, \mu_B)$ dependent EoS - consistent with the lQCD data \cite{Aoki:2009sc, Cheng:2007jq} - interpreted in terms of quasiparticles with $(T, \mu_B)$- dependent masses and widths, strong coupling, partonic interaction cross sections \cite{Moreau:2019vhw, Soloveva:2019xph} as well as $(T, \mu_B)$- dependent dilepton production cross sections from QGP channels. 
This scenario has not been tested yet for dilepton production which is a goal of the present study.
\end{enumerate}

We recall that the DQPM transport coefficients show a visible but smooth dependence on $\mu_B$ (cf. Refs. \cite{Moreau:2019vhw, Soloveva:2019xph, Soloveva:2020ozg, Fotakis:2021diq, Soloveva:2021quj}).
In particular - as shown in Figs. 9 and 10 of Ref. \cite{Soloveva:2021quj} - 
the shape of the DQPM speed of sound $c_s(T)$ (which reproduces the lQCD data \cite{Borsanyi:2013bia,HotQCD:2014kol}
for $\mu_B=0$) moves towards lower $T$ with increasing $\mu_B$. This reflects the modification of the QGP dynamics for decreasing collisional energy, where the $\mu_B$ dependence becomes more pronounced.
It is visible for example in the modification of flow coefficients for strange mesons and antibaryons as shown in Ref. \cite{Soloveva:2020ozg}.

\begin{figure}[h!]
    \centering
\includegraphics[width=0.90\linewidth]{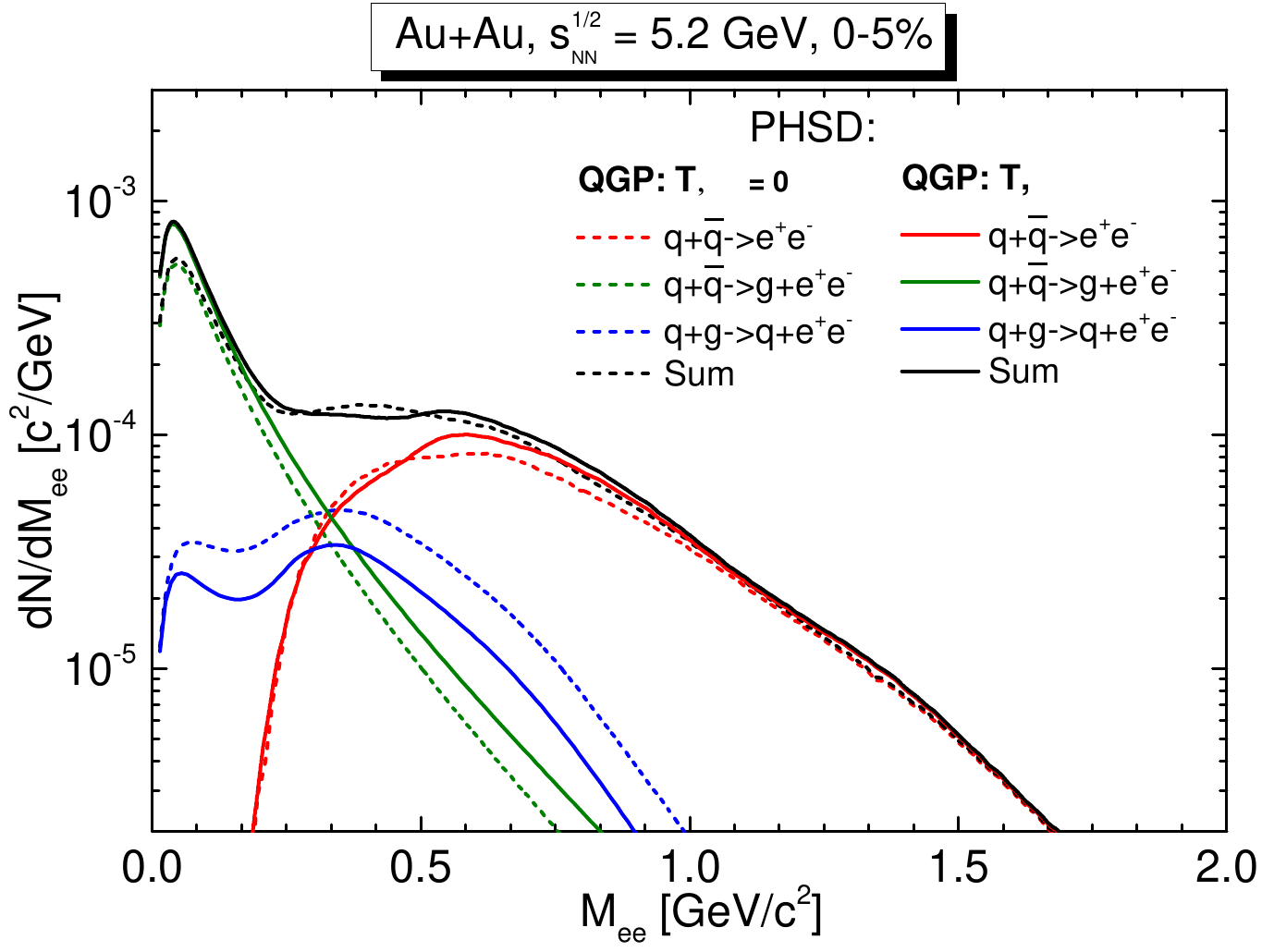}
\includegraphics[width=0.90\linewidth]{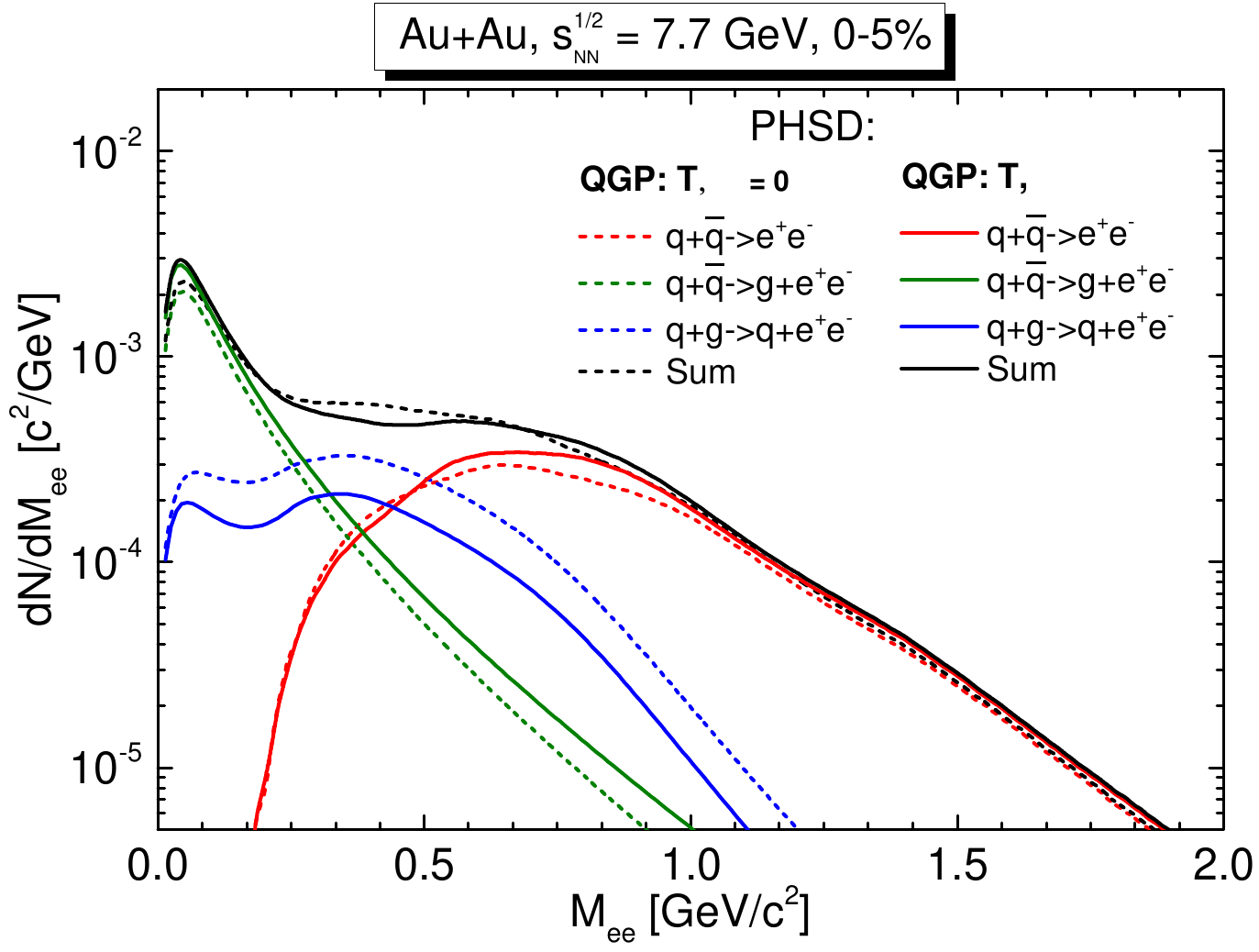}
\includegraphics[width=0.90\linewidth]{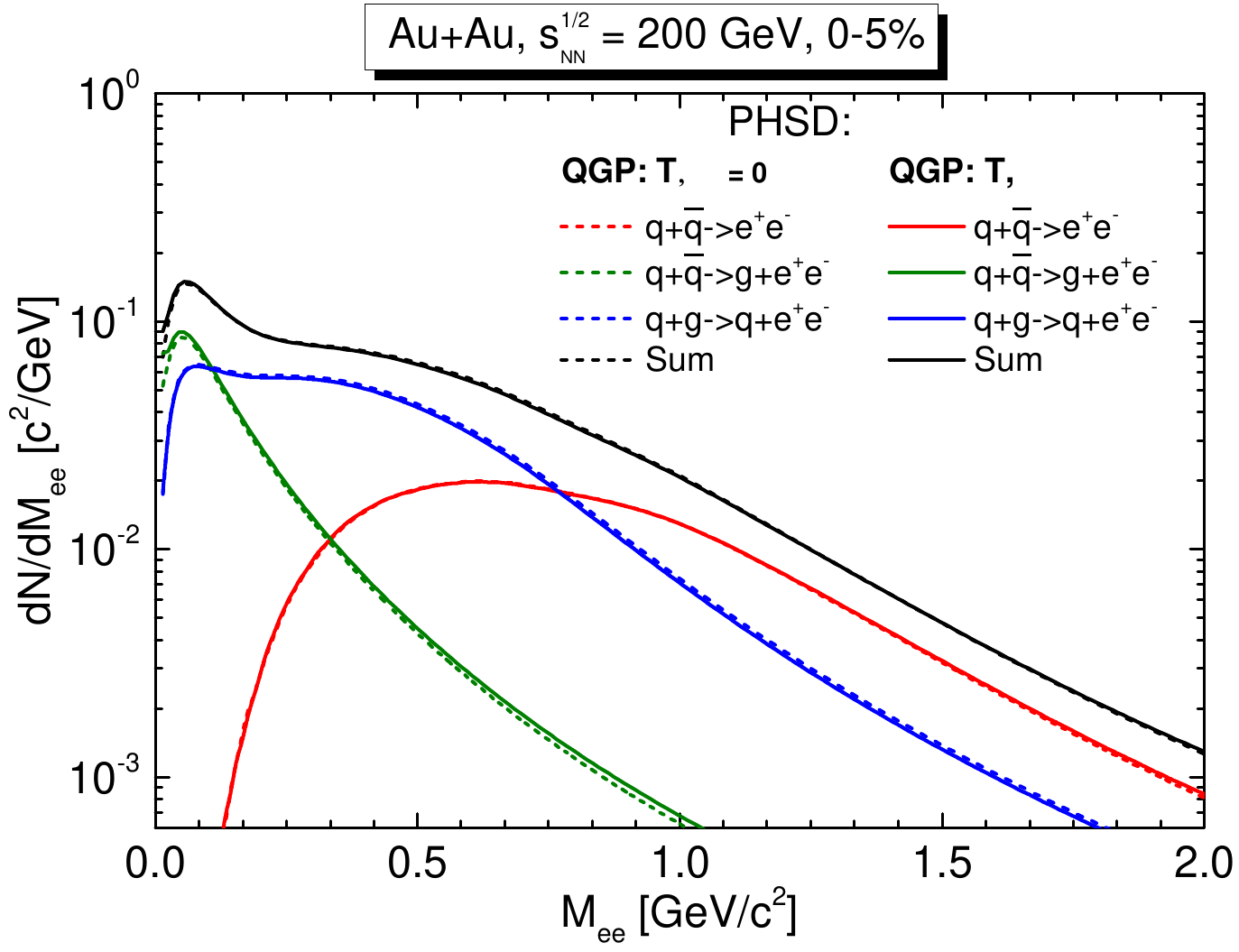}
    \caption{Invariant mass spectra of dileptons from QGP channels for 0-5 \% central Au+Au collisions at $\sqrt{s_{NN}}=5.2$ (upper),  7.7 (midle) and 200 GeV (lower). The color lines show the individual channels: red --  $q\bar{q} \to e^+ e^-$, green -- $q\bar{q} \to g  e^+ e^-$,  blue --  the sum of Compton scatterings: $qg \to q e^+ e^-, \ \ \bar{q}g \to \bar{q} e^+ e^-$, black -- sum of all QGP channels. 
    The solid lines display QGP dilepton yields calculated  by accounting for the $(T,\mu_B)$  dependence of parton quasiparticle properties and cross sections while the dashed lines display the results including only the $T$-dependence (calculated for $\mu_B=0$). 
   }
    \label{mu_temperature_AuAu}
\end{figure}


Fig.~\ref{mu_temperature_AuAu} shows the invariant mass spectra of dileptons (in $4\pi$) originating from partonic  channels in Au+Au collisions at 3 energies
$ \sqrt{s_{NN}} = 5.2, \ 7.7 \, \mathrm{GeV} $  and $ \sqrt{s_{NN}} = 200 \, \mathrm{GeV}$. 
The individual QGP channels are shown by different color lines:  red --  $q\bar{q} \to e^+ e^-$, green -- $q\bar{q} \to g  e^+ e^-$,  blue --  the sum of Compton scatterings: $qg \to q e^+ e^-, \ \ \bar{q}g \to \bar{q} e^+ e^-$, black -- sum of all QGP channels. 
The solid lines display QGP dilepton yields calculated  for the "DQPM $(T, \mu)$" scenario, i.e. by accounting for the $(T,\mu_B)-$  dependence of the parton quasiparticle properties (i.e. masses and widths as well as strong coupling) and their interaction cross sections and dilepton emission cross sections while the dashed lines display the results for the "DQPM $(T, \mu_B=0)$" scenario.

As illustrated in Fig. \ref{3D-En-200GeV_muB}  at higher energy $(\sqrt{s_{NN}} = 200 \, \mathrm{GeV})$, the baryon chemical potential approaches zero $(\mu_B \approx 0)$ in the central region due to an almost symmetric production of particles and antiparticles in the system. Consequently, the dilepton spectra exhibit a minimal variation between the two scenarios (with and without $\mu_B$) - cf. lower plot in Fig.~\ref{mu_temperature_AuAu}. In this high energy regime the cross-sections are primarily governed by the temperature dependence. This is consistent with the findings in the previous PHSD study in Ref. \cite{Moreau:2019vhw} (see Fig. 15).

In contrast, with decreasing energy  $(\sqrt{s_{NN}} = 5.2, \ 7.7 \, \mathrm{GeV})$ the influence of the baryon chemical potential increases, however, the total QGP dilepton yield is only slightly modified.  In particular, the dilepton yield from quark-gluon Compton scattering  $q g \to g + e^+e^-$ is reduced, while $q\bar q$ annihilation processes are rather similar for both cases. This is due to the fact that the $\mu_B$-dependent EoS (i.e. "DQPM $(T, \mu)$") changes the "chemical decomposition" of QGP partons ($q, \bar q, g$) as well as their properties and interactions. 
The gluons are dominantly produced by quark-antiquark fusion ($q\bar q \to g$) during the QGP phase.  The $(T, \mu_B)$- dependent EoS leads to a reduction of gluons in the system compared to the $\mu_B=0$ case.  This is a consequence of the reduction of parton masses with increasing $\mu_B$ which lead to the lowering of the invariant energy $\sqrt{s}$ of $q\bar q$ collisions. Moreover, the strong coupling $g(T,\mu_B)$ is also reduced with $\mu_B$. Consequently, the cross section for $q\bar q \to g$ processes is reduced, too.
Due to that the dilepton production from $qg, \ \bar qg$ scattering is reduced at low $\sqrt{s_{NN}}$. On the other hand,  the contribution of $q g, \ \bar q g$ channels to the total dilepton yield is decreasing with decreasing $\sqrt{s_{NN}}$. The dilepton channel - $q \bar q \to g e^+e^-$  - is only slightly enhanced at larger $M_{ee}$ due to a decrease of gluon masses. Consequently,  the total QGP yield changes very slightly even at low bombarding energies where $\mu_B$ is large.
Thus, the total sum of all partonic channels shows only a weak dependence on $\mu_B$ at all $\sqrt{s_{NN}}$.

\begin{figure*}[ht!]
    \centering
\includegraphics[width=0.329\linewidth]{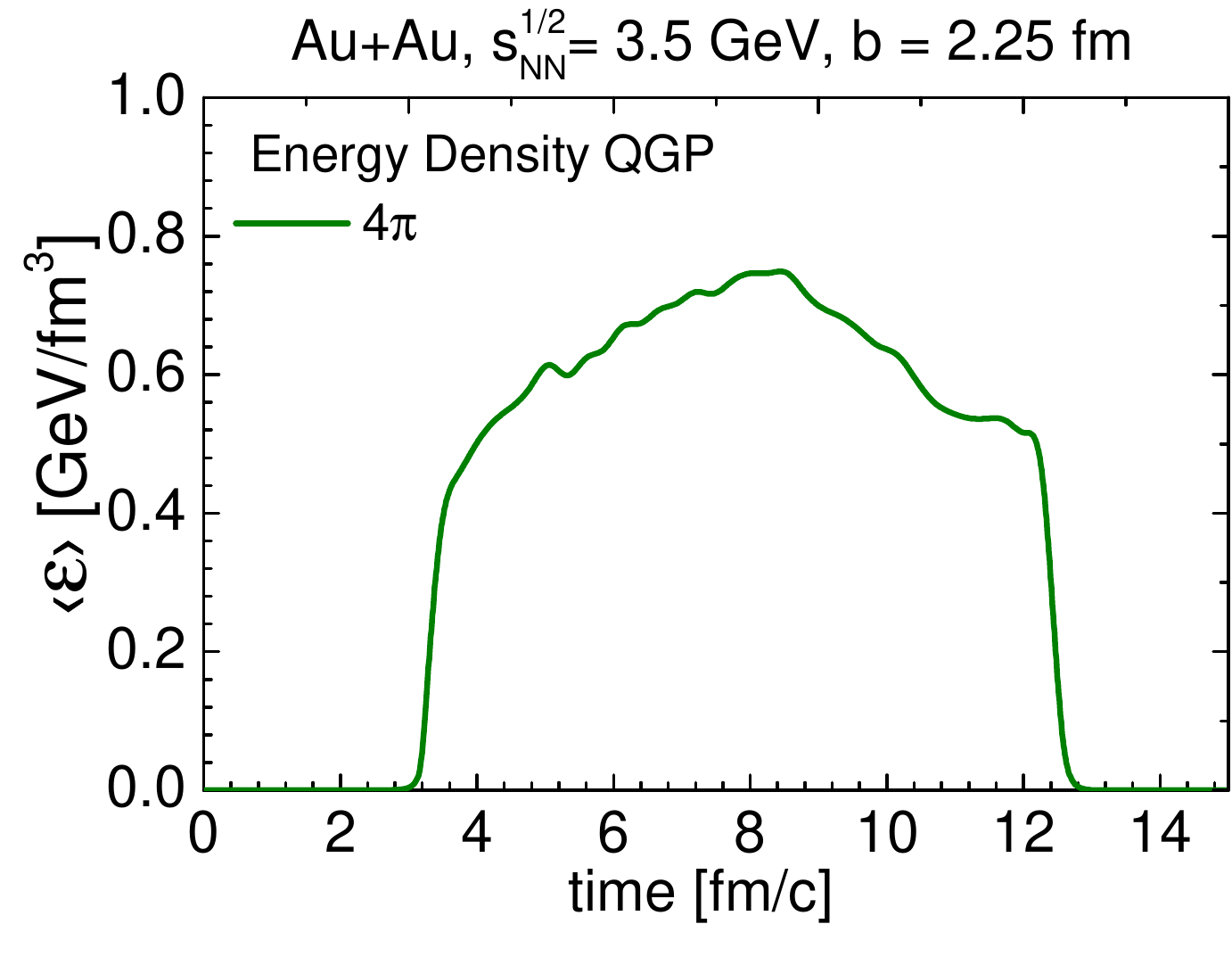}
\includegraphics[width=0.329\linewidth]{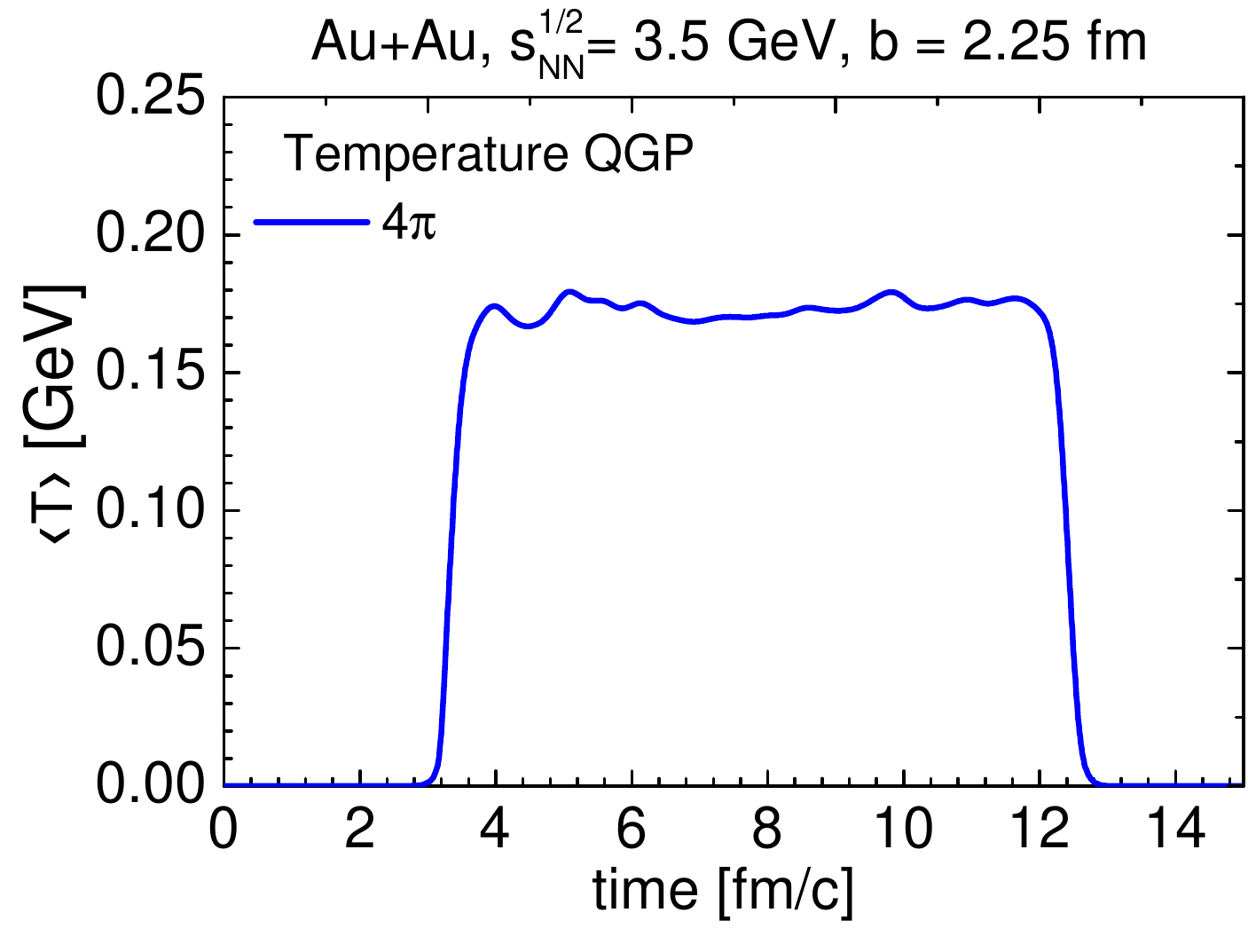}
\includegraphics[width=0.329\linewidth]{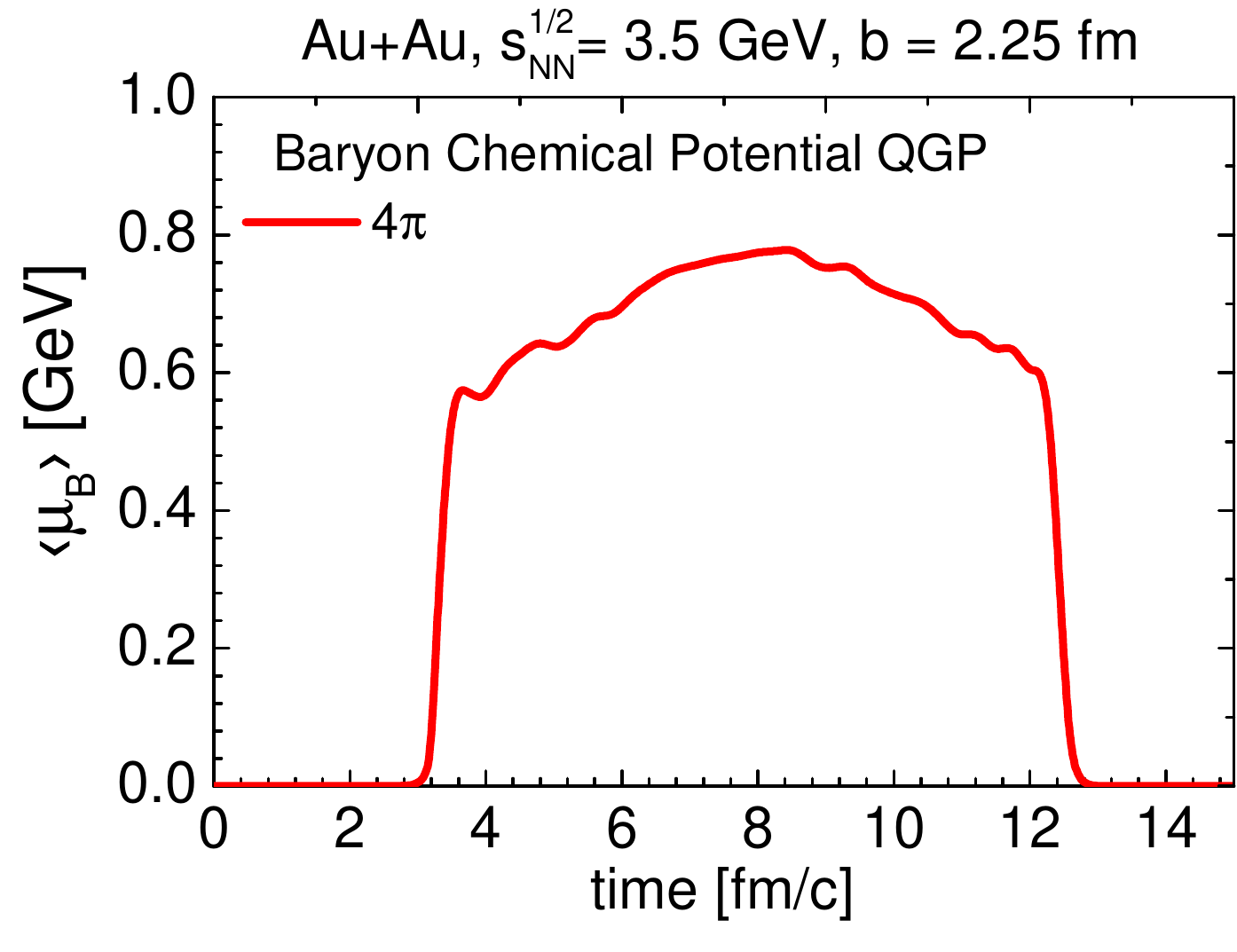} \\
\includegraphics[width=0.329\linewidth]{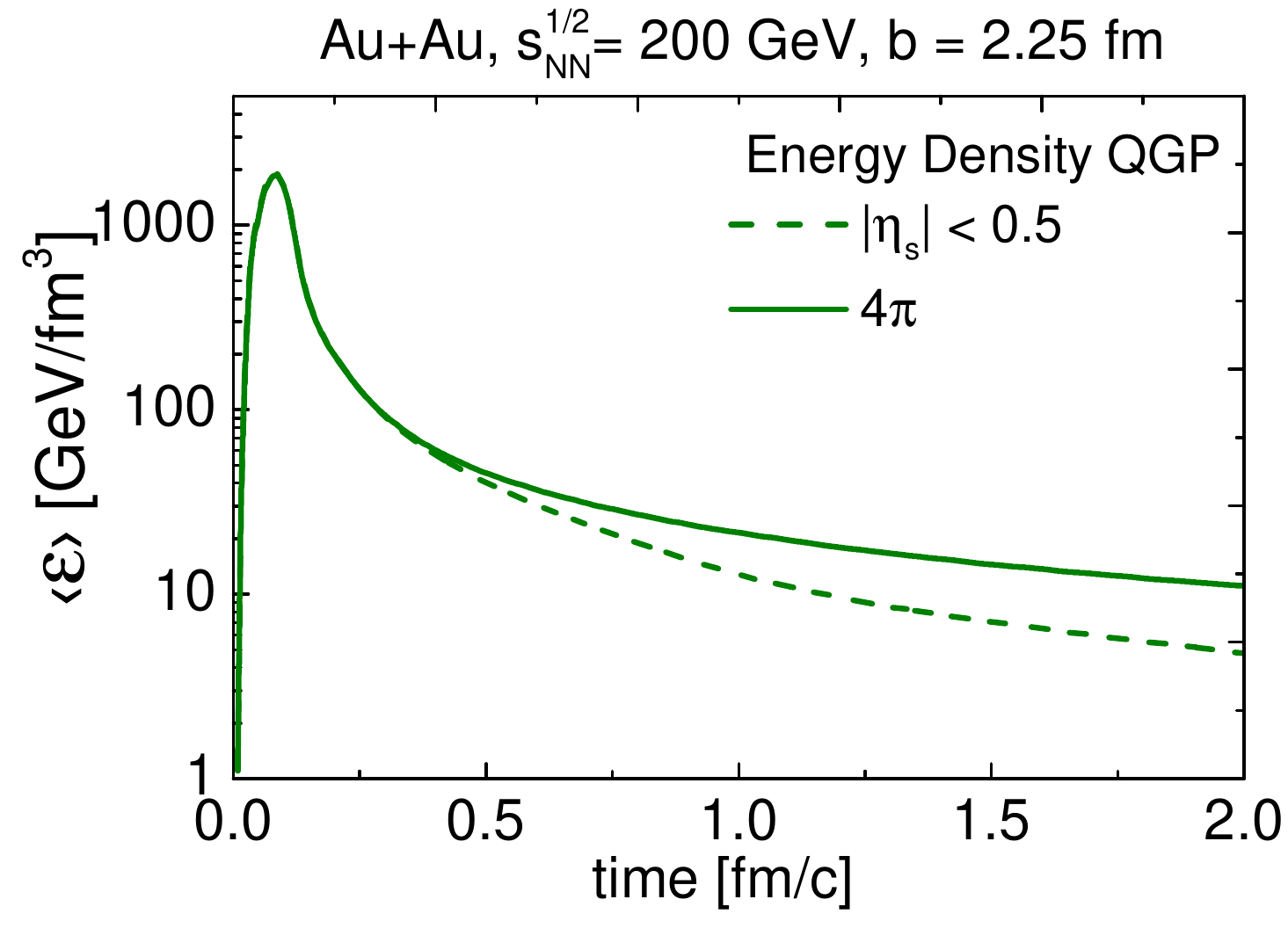}
\includegraphics[width=0.329\linewidth]{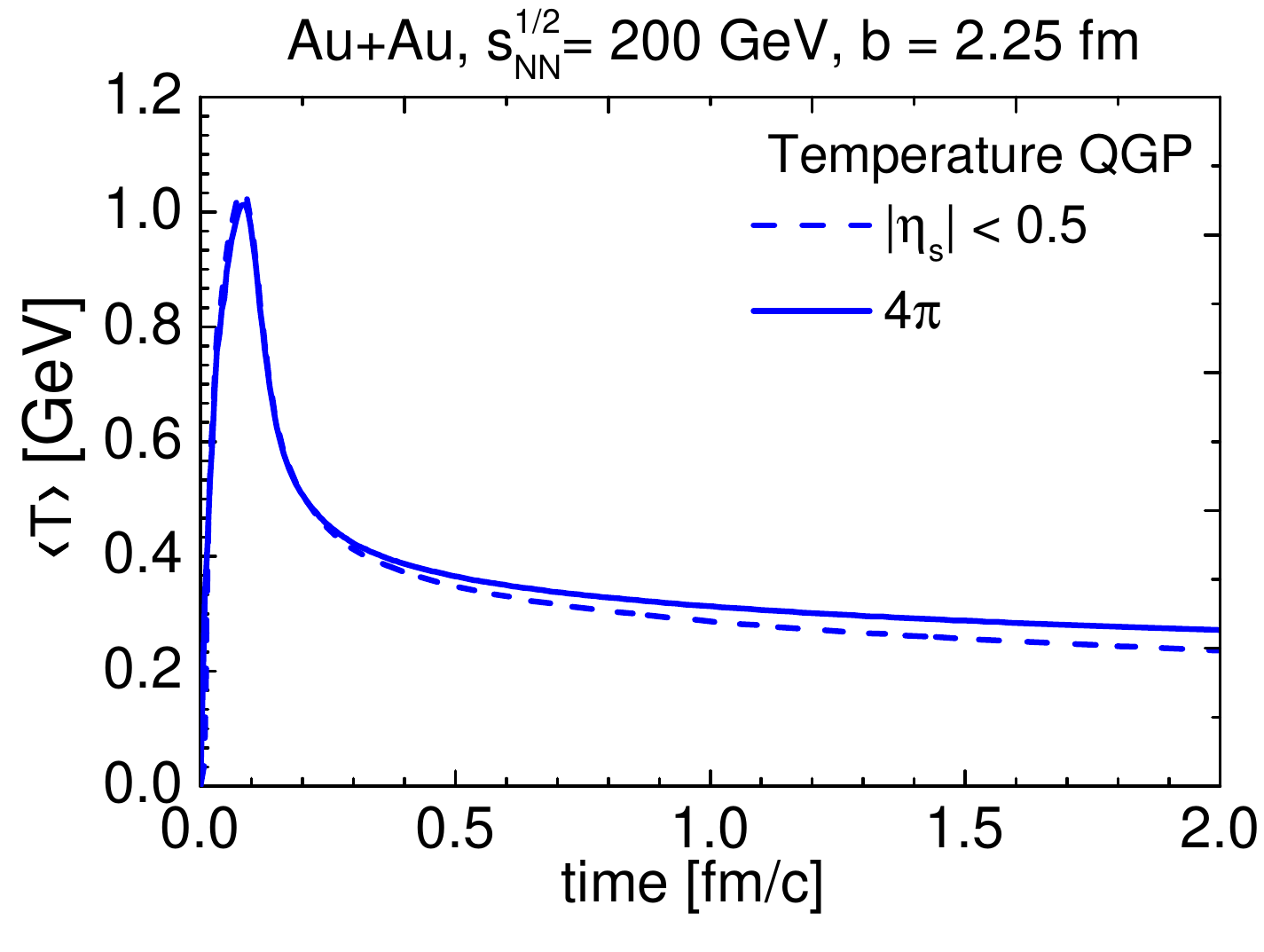}
\includegraphics[width=0.329\linewidth]{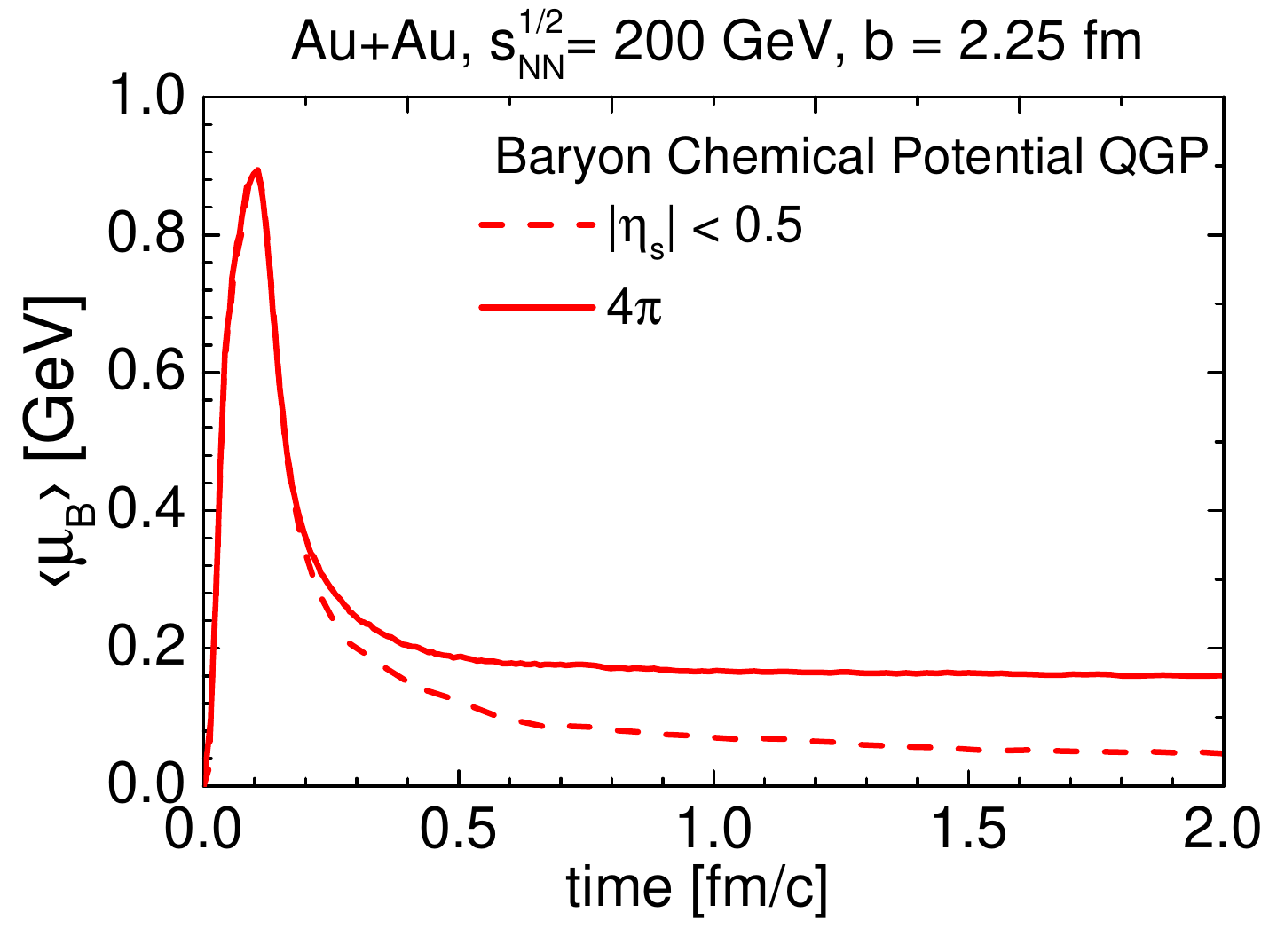}
    \caption{Averaged energy density (left column), temperature (middle column) and baryon chemical potential (right column) from QGP  versus time for central ($b=2.25$ fm) Au+Au collisions at $\sqrt{s_{NN}}=3.5 $ GeV (upper panel) and 200 GeV (lower panel). The solid lines correspond to the full space while the dashed lines stand for the spacial rapidity $|\eta_s|<0.5$.  }
    \label{3D-En-200GeV_AveTmuB}
\end{figure*}
 While Figs. \ref{3D-En-200GeV}, \ref{3D-En-35GeV}, \ref{3D-En-200GeV_muB}, \ref {3D-En-35GeV_muB}  show the time evolution of energy densities and $\mu_B$ in the whole system, it is interesting to  illustrate which QGP matter we probe by dileptons emitted from partonic processes. For that purpose  we present in  Fig. \ref{3D-En-200GeV_AveTmuB} the averaged energy density $\langle \varepsilon \rangle$ (left column), temperature $\langle T \rangle$ (middle column) and baryon chemical potential $\langle \mu_B\rangle$ (right column) - taken at each production point of QGP dileptons -  versus time for central ($b=2.25$ fm) Au+Au collisions at $\sqrt{s_{NN}}=3.5 $ GeV (upper panel) and 200 Gev (lower panel).  The solid lines correspond to the full space while the dashed lines for 200 GeV stand for the spacial rapidity region $|\eta_s|<0.5$.  For the latter case  all quantities decrease with time due to the rapid expansion of the system, i.e. the matter leaves the central zone, while for the $"4\pi"$ case (i.e. integrating over the whole space) the  $\langle \varepsilon \rangle$ and $\langle T \rangle$ decrease only slightly due to the inclusion of the expanding matter towards target/projectile regions. For the same reason the  $\langle \mu_B\rangle$ stays approximately constant for the considered time interval.  As seen from the upper row of Fig. \ref{3D-En-200GeV_AveTmuB}, at 3.5 GeV the partonic dileptons probe the matter slightly above the critical energy density $\varepsilon_C$ and $T_C$ and at large $\mu_B$ ($> 0.6$ GeV).

\section{Comparison of the Invariant Mass Spectra of Dielectrons from the PHSD with Experimental Data}\label{sec7}

In this section we examine the invariant mass spectra of dielectrons obtained from the Parton-Hadron-String Dynamics approach and compare them with experimental data from Au+Au collisions at $\sqrt{s_{NN}} = 7.7$ to 200 GeV provided by the STAR collaboration \cite{STAR:2023wta,Han:2024nzr,STAR:2015tnn}, as well as p+p and Pb+Pb collisions from the ALICE collaboration from $\sqrt{s_{NN}} = 2.76$ to 13 TeV \cite{Gunji:2017kot,ALICE:2018fvj,ALICE:2020umb,Meninno:2020mjd,ALICE:2018ael,ALICE:2023jef,Gunji:2017kot}. Furthermore, we include  p+p and p + Nb collisions at $E_{kin}=$3.5 AGeV, as well as Ar + KCl collisions at $E_{kin}=$1.76 AGeV and Au + Au collisions at $E_{kin}=$1.23 AGeV from the HADES collaboration \cite{HADES:2012sir,HADES:2011jqb,HADES:2012sui,HADES:2011nqx,HADES:2019auv}. 

It is important to note that the experimental datasets from STAR, HADES, and ALICE differ in centrality and acceptance criteria. 
The STAR data correspond to minimum-bias Au+Au collisions with transverse momenta $p_T \geq 0.2$ GeV/c and pseudo-rapidities $|\eta_e| < 1.0$.
In contrast, the ALICE data use a wide range of cuts depending on the collision system, with electrons and positrons having transverse momenta $0.075< p^{e}_T < 10 $  GeV/c in the widest range, transverse momenta cut for the pair $e^+e^-$ of $p^{ee}_T < 10$ GeV/c and pseudo-rapidities $|\eta_e| < 0.8$. 
Furthermore, for the HADES collaboration we use the corresponding filter and mass/momentum resolution to get the right acceptance.  
We note that our calculations for the dilepton yields from vector mesons include the collisional broadening scenario. Moreover, for the estimates of the dilepton yields from the QGP channels we have included of the $(T,\mu_B)$-dependence of partonic cross sections.

\subsection{p+p, p+Nb,Ar+KCL and Au+Au collisions in HADES experiment at SIS18 energies}
\begin{figure*}[t]
    \centering
\includegraphics[width=0.49\linewidth]{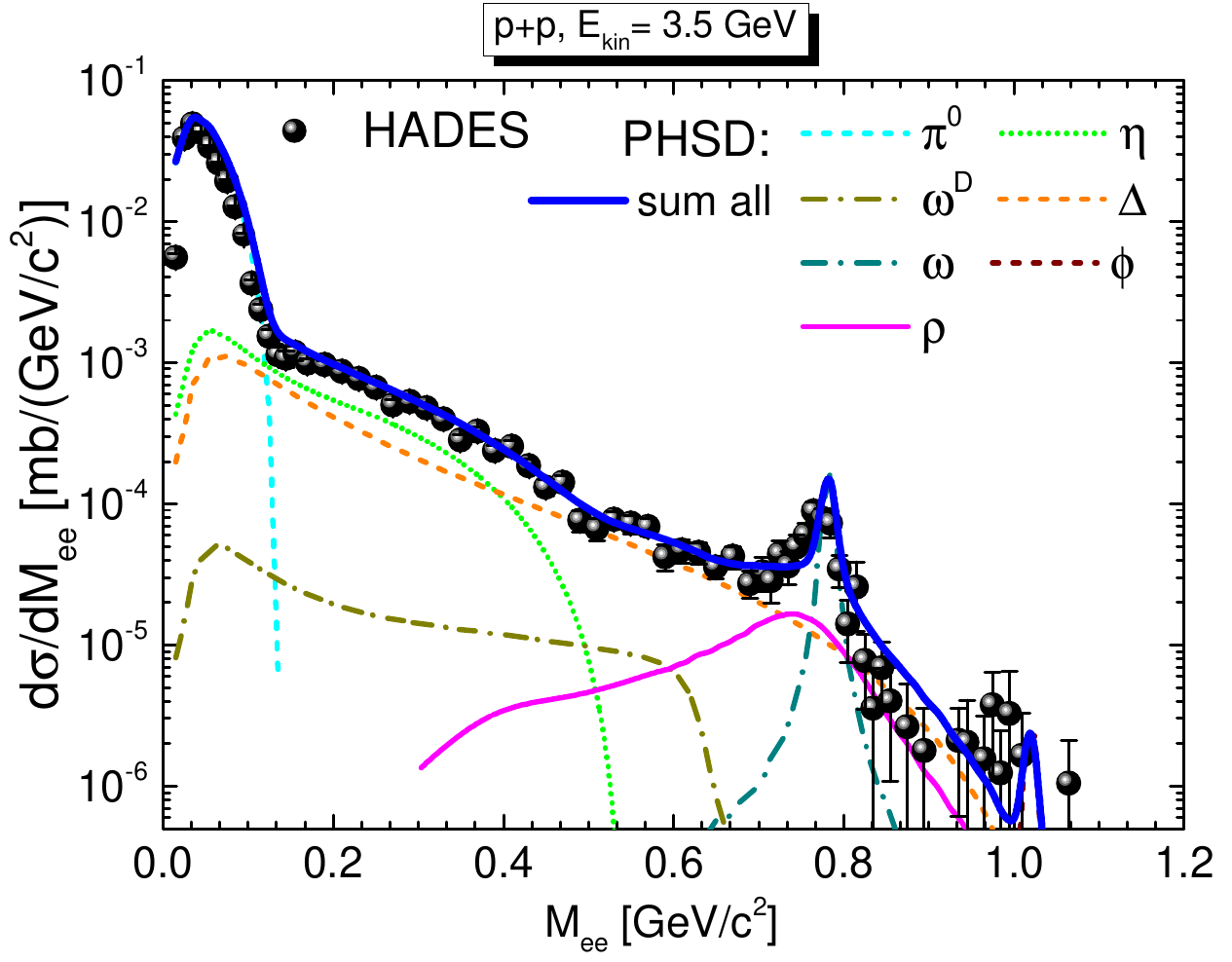} \includegraphics[width=0.49\linewidth]{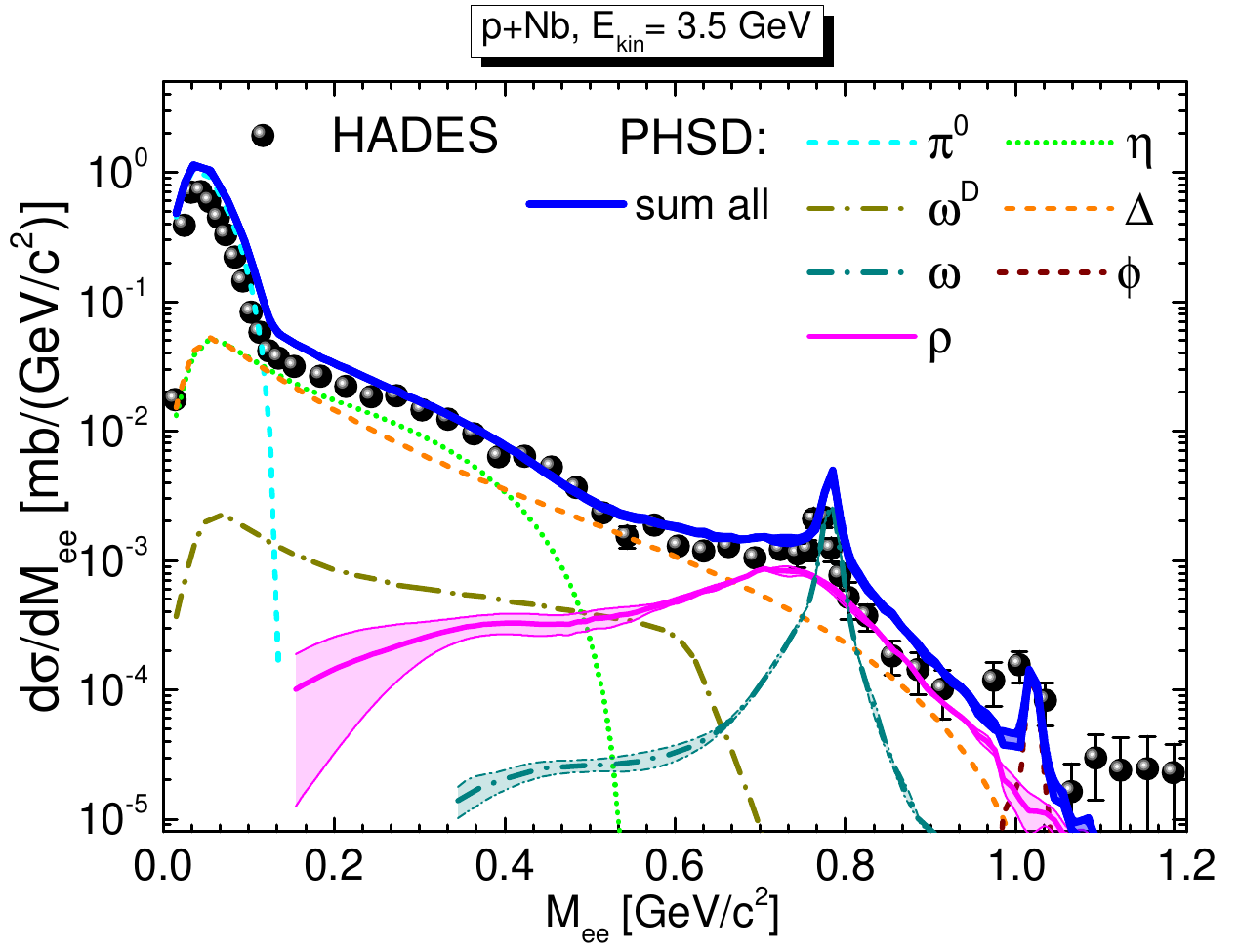}
\includegraphics[width=0.49\linewidth]{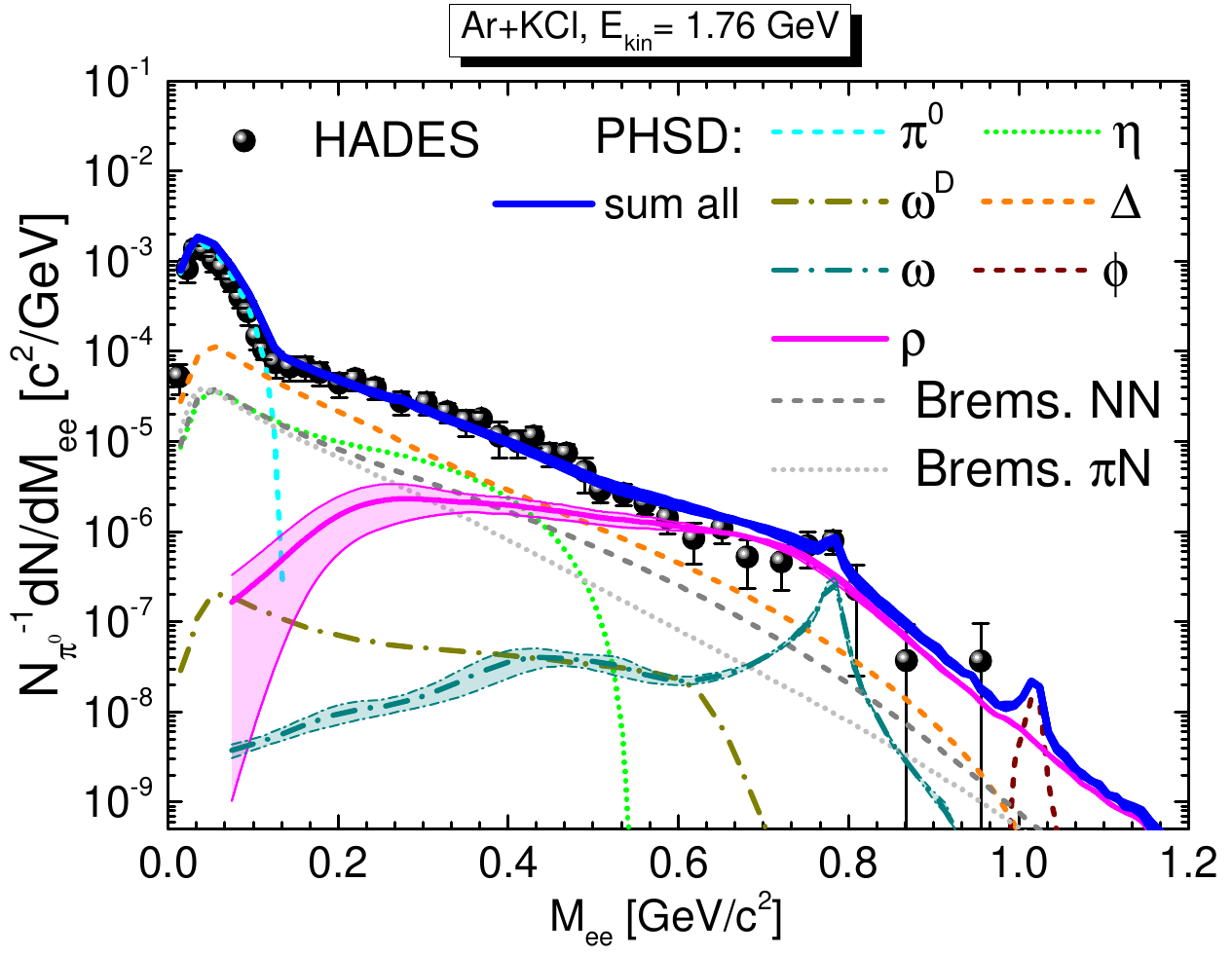}
\includegraphics[width=0.49\linewidth]{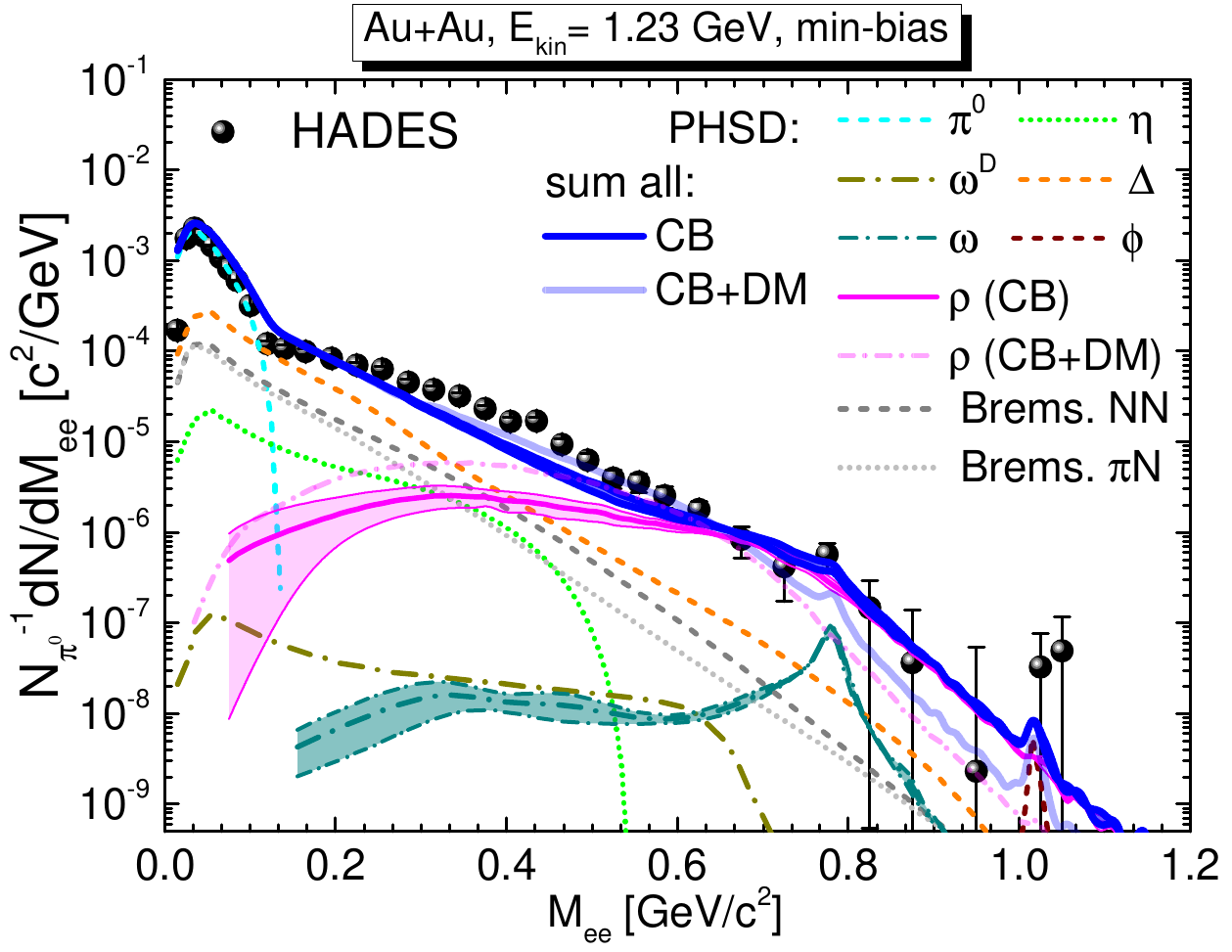}
    \caption{Invariant mass spectra of dileptons from PHSD in  p+p (upper left) and p + Nb reactions
(upper right) at 3.5 GeV beam energy and for the mass differential dilepton spectra $dN/dM_{ee}$—normalized to the $\pi^0$ multiplicity—for Ar + KCl collisions at 1.76 AGeV (lower left) and for Au + Au collisions at 1.23 AGeV (lower right) in comparison to the experimental measurements by the HADES collaboration. The solid dots present the HADES data for p+p \cite{HADES:2012sir}, for p + Nb \cite{HADES:2011jqb,HADES:2012sui}, for Ar + KCl \cite{HADES:2011nqx} and for Au + Au \cite{HADES:2019auv}, respectively. The total yield is displayed in terms of the blue lines while the different contributions are specified in the legends. The theoretical calculations are passed through the corresponding HADES acceptance filter and mass/momentum resolution. 
The in-medium modifications of the vector mesons  in the  collisional broadening (CB) scenario are shown by shaded bands (magenta for $\rho$ and cyan for $\omega$ mesons) which indicate the variation of the broadening coefficients $\alpha_V$ in Eq. (\ref{eq:width_density}).
The results for the collisional broadening + dropping mass (CB+DM) scenario for the $\rho$- meson spectral function for Au+Au at 1.23 A GeV (lower right plot) is shown by the dash-doted line and the sum of all contributions by the light blue line.
 }
    \label{mass_spectra_SIS18}
\end{figure*}

\begin{figure}[h]
    \centering
\includegraphics[width=0.99\linewidth]{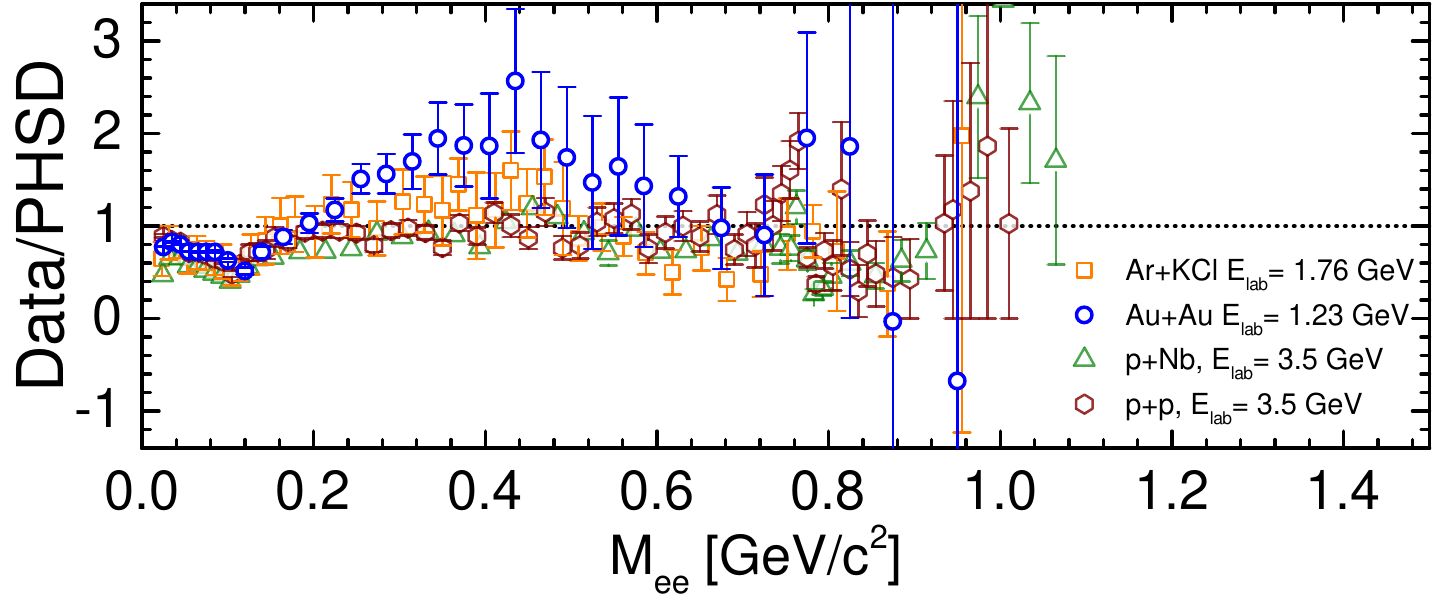}
    \caption{Ratio of the HADES data to the PHSD calculations (shown in Fig. \ref{mass_spectra_SIS18}) for p+p \cite{HADES:2012sir}, for p + Nb \cite{HADES:2011jqb,HADES:2012sui}, for Ar + KCl \cite{HADES:2011nqx} and for Au + Au \cite{HADES:2019auv}.}  
    \label{ratio_SIS}
\end{figure}

Fig. \ref{mass_spectra_SIS18} presents the PHSD calculations for the differential cross section $d\sigma/dM_{ee}$ for $e^+e^-$ pair production. The results are displayed for p+p collisions (top left panel) and p+Nb reactions (top right panel) at a beam energy of $E_{\text{kin}}=3.5$~GeV. Moreover, the dilepton invariant mass spectra, $dN/dM_{ee}$, normalized to the $\pi^0$ multiplicity, are shown for Ar+KCl collisions at $E_{\text{kin}}=1.76$~AGeV (bottom left panel) and for Au+Au collisions at $E_{\text{kin}}=1.23$~AGeV (bottom right panel). These results are compared to the experimental measurements provided by the HADES Collaboration \cite{HADES:2019auv}.
In order to quantify the deviations of theoretical calculations from experimental data we show in Fig. \ref{ratio_SIS}  the ratio of the HADES data to the PHSD calculations.

For p+p and  p+Nb collisions (top panels) at $E_{\text{kin}} = 3.5 \, \text{GeV}$ the spectra are dominated by contributions from $\pi^0$, $\eta$ and $\Delta$ Dalitz decays at low invariant masses ($M_{ee} < 0.6 \, \text{GeV}/c^2 $). The $\rho$ meson contribution becomes more prominent in the mass region ($ 0.6 < M_{ee} < 0.8 \, \text{GeV}/c^2$). The PHSD model provides a good description of the HADES data across the entire invariant mass range, with slightly higher yields for p+Nb compared to p+p due to nuclear effects.

For Ar+KCl collisions (bottom left panel) at $ E_{\text{kin}} = 1.76 \, \text{AGeV}$, 
additionally to the $\pi^0$, $\eta$ and $\Delta$ Dalitz decays, the dilepton spectra display a visible contribution from $\rho$ meson decays in the intermediate mass region  ($ M_{ee} > 0.4 \, \text{GeV}/c^2$). The inclusion of bremsstrahlung processes ($NN$ and $\pi N$) further enhances the yield at lower invariant masses ($ M_{ee} < 0.3 \, \text{GeV}/c^2$). 
For Au+Au collisions (bottom right panel) at $ E_{\text{kin}} = 1.23 \, \text{AGeV}$, the dilepton spectra exhibit a similar trend to Ar+KCl but with significantly higher yields due to the increased system size even if the collision energy is slightly lower. The $\rho$ meson contribution remains dominant in the intermediate mass region, while bremsstrahlung processes and $\Delta$ Dalitz decay play a key role at lower masses. 

We note that the vector meson spectral functions are modeled within the collisional broadening scenario 
according to Eq. (\ref{eq:width_density}).  The shaded bands indicate the results from variations of the “broadening coefficient”:
 $\alpha^\rho_{coll} = 70 - 150$ MeV for the $\rho$-mesons,  $\alpha^\omega_{coll} = 40 - 70$ MeV for $\omega$-mesons and $\alpha^\phi_{coll} = 25 - 40$ MeV  for $\phi$- mesons. 
 The PHSD model reproduces the general behavior of the HADES data for all systems (in line with our early calculations \cite{Bratkovskaya:2007jk,Bratkovskaya:2013vx,Schmidt:2021hhs}), albeit with small deviations at $M_{ee} \sim 0.4$ GeV$/c^2$ for Au+Au. In order to explore if this can be attributed to the modeling of the in-medium effects, we investigate another scenario for the $\rho$- meson spectral function such as the collisional broadening +  dropping mass scenario (CB+DM) in line with the early PHSD studies \cite{Bratkovskaya:2007jk,Bratkovskaya:2013vx}.
In this combined scenario additionally to the collisional broadening of the width (with  $\alpha^\rho_{coll}=150$ MeV), a density dependent pole mass $M_V^{\ast}(\rho)$ of the spectral function  is incorporated:
\begin{align}
M_V^{\ast}(\rho) = \frac{M_0}{1 + \alpha \rho/\rho_0}, 
\label{eq:mass_density}
\end{align}
where $M_0$ is the $\rho$- meson vacuum mass and the parameter $\alpha$ determines the magnitude of the mass shift 
in line with the Hatsuda/Lee suggestion \cite{Hatsuda:1991ez} or Brown/Rho scaling \cite{Brown:1991kk}.
The CB+DM scenario leads to a stronger modification of the $\rho$-meson spectral function due to a shift of the pole mass and fills the gap between the calculations and experiment at $M_{ee}\sim 0.4$ GeV$/c^2$ (light blue line) as well as in the ratio shown in Fig. \ref{ratio_SIS}. On the other hand this scenario leads to a reduction of the dilepton yield for $M_{ee} > 0.7$ GeV$/c^2$ compared to the CB scenario. However, the large experimental errorbars for large $M_{ee}$ do not allow to distinguish the scenarios. We note that a more realistic modeling of the vector meson in-medium spectral functions  and in-medium interaction cross sections based on a G-matrix approach - incorporated in the microscopic off-shell transport model - would be required for a consistent theoretical description of the in-medium effects.

The comparison of PHSD results with the HADES experimental data shows that the PHSD model successfully reproduces the invariant mass spectra of dileptons across various collision systems and beam energies. This agreement aligns with previous studies \cite{Bratkovskaya:2007jk, Bratkovskaya:2013vx}, which investigated collision systems such as p+p, p+Nb, C+C and Ar+KCl,  and with \cite{Schmidt:2021hhs} for Au+Au  at SIS18 energies. 
 
According to the PHSD results (reported first in 2013 in Ref. \cite{Bratkovskaya:2013vx}) a strong in-medium enhancement of the dilepton yield in A+A versus p+p -- as observed by the HADES Collaboration -- which increases with the system size,  is attributed to:
1) multiple $\Delta$-resonance regeneration, i.e. dilepton emission from intermediate $\Delta$’s which are part of the reaction cycles  $\Delta \leftrightarrow \pi N$ and $\Delta  N \leftrightarrow NN$;
2) $pN$ bremsstrahlung based on boson-exchange model calculations \cite{Kaptari:2005qz};
3) collisional broadening of the $\rho, \omega, \phi$ meson spectral functions.

Further studies of the dilepton spectra at low energies will allow to probe the dilepton emissivity from the dense hadronic medium and study their properties and interactions.
We note that reliable data on p+p (and p+n) collisions are needed in order to interpret the A+A measurements in a consistent way.

\subsection{Au+Au collisions at RHIC energies}

\begin{figure*}[h!]
    \centering
\includegraphics[width=0.49\linewidth]{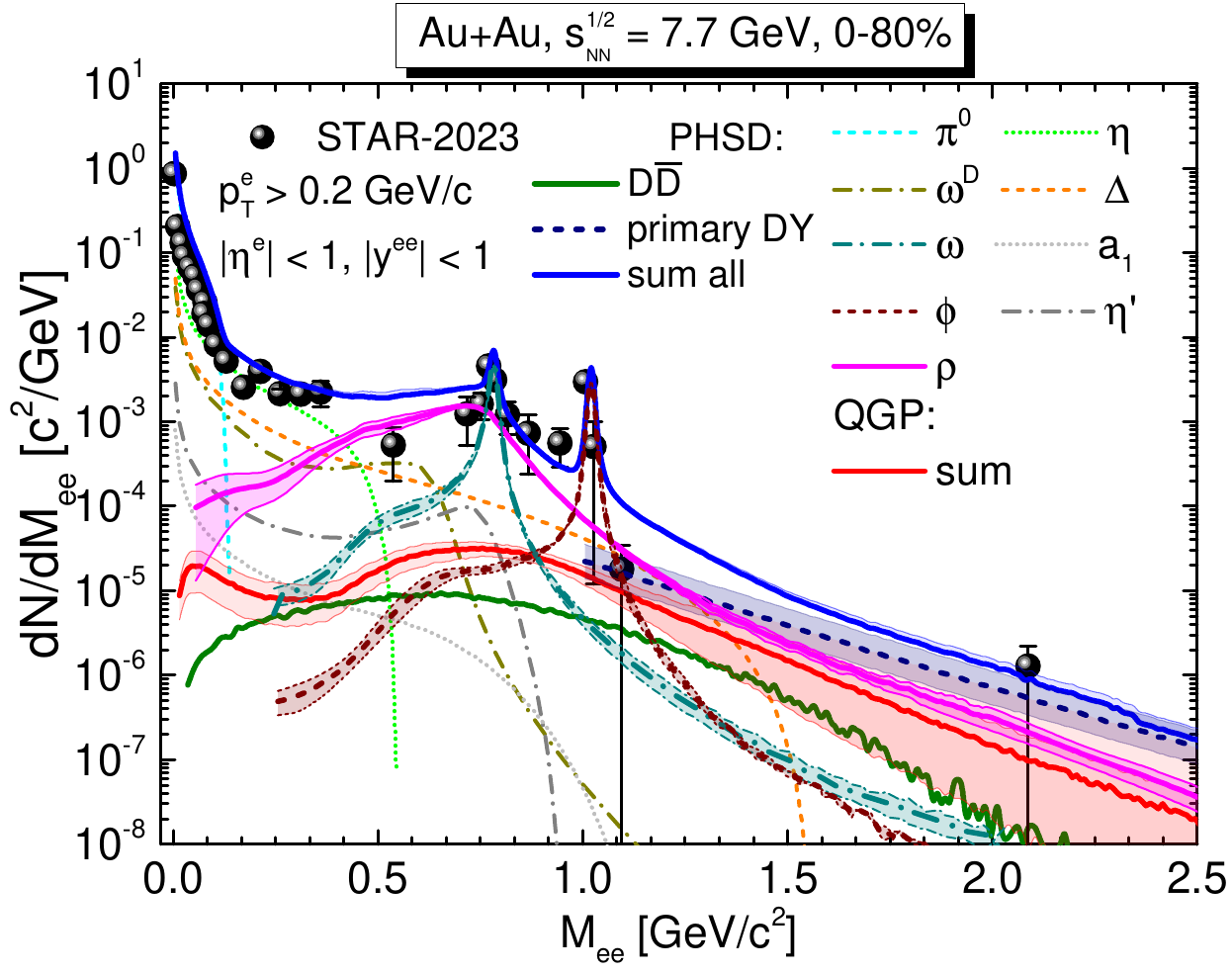}
\includegraphics[width=0.49\linewidth]{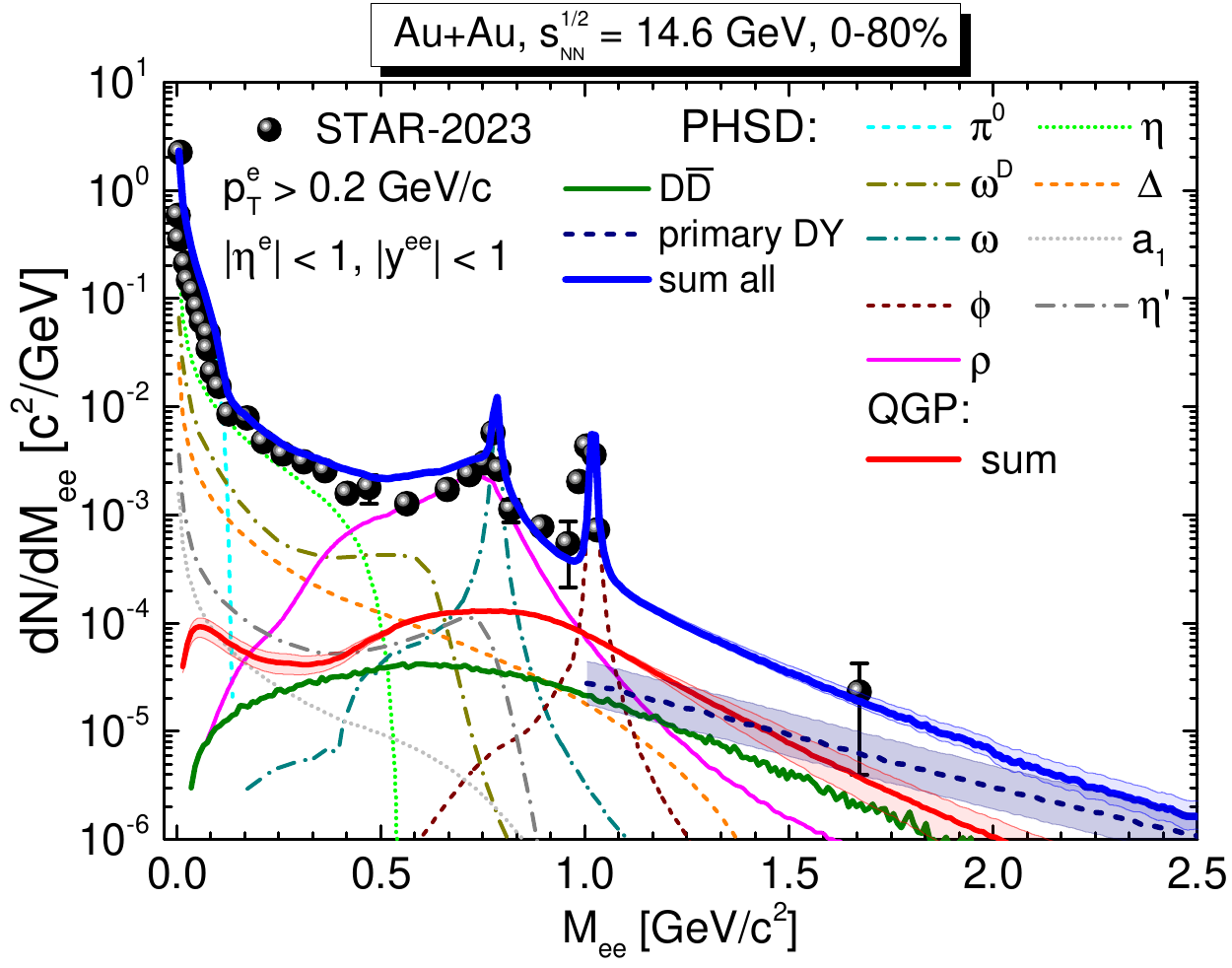}
\includegraphics[width=0.49\linewidth]{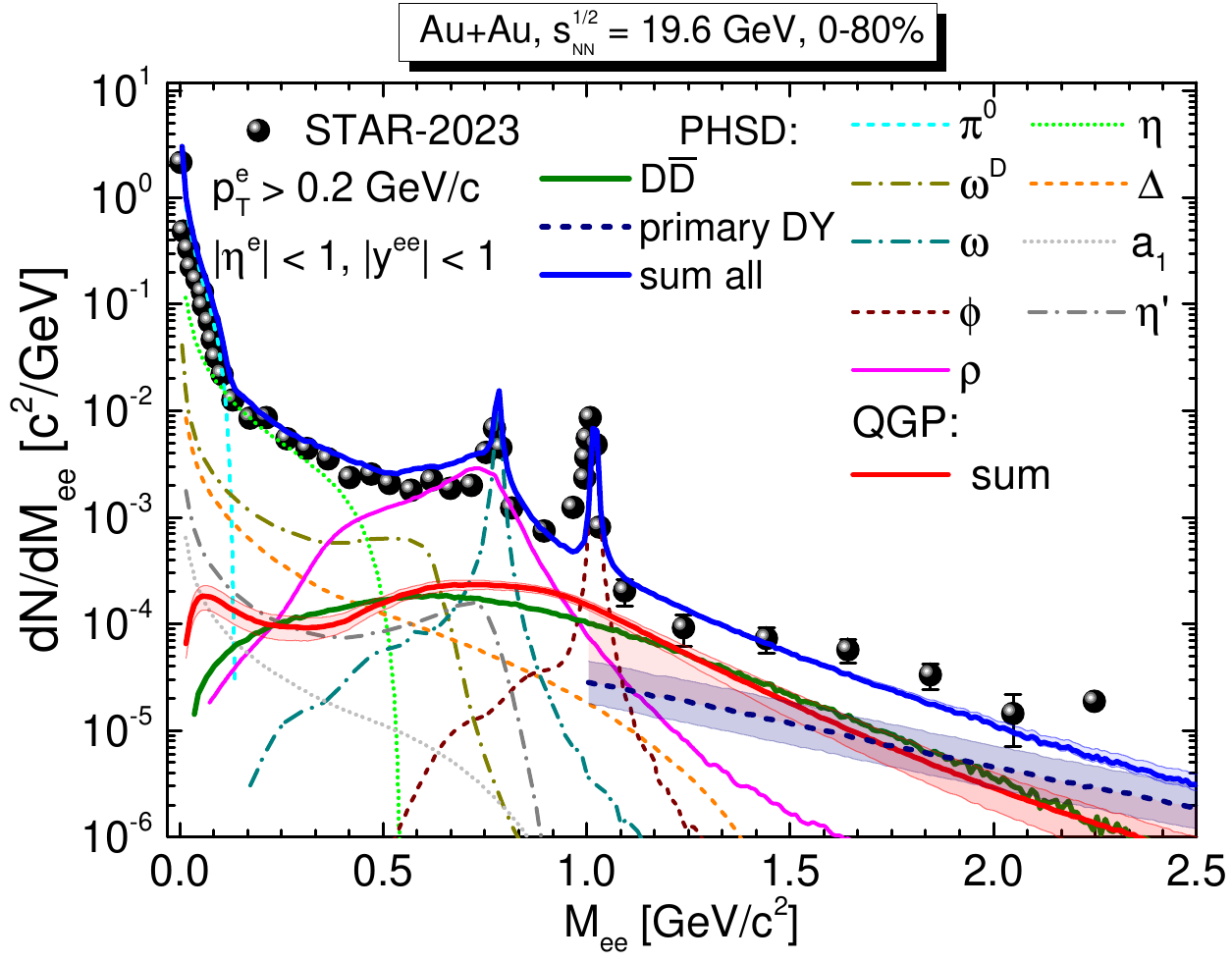}
\includegraphics[width=0.49\linewidth]{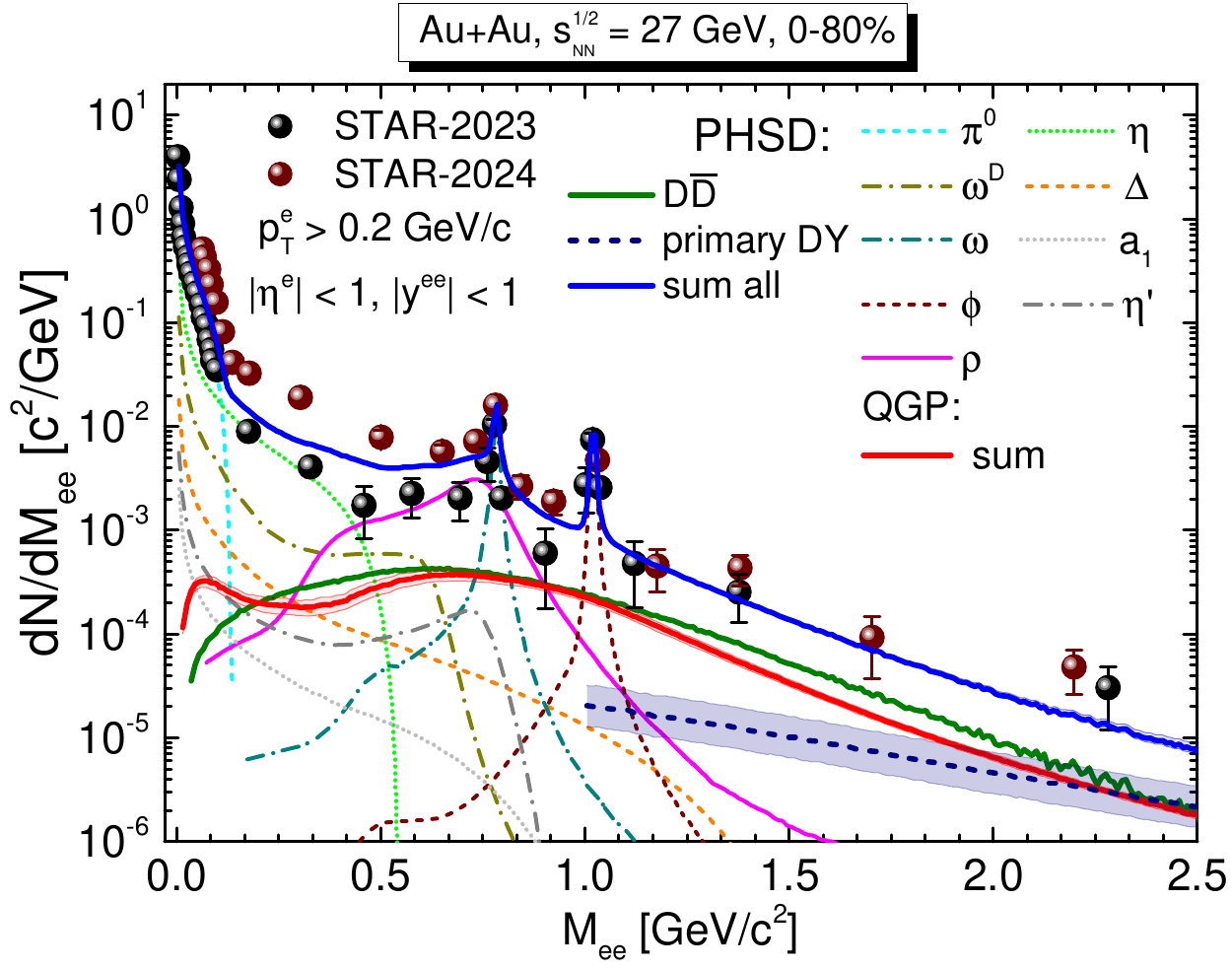}
\includegraphics[width=0.49\linewidth]{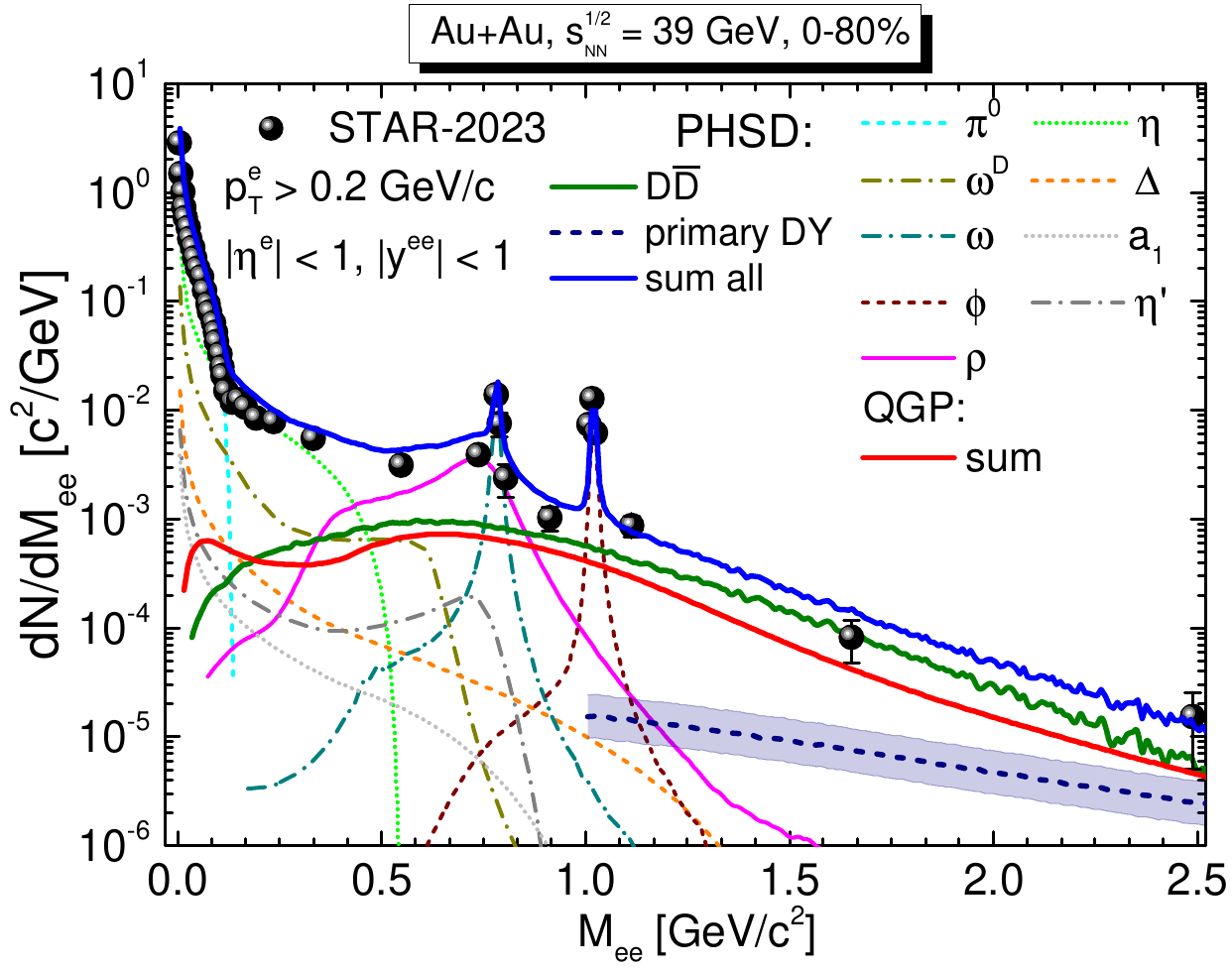}  \includegraphics[width=0.48\linewidth]{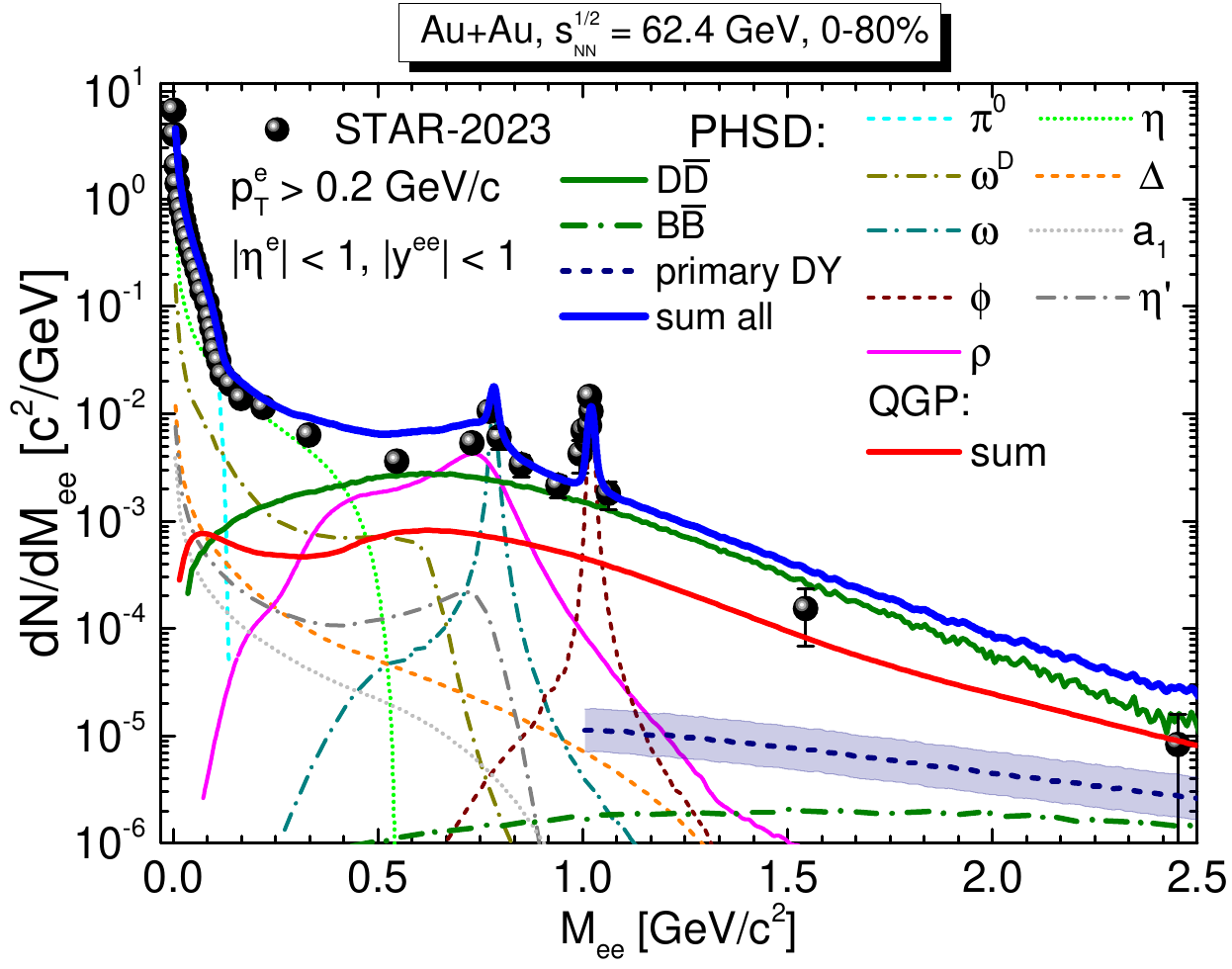}
    \caption{Invariant mass spectra of dileptons from PHSD in comparison to STAR data for Au+Au collisions for $0-80$\% centrality at $\sqrt{s_{NN}}=7.7, \, 14.5, \, 19.6, \, 27, \, 39, \, 62.4$ GeV.
    The total yield is shown by the blue lines while the different contributions are specified in the legends.
    The solid dots represent the STAR data:
    black \cite{STAR:2023wta,Han:2024nzr} and brown \cite{STAR:2024bpc}. The theoretical calculations are passed through the corresponding STAR acceptance filter and mass/momentum resolution.     }
    \label{mass_spectra_RHIC}
\end{figure*}
\begin{figure}[h!]
    \centering
     \includegraphics[width=0.99\linewidth]{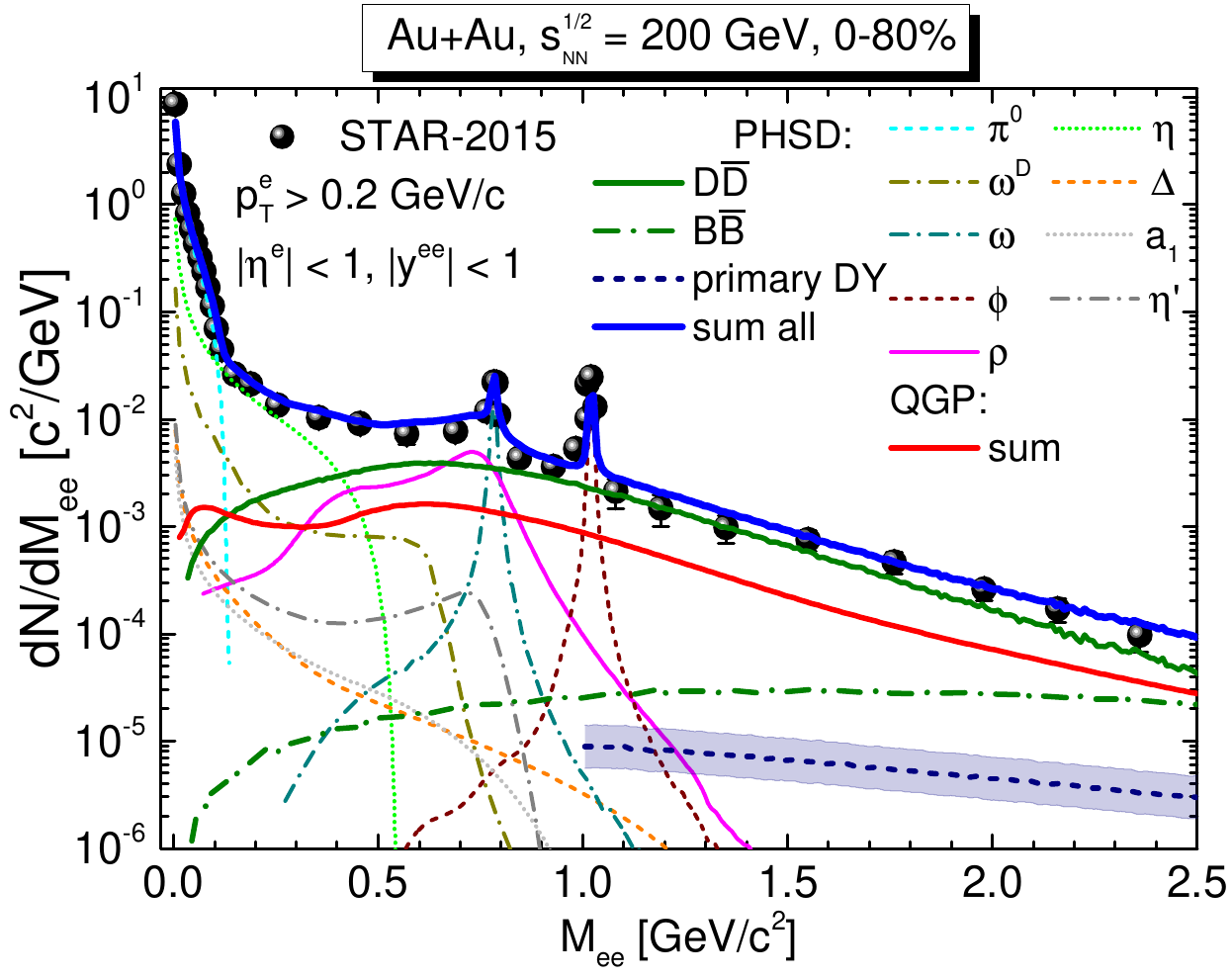}
    \caption{Invariant mass spectra of dileptons from PHSD in comparison to STAR data for Au+Au collisions for $0-80$\% centrality at $\sqrt{s_{NN}}=200$ GeV  \cite{STAR:2015tnn}. The total yield is shown by the blue lines while the different contributions are specified in the legends. The solid dots represent the STAR data. The theoretical calculations are passed through the corresponding STAR acceptance filter and mass/momentum resolution.   }
    \label{mass_spectra_RHIC_200}
\end{figure}
%
\begin{figure}[h!]
    \centering
\includegraphics[width=0.99\linewidth]{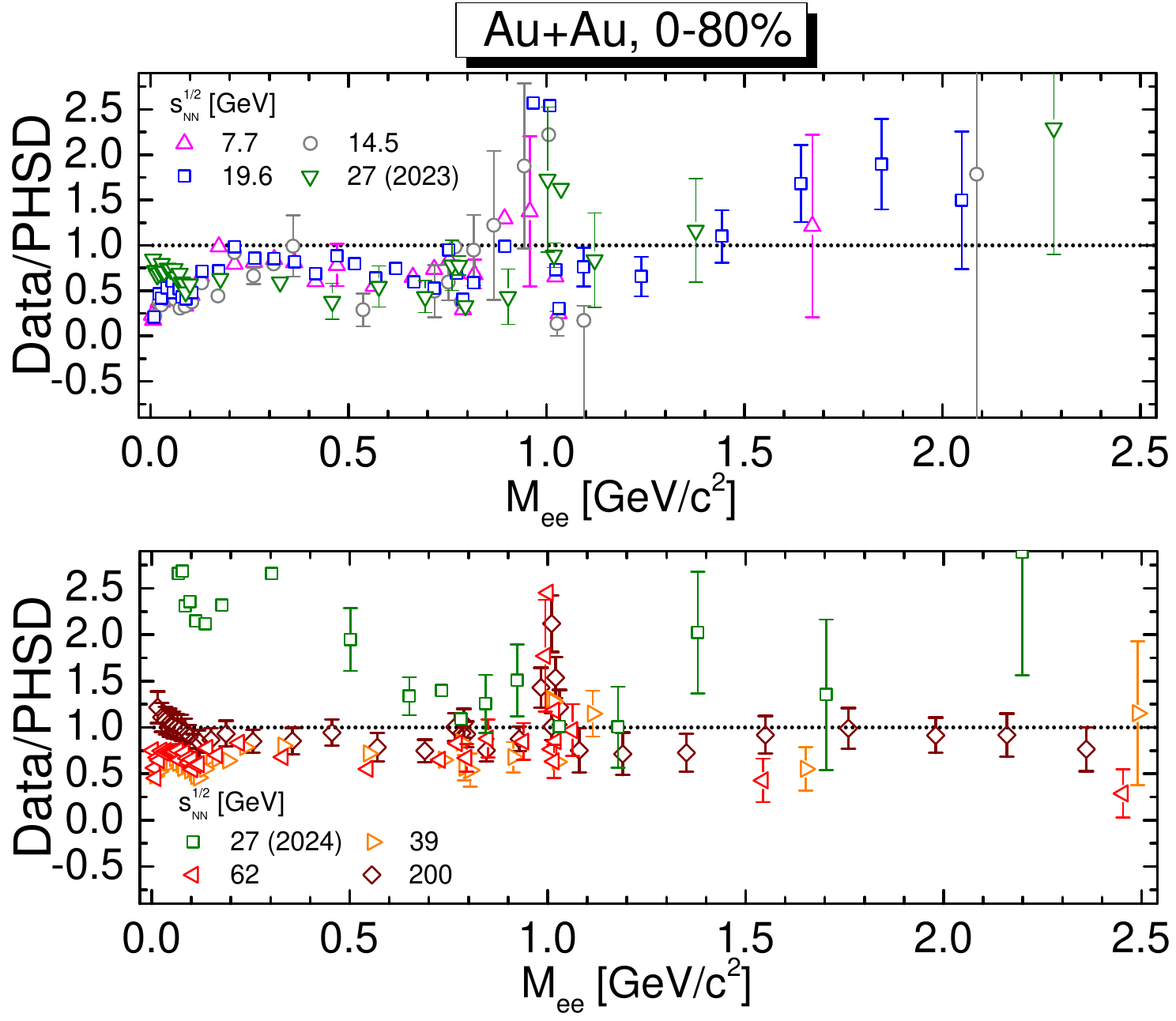}
    \caption{Ratio of the STAR data to the PHSD calculations (shown in Figs. 
    \ref{mass_spectra_RHIC}, \ref{mass_spectra_RHIC_200}) for Au+Au collisions for 0-80\% centrality from   $\sqrt{s_{NN}}=7.7, \, 14.5, \, 19.6, \, 27$ GeV \cite{STAR:2023wta,Han:2024nzr}(top panel) and $\sqrt{s_{NN}}= 27, \, 39, \, 62.4, \, 200$ GeV \cite{STAR:2015tnn,STAR:2024bpc} (bottom panel).  }
    \label{ratio_RHIC}
\end{figure}

Now we step to high energies and show in Figs. \ref{mass_spectra_RHIC} and \ref{mass_spectra_RHIC_200} the invariant mass mass spectra of dileptons $dN/dM_{ee}$ for 0-80\% central Au+Au collisions at  $\sqrt{s_{NN}}=$ 7.7, \, 14.5, \, 19.6, \, 27, \, 39, \, 62.4, and 200~GeV. These results are compared to the corresponding experimental data from the STAR Collaboration \cite{STAR:2015tnn,STAR:2023wta,Han:2024nzr,STAR:2024bpc}. 
%
 Fig. \ref{ratio_RHIC} presents the ratio of the STAR data to the PHSD calculations.


The following dilepton production channels are accounted for:\\
-- {\it Hadronic decays:} contributions from  Dalitz decays of mesons $\pi^0, \eta, \omega, \eta',  a_1$ and $\Delta$ resonance and direct decays of vector mesons $\rho, \omega, \phi$ are displayed by dashed and dotted lines (see legends).   \\
-- {\it QGP radiation:} the red band represent the sum of dilepton yield from the partonic channels: $q\bar q$ annihilation  $q\bar{q} \to e^+ e^-$, $q\bar{q} \to g  e^+ e^-$ and Compton scattering $qg \to q e^+ e^-, \ \ \bar{q}g \to \bar{q} e^+ e^-$.  As pointed out in Section III, the QGP dileptons include the contributions from i) "thermal" QGP partonic scattering indicated  by the dotted red lines on the low red band as well as ii) "non-equilibrium" scattering of leading quarks/antiquarks from string ends with the partons from the QGP. The sum of both contributions are indicated by the dotted red lines on the upper red band. The solid red lines show the averaged QGP contribution. We  note that all PHSD calculations are performed including the $(T,\mu_B)$ dependence of the QGP dilepton production, i.e. within "DQPM $(T, \mu_B)$". While at low energies the "non-equilibrium" partonic contribution is important (as follows from the width of the red band), the QGP band shrinks with increasing energies due to a rapid increase of the QGP volume and dominant dilepton emission from the "thermal" QGP.
\\
-- {\it Correlated charm and beauty decays:} contributions from correlated $D\bar{D}$ pairs and  $B\bar{B}$ pairs  are shown by green solid and dashed lines, respectively.  \\
-- {\it Primary Drell-Yan process:} the contribution from the initial Drell-Yan process is shown by the dark-blue dashed lines. The gray band shows the uncertainties in the calculation of Drell-Yan processes related to the different PDF's (parton distribution functions) used in the calculations (cf. Appendix A). \\
The total sum of all contributions is shown by solid blue lines, the blue bands include uncertainties from the primary Drell-Yan and QGP calculations as explained above. \\
We note that our calculations of dileptons do not include the contributions from direct decay of hidden charm mesons  $J/\Psi, \Psi'$, thus, we restrict our calculations to $M_{ee} < 2.5 \, \text{GeV}/c^2$. 

\begin{figure}[h!]
    \centering
     \includegraphics[width=0.95\linewidth]{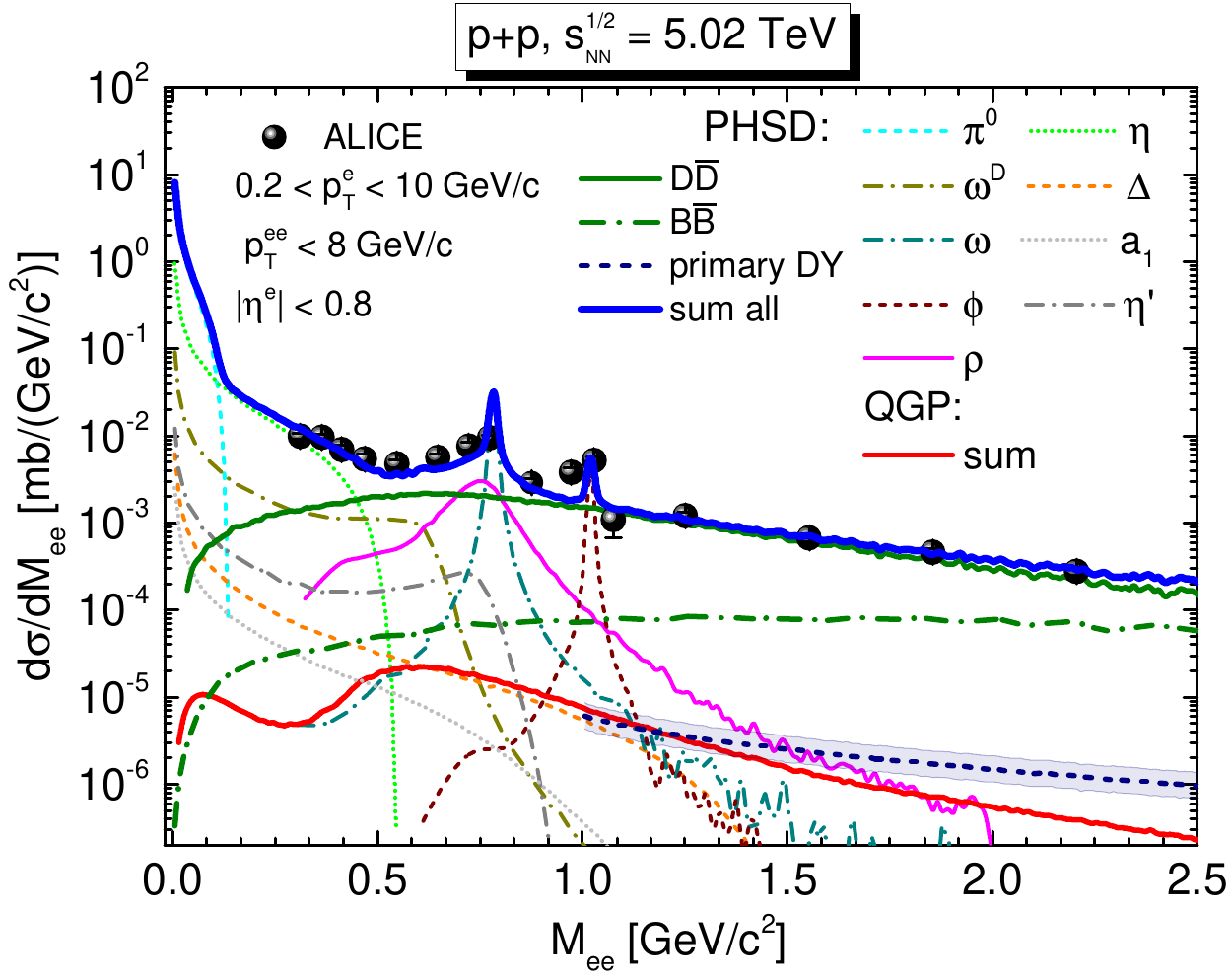}
     \includegraphics[width=0.95\linewidth]{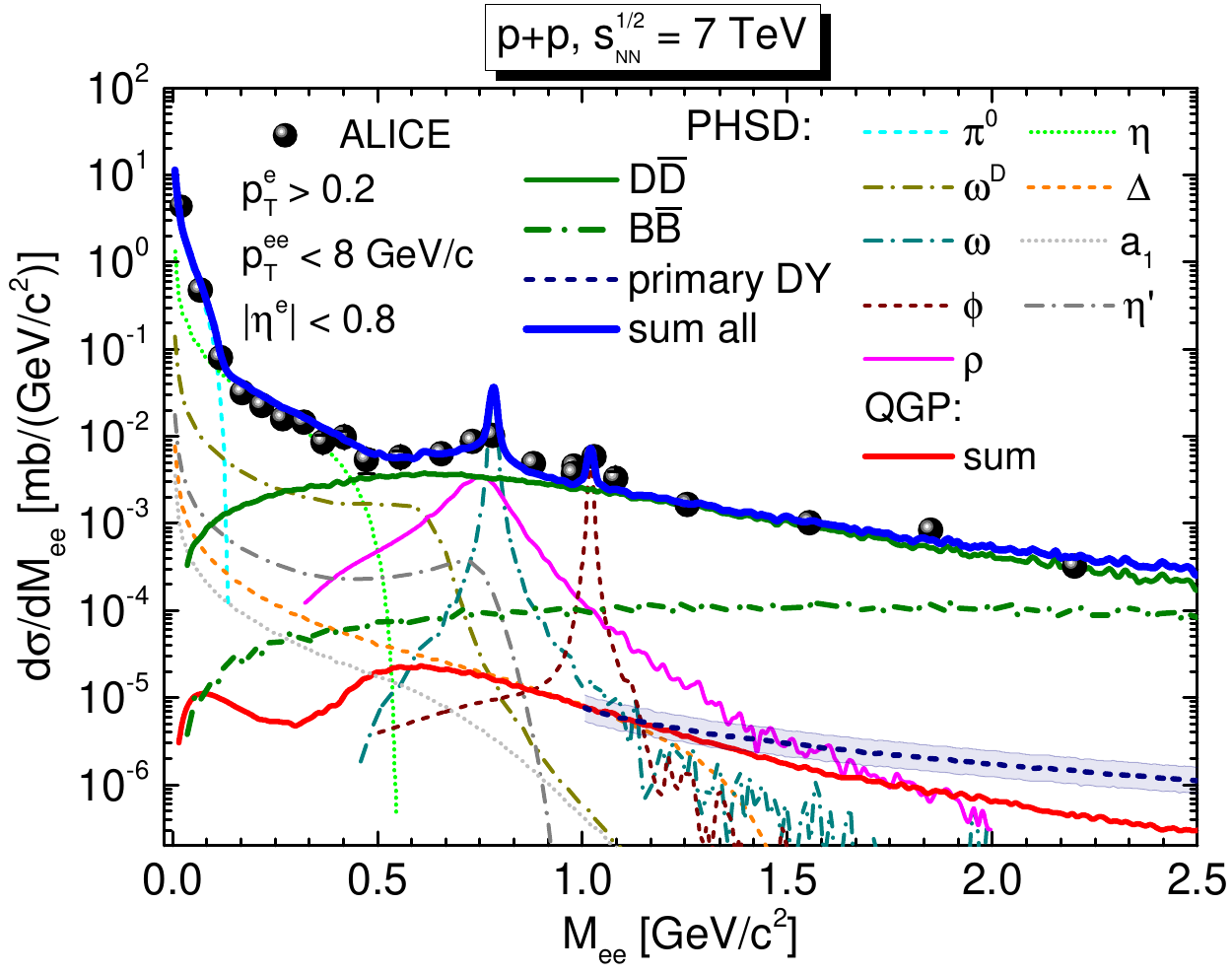}
     \includegraphics[width=0.95\linewidth]{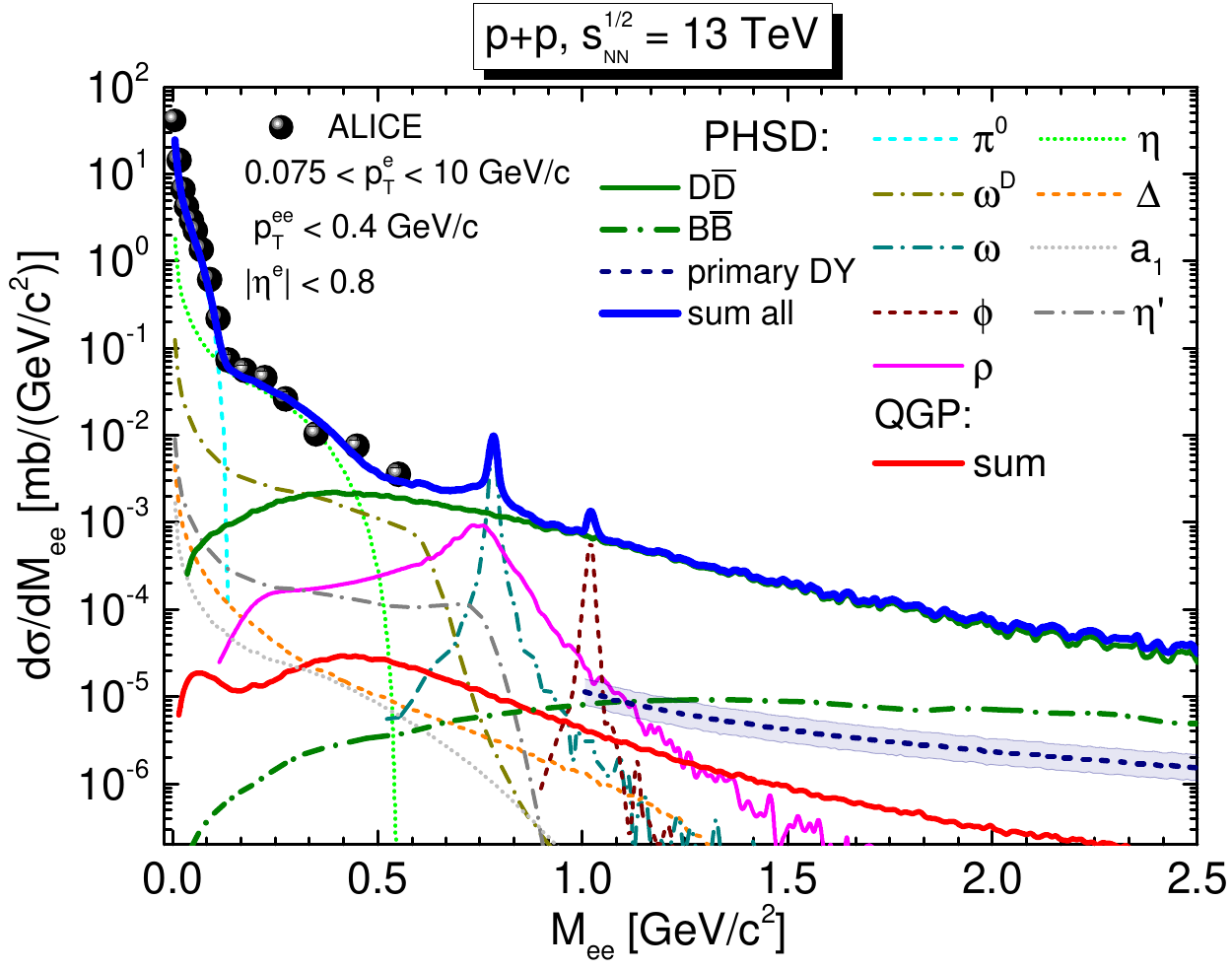}     
    \caption{ Differential cross section of dilepton production $d\sigma/dM_{ee}$  from PHSD for p+p collisions at 5.02 TeV (upper), 7 TeV (middle) and 13 TeV (lower) in comparison to the experimental measurements by the ALICE collaboration. The solid dots present the ALICE data for \cite{Gunji:2017kot,ALICE:2018fvj,ALICE:2020umb,Meninno:2020mjd}. The theoretical calculations are passed through the corresponding ALICE acceptance filter and mass/momentum resolution.     }
    \label{mass_spectra_LHC_pp}
\end{figure}

As follows from the figures,  in the low mass region ($M_{ee} < 0.4 \, \text{GeV}/c^2$) the dileptons are mainly produced by the Dalitz decays of mesons at all energies. In the intermediate  mass region ($ 0.4 < M_{ee} < 1.0 \, \text{GeV}/c^2$), the $\rho$ meson contribution dominates over other channels for energies up to $\sqrt{s_{NN}}= 39$ GeV. One can see a clear broadening of vector meson spectral functions, which is mandatory for a consistent description of the experimental data since the $\rho$ meson contribution is the main channel at ($ 0.4 < M_{ee} < 0.7 \, \text{GeV}/c^2$).
We note that we have shown the contributions of vector mesons - calculated for a variety of $\alpha_{coll}$  - by the shaded bands for Au+Au at 7.7 GeV explicitly for illustration of uncertainnesses. Since the main spreading of the bands is at low $M_{ee}$ and  practically doesn't affect the total contributions, we omit showing the bands for high energies. 

At higher RHIC energies $\sqrt{s_{NN}}= 62.4$ and 200 GeV the contributions from the QGP and correlated charm $D\bar D$ are dominant; the yield from $D\bar D$  channels grows with energy due to the increase of charm production \cite{Song:2018xca}. The contribution from correlated $B\bar B$ production is subdominant at RHIC energies due to the low production cross section of beauty quarks.
Dileptons from correlated charm overshine the QGP dileptons above $\sqrt{s_{NN}}= 39$ GeV. 
We note that our calculations with "DQPM $(T, \mu_B)$" at RHIC energies are consistent with the previous PHSD study in Ref. \cite{Song:2018xca} (related to the "DQPM $(T, \mu_B=0)$" scenario).

Thus, the comparison of PHSD calculations and STAR experimental data demonstrates the capability of the PHSD model to describe the dielectron spectra across the entire RHIC energy range. The contributions from hadronic decays, QGP radiation, and open charm decays evolve systematically with increasing $\sqrt{s_{NN}} $, highlighting the transition from a hadron-dominated regime at lower energies to an increased significance of QGP and charm contributions at higher energies.

\subsection{p+p and Pb+Pb collisions  at LHC energies}

\begin{figure}[h!]
    \centering
\includegraphics[width=0.93\linewidth]{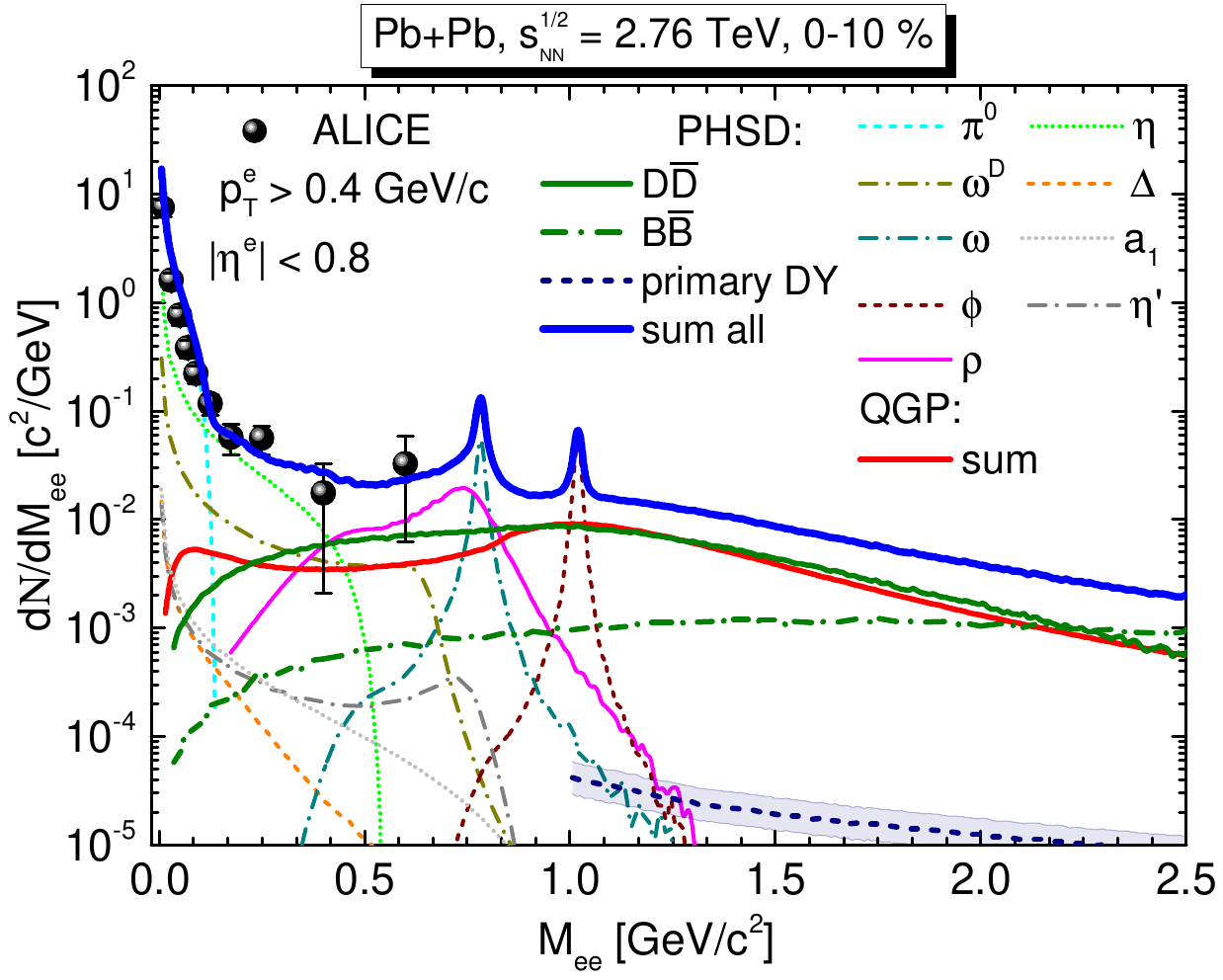}
\includegraphics[width=0.93\linewidth]{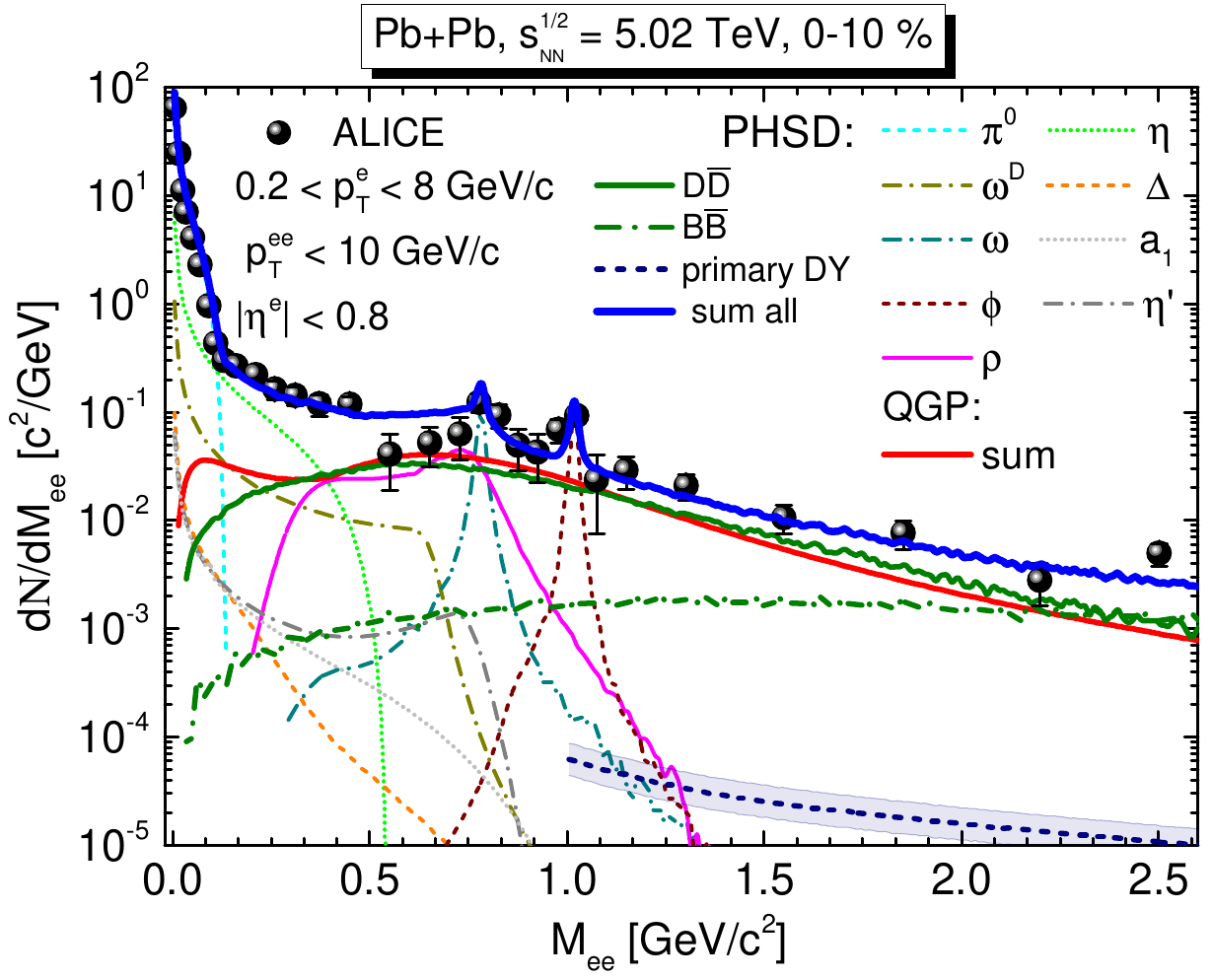} \includegraphics[width=0.93\linewidth]{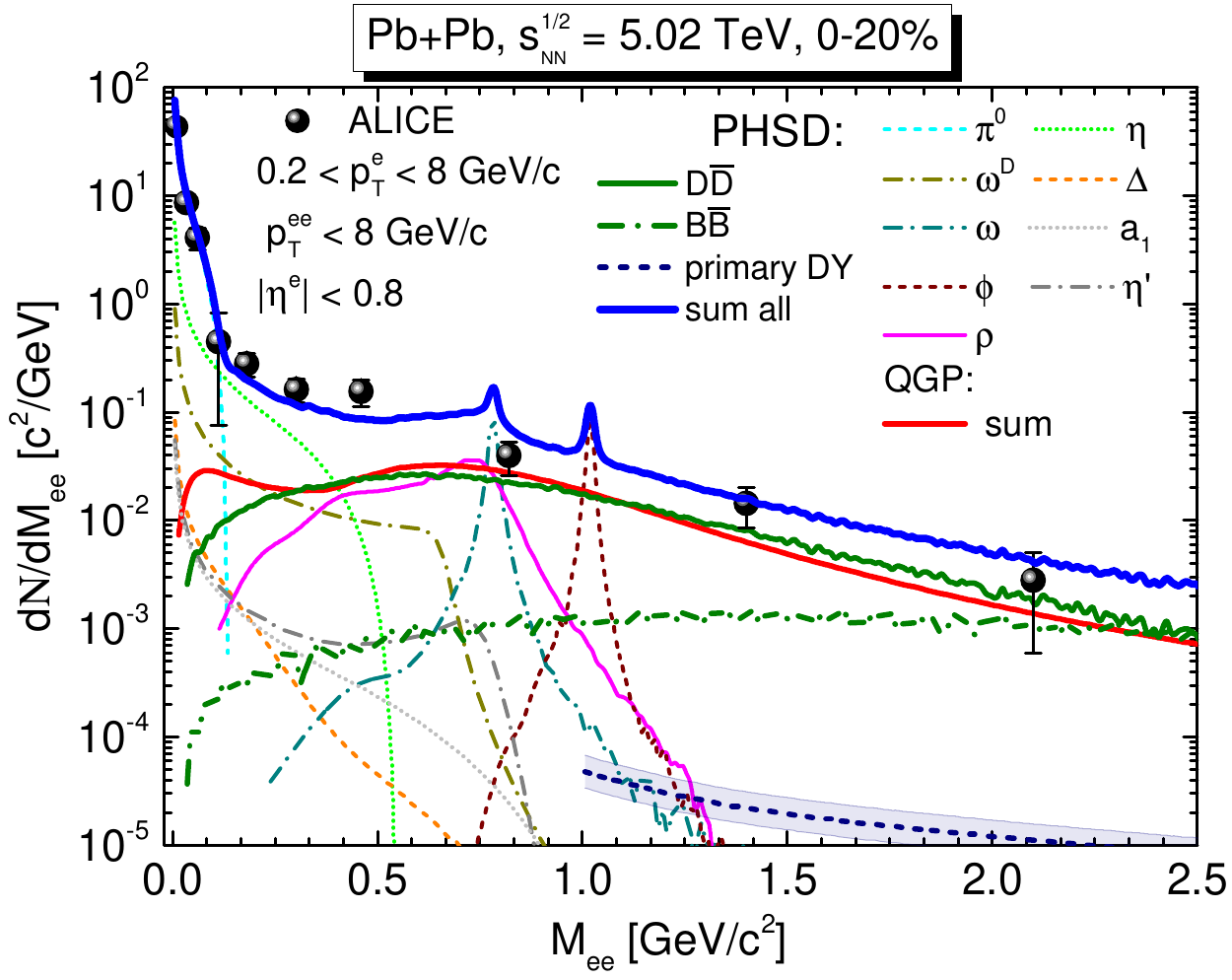}
    \caption{ Invariant mass spectra of dileptons from PHSD  $dN/dM_{ee}$  for Pb+Pb
 at  $\sqrt{s_{NN}}=$2.76 TeV  (upper) and 5.02 TeV  for 0-10\% central (middle) and 0-20\% central (lower) Pb+Pb collisions in comparison to the experimental measurements by the ALICE collaboration. The solid dots present the ALICE data \cite{ALICE:2018ael,ALICE:2023jef,Gunji:2017kot,Meninno:2020mjd}. The theoretical calculations are passed through the corresponding ALICE acceptance filter and mass/momentum resolution.     }
\label{mass_spectra_LHC_PbPb}
\end{figure}
\begin{figure}[h]
    \centering
\includegraphics[width=0.99\linewidth]{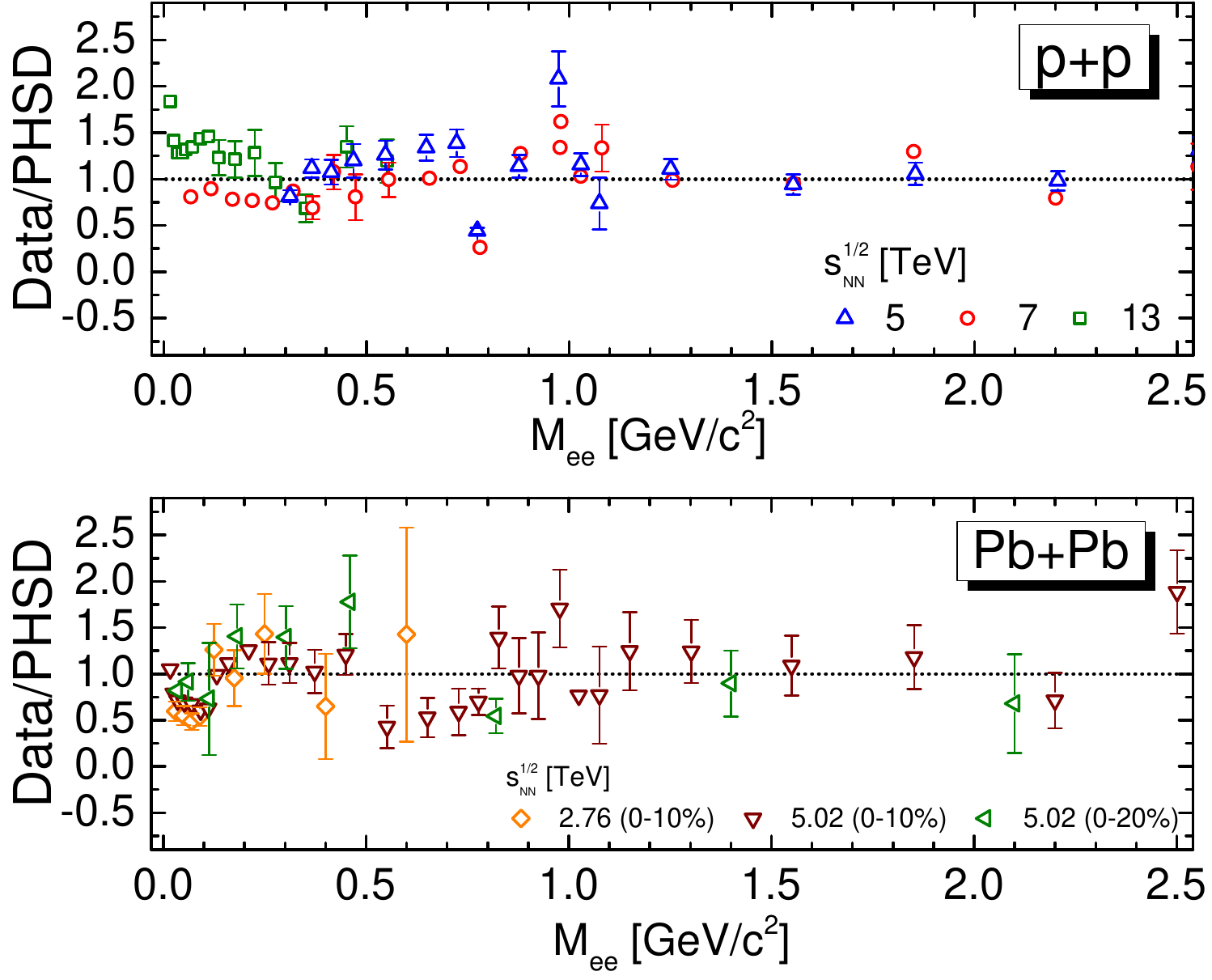}
    \caption{Ratio of the ALICE data to the PHSD predictions for p+p collisions (shown in Fig. \ref{mass_spectra_LHC_pp} from   $\sqrt{s_{NN}}=5.02, \, 7, \, 13$ TeV \cite{Gunji:2017kot,ALICE:2018fvj,ALICE:2020umb,Meninno:2020mjd} (top panel) and Pb+Pb collisions shown in \ref{mass_spectra_LHC_PbPb} at $\sqrt{s_{NN}}=$2.76 TeV  (0-10\%) and 5.02 TeV  for 0-10\% central  and 0-20\% central collisions \cite{ALICE:2018ael,ALICE:2023jef,Gunji:2017kot,Meninno:2020mjd}(bottom panel).    }
    \label{ratio_LHC}
\end{figure}

Fig. \ref{mass_spectra_LHC_pp}  shows the differential cross section of dilepton production $d\sigma/dM_{ee}$  from PHSD for p+p collisions at  $\sqrt{s_{NN}}=$5.02 TeV (upper), 7 TeV (middle) and 13 TeV (lower) in comparison to the experimental measurements by the ALICE collaboration. The solid dots present the ALICE data for \cite{Gunji:2017kot,ALICE:2018fvj,ALICE:2020umb,Meninno:2020mjd}.
Fig. \ref{mass_spectra_LHC_PbPb} presents the PHSD results for the invariant mass spectra of dileptons from PHSD  $dN/dM_{ee}$  for Pb+Pb  at 2.76 TeV  (upper) and 5.02 GeV  for 0-10\% central (middle) and 0-20\% central (lower) Pb+Pb collisions in comparison to the experimental measurements by the ALICE collaboration \cite{ALICE:2018ael,ALICE:2023jef,Gunji:2017kot,Meninno:2020mjd}.
The color coding is the same as in the previous subsection VI.A.
Fig. \ref{ratio_LHC} presents the ratio of the LHC data to the PHSD calculations.

The dilepton spectra from p+p collisions are dominated by the  hadronic Dalitz and direct decays ($ \pi^0, \eta, \rho, \omega, \phi, \Delta, a_1, \eta' $) at low invariant masses, by the open heavy-flavor decays ($ D\bar{D} $ and $ B\bar{B} $, green lines) which  become increasingly significant at intermediate and higher invariant masses ($ M_{ee} > 0.6 \, \text{GeV}/c^2 $),  Drell-Yan process (dark-blue dashed lines), that contributes to the continuum but remains subdominant, and  QGP radiation (red lines) whose contribution remains small compared to other sources at LHC energies, however, are non-zero even in p+p collisions due to small "droplets" of QGP created in  high multiplicity p+p events. As shown in Ref. \cite{Kireyeu:2020wou}, these QGP "droplets" play a minor role compared to the hadronic spectra in p+p collisions.
The total dielectron yield, depicted by the solid blue line, shows a good agreement with ALICE measurements  across the entire mass range. Open charm ($ D\bar{D} $) and bottom ($ B\bar{B} $) decays dominate the higher-mass region, particularly at $\sqrt{s_{NN}}= 13 \, \text{TeV} $, where heavy-flavor production is enhanced.
It is interesting to note that if the correlated charm and bottom contribution as well as primary Drell-Yan contribution could be accurately subtracted from the total dilepton spectra, one can get access to the  radiation from QGP droplets in p+p reactions which is impossible by hadronic observables.

In Pb+Pb collisions, Fig. \ref{mass_spectra_LHC_PbPb},  the contribution from the QGP thermal radiation (red lines) becomes more prominent at intermediate masses, reflecting the hot and dense medium created in Pb+Pb collisions.  Open heavy-flavor decays ($ D\bar{D} $ and $ B\bar{B} $) contribute significantly in the intermediate-to-high mass region, particularly at $ 5.02 \, \text{TeV} $, where charm and bottom quark production are enhanced.  
We note that the experimental data for Pb+Pb collisions at 5.02 TeV and 0-10\% centrality align well with the PHSD calculations and predictions in Ref. \cite{Song:2018xca}.
The total dielectron spectra (solid blue line) describe the ALICE experimental data well. At higher masses, the contributions from $ D\bar{D} $ and $ B\bar{B} $ decays dominate, highlighting the significant role of heavy-flavor mesons in this region.

Summarizing, the comparison between PHSD predictions and ALICE data for p+p and Pb+Pb collisions at LHC energies demonstrates that the model accurately captures the underlying physics of dielectron production. The spectra reflect the contributions from hadronic decays, open heavy-flavor decays, and QGP thermal radiation, with the latter being more pronounced in Pb+Pb collisions due to the presence of a hot and dense medium.

\subsection{Predictions for p+p and Au+Au collisions at SIS100 energies}

\begin{figure*}[t!]
    \centering
     \includegraphics[width=0.49\linewidth]{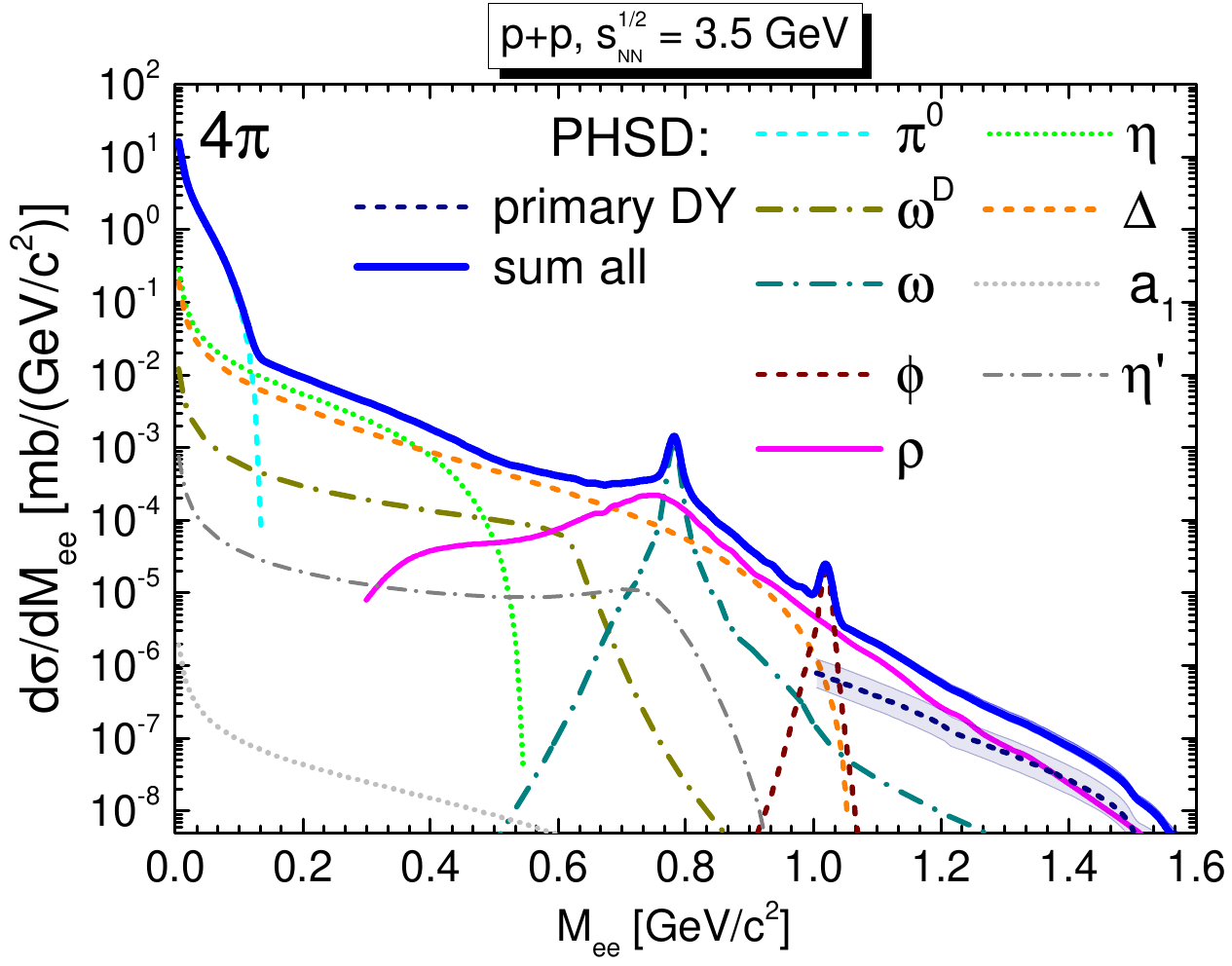}
     \includegraphics[width=0.49\linewidth]{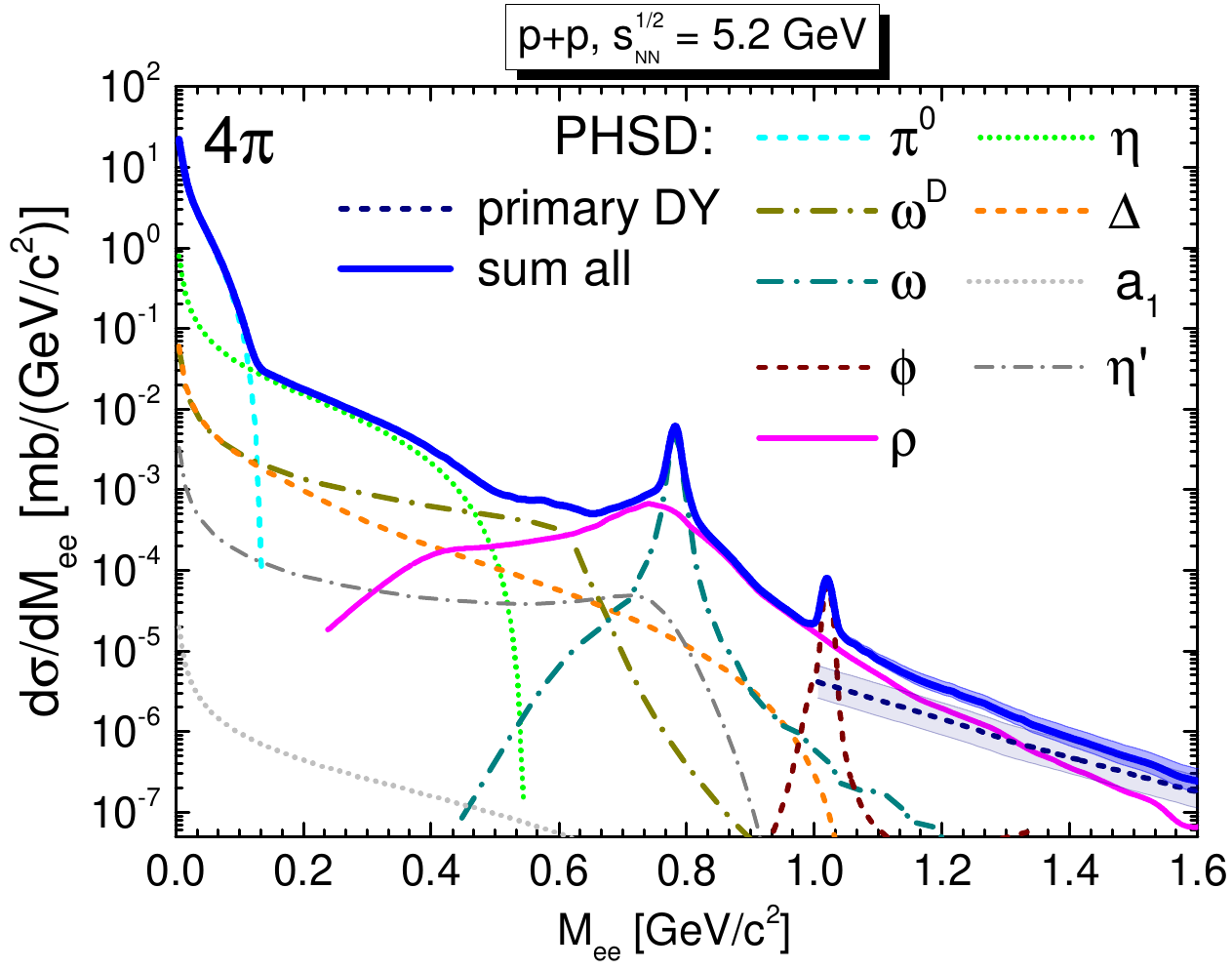}
          \includegraphics[width=0.49\linewidth]{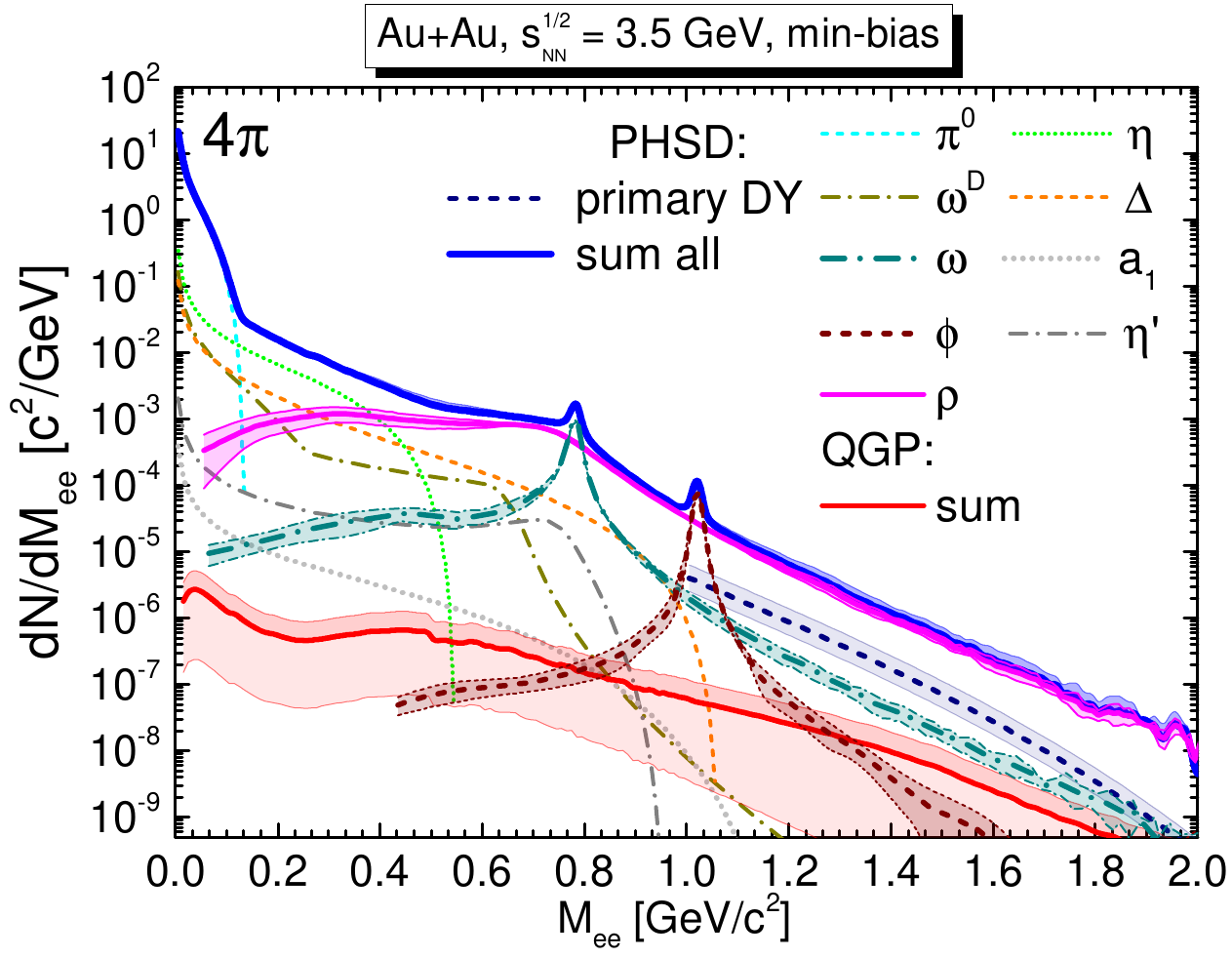}
     \includegraphics[width=0.49\linewidth]{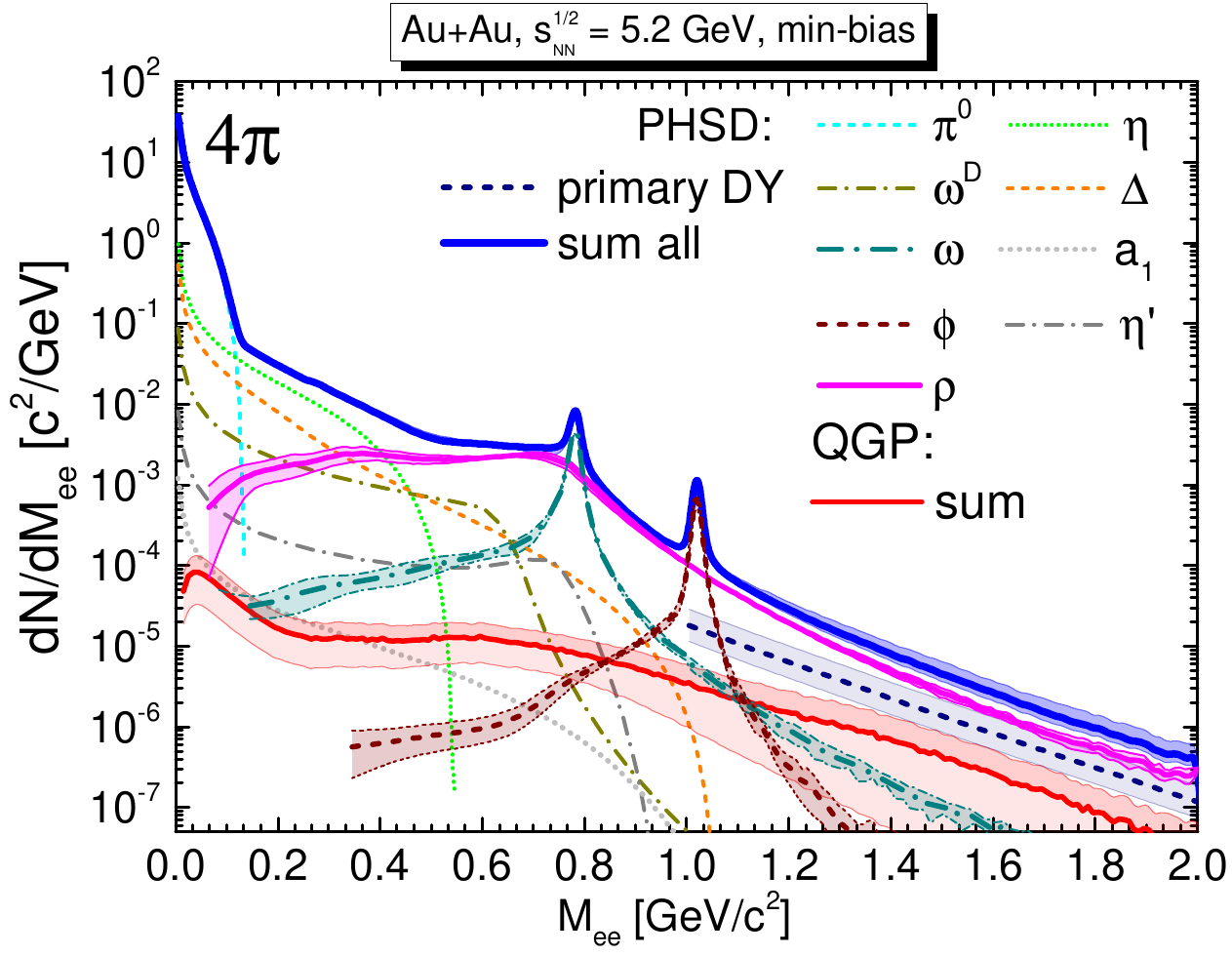}
    \caption{
    Invariant mass spectra $d\sigma/dM_{ee}$ of dileptons from PHSD for  p+p  at  $\sqrt{s_{NN}}=$3.5 GeV (upper left) and 5.2 GeV (upper right)  and for the mass differential dilepton spectra $dN/dM_{ee}$ for Au + Au collisions for minimal bias at  $\sqrt{s_{NN}}=$3.5  GeV (lower left) and  5.2  GeV (lower right).  The total yield is displayed in terms of the blue lines while the different contributions are specified in the legends. 
}
    \label{mass_spectra_SIS100}
\end{figure*}

Based on a good reproduction of experimental data by the PHSD in a wide energy range from SIS18, $\sqrt{s_{NN}}$ = 2.41 GeV, to the LHC energy of $\sqrt{s_{NN}} = 13$~TeV, we provide predictions for dielectron production for the future CBM experiment at SIS100 energies at FAIR. We present the invariant mass spectra for  p+p and Au+Au collisions at $\sqrt{s_{NN}} =$  3.5 and 5.2 GeV.

The top row of Fig. \ref{mass_spectra_SIS100} shows the dielectron invariant mass spectra $d\sigma/dM_{ee}$  for p+p collisions at beam energies of $ \sqrt{s_{NN}} = 3.5 \, \text{GeV} $ (top left) and $ 5.2 \, \text{GeV} $ (top right) in the full phase space (i.e. without any cuts). 
Similar to  the previous  cases in the low invariant mass region ($ M_{ee} < 0.5 \, \text{GeV}/c^2 $) the hadronic decays, particularly from $ \pi^0 $ and $ \eta $ mesons are dominant;  
in the intermediate mass region ($ 0.5 < M_{ee} < 1.0 \, \text{GeV}/c^2 $): contributions from the $ \rho $-meson (solid magenta line) become prominent.  At higher invariant masses ($ M_{ee} > 1.0 \, \text{GeV}/c^2 $) the Drell-Yan process (dark-blue dashed lines) is an important contribution. 
The total dielectron yield, shown by a solid blue line, combines all contributions and reflects the key processes expected in p+p collisions at SIS100 energies.

The bottom row of Fig. \ref{mass_spectra_SIS100} presents the invariant mass spectra for $ \text{Au+Au} $ collisions at $ \sqrt{s_{NN}} = 3.5 \, \text{GeV} $ (bottom left) and $ 5.2 \, \text{GeV} $ (bottom right) for minimum bias events. Key observations include:  
Enhanced low-mass dilepton production due to hadronic decays, particularly from $ \pi^0 $, $ \eta $, and baryonic resonances ($ \Delta $).  
Intermediate mass region: the contribution from the $ \rho $-meson becomes more pronounced, reflecting its role as a mediator of in-medium effects in heavy-ion collisions.  
The shaded bands show the contributions of vector mesons calculated for a variety of $\alpha_{coll}$. 

QGP thermal radiation: at 3.5 and $ 5.2 \, \text{GeV} $, the QGP contribution (red lines) starts to emerge in the intermediate mass region, indicating the potential formation of a hot and dense medium at SIS100 energies.  
Higher masses: the Drell-Yan process remains subdominant, but is higher than the QGP contribution and with a slight increase at higher invariant masses compared to p+p collisions.  

The total dielectron yield (solid blue line) highlights the expected increase in dielectron production for heavy-ion collisions relative to p+p systems, due to the higher particle density and enhanced hadronic contributions. Moreover, the red bands show the contribution from partonic processes. As discussed in Sections IV, V at $\sqrt{s_{NN}} =$ 3.5 GeV we probe already matter with the energy densities  above the critical $\varepsilon_C$ and at large baryon chemical potential $\mu_B$ - see Figs. \ref{3D-En-35GeV}, \ref{3D-En-35GeV_muB} and \ref{3D-En-200GeV_AveTmuB}, such  
that the contribution of dileptons from QGP "hot spots" becomes essential at intermediate masses 
which leaves hope for experimental access to this contribution (together with the $\rho$ meson dileptons) if the primary Drell-Yan contribution could be subtracted by corresponding measurements in p+p collisions.

The predictions for p+p and $ \text{Au+Au} $ collisions at $\sqrt{s_{NN}} = 3.5 \, \text{GeV} $ and $ 5.2 \, \text{GeV} $ demonstrate the sensitivity of dielectron production to various hadronic and partonic processes, including QGP radiation, highlighting the potential of SIS100 to probe the properties of dense nuclear matter in a previously unexplored energy regime.

\section{Dilepton Excess Yields as Probes of In-Medium Effects and the QGP}\label{sec5}

\begin{figure}[h]
    \centering
\includegraphics[width=0.99\linewidth]{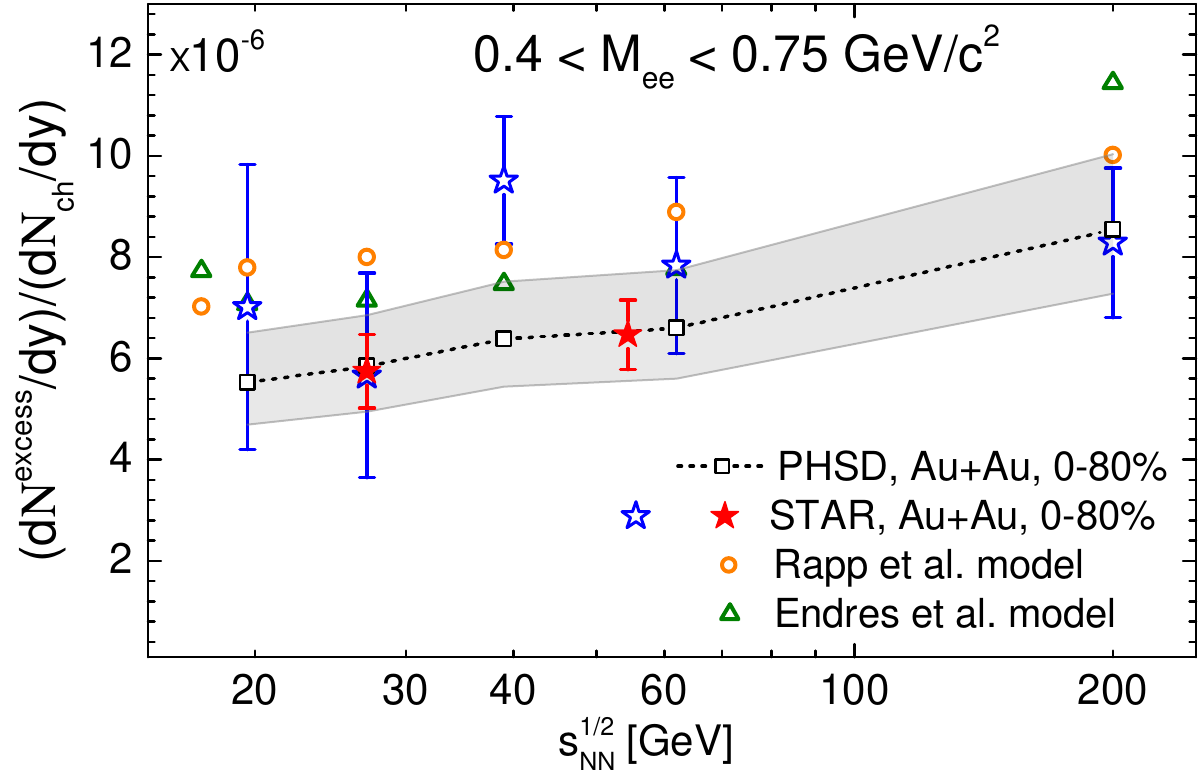}
    \caption{Collision energy dependence of the integrated dilepton excess yields for $0.4 < M_{ee} < 0.75$ GeV/c$^2$, normalized by $dN_{ch}/dy$ \cite{STAR:2023wta}, for 0-80 \% central Au + Au collisions. Black squares (corrected by the black dashed line) represent the  PHSD results including theoretical uncertainties in centrality determination for $dN_{ch}/dy$ shown by a shaded gray.  The STAR acceptance has been applied. 
    The experimental data from  the STAR Collaboration  are shown by the blue stars \cite{STAR:2023wta} and red stars \cite{STAR:2024bpc}. 
    The theoretical calculations from Rapp et al.  (orange circles) \cite{Rapp:2014hha} and Endres et al. (olive triangles) \cite{Endres:2016tkg} are also presented for comparison.  }
    \label{excess_yield}
\end{figure}

\begin{figure}[h]
    \centering
\includegraphics[width=0.99\linewidth]{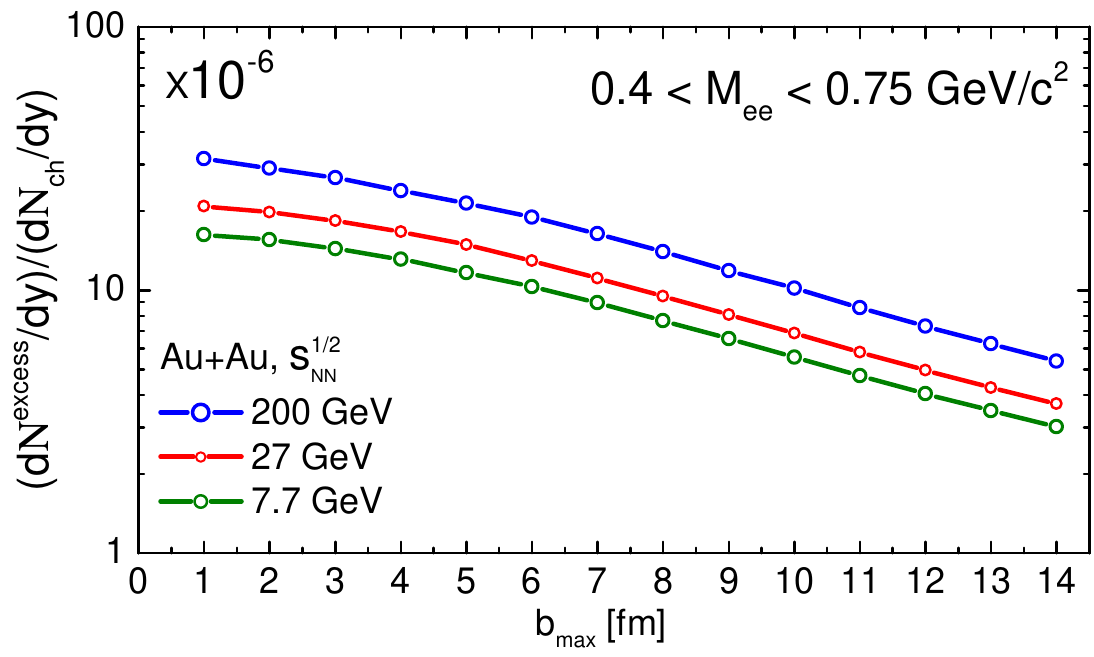}
\includegraphics[width=0.99\linewidth]{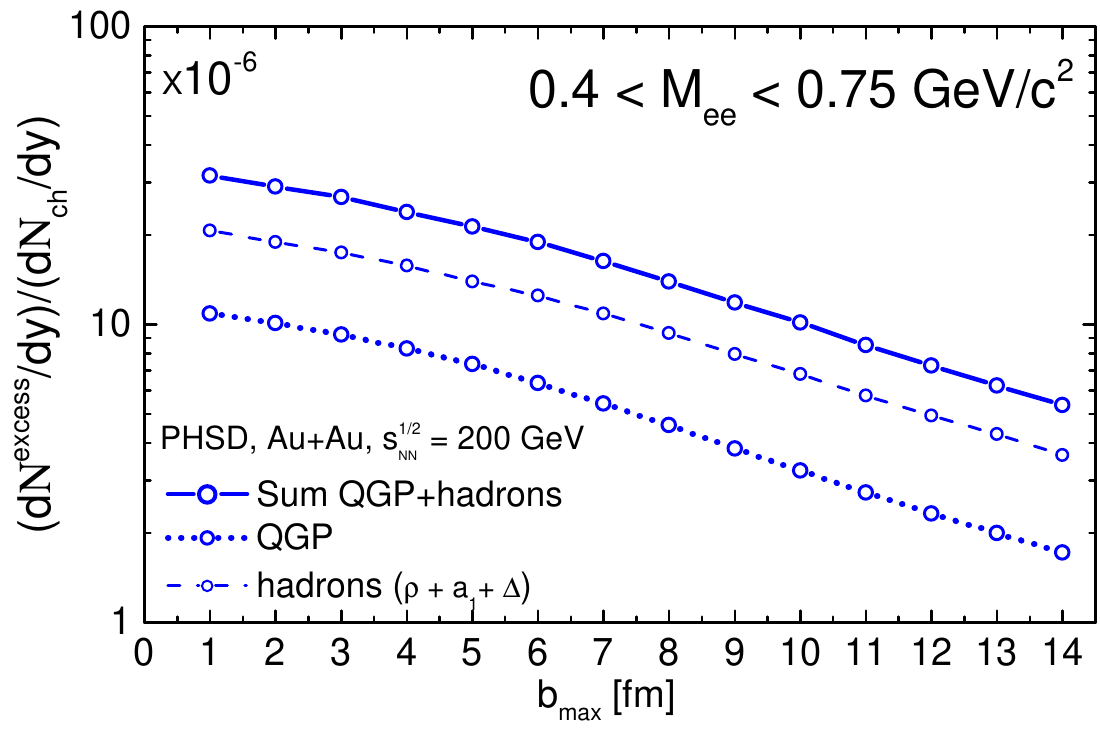}
    \caption{    
    Centrality dependence of the dielectron excess yield from PHSD, normalized by $dN_{ch}/dy$ \cite{STAR:2023wta},  integrated for $0.4 < M_{ee} < 0.75$ GeV/c$^2$  in the impact parameter interval $[0, b_{max}]$.
    Top panel: for Au + Au collisions at $\sqrt{s_{NN}}=$ 7.7, 27 and 200 GeV as a function of the maximal impact parameter $b_{max}$. Bottom panel: for Au + Au collisions at $\sqrt{s_{NN}}=200$ GeV where the solid blue line displays the sum of the QGP contribution (dotted blue) and hadronic channels, which include $\rho$, $a_1$, and $\Delta$ decays (dashed blue).  
}\label{excess_yield_bmax}
\end{figure}

\begin{figure}[h]
    \centering
\includegraphics[width=0.99\linewidth]{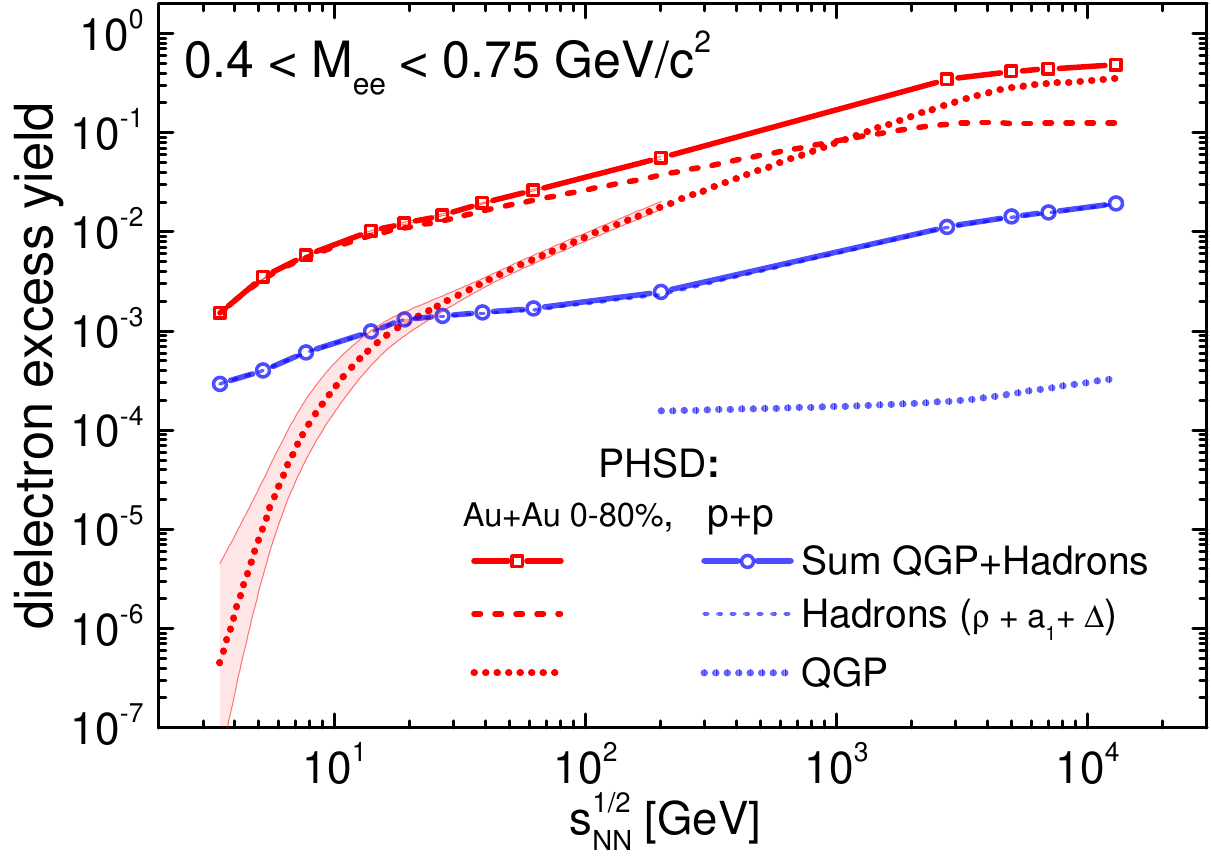}
   \caption{PHSD results for the dielectron excess yield integrated in the mass range $0.4 < M_{ee} < 0.75$ GeV/c$^2$ as a function of the center of mass energy $\sqrt{s_{NN}}$ from 3.5 GeV to 13 TeV, for p+p (red)   and Au + Au (blue) collisions. Solid lines display the sum of the QGP contribution (dotted) and hadronic sources, which include $\rho$, $a_1$, and $\Delta$ decays (dashed).  }
    \label{excess_AuAu_pp}
\end{figure}

The study of dilepton production in heavy-ion collisions provides information on the properties of the hot and dense medium formed in these events. In particular, the low-mass region ($ 0.4 < M_{ee} < 0.75 \, \text{GeV}/c^2 $) is sensitive to in-medium modifications of vector mesons, such as the $ \rho $-meson, due to interactions with the surrounding hadronic environment. This phenomenon manifests as an excess yield of dielectrons compared to the expected contributions from hadronic decays at freeze-out. 
By analyzing this excess, we gain insights into the spectral properties of the $\rho$ meson in the hot and dense medium as well as  the thermal dileptons from the QGP. 
This section presents the integrated dilepton excess yields as a function of collision energy and centrality, comparing PHSD model predictions with experimental data and other theoretical calculations.
  
The physical sources contributing to the background of $\rho$ meson and QGP thermal dileptons include the hadronic decays and the initial Drell-Yan. Specifically, the relevant hadronic channels are:  $ \pi^0,\eta, \eta' \to \gamma e^-e^+ $, $ \omega \to \pi^0e^-e^+ $, $ \omega \to e^-e^+ $, and $ \phi \to e^-e^+ $ and correlated charm $ D\bar{D} \to e^-e^+X $. These channels collectively form the "cocktail" background,  which is subtracted from the measured dilepton yield  \cite{STAR:2023wta}.

Fig. \ref{excess_yield} displays the dilepton excess yields integrated in $0.4 < M_{ee} < 0.75$ ${\rm GeV/c^2}$ and normalized by $dN_{ch}/dy$ for 0-80 \% central Au+Au collision from $\sqrt{s_{NN}}=$ 19.6 to 200 GeV. 
The excess yield from the PHSD includes the $ \rho $-meson contribution, the QGP contribution, dileptons from Dalitz decays of the $a_1$ mesons and $ \Delta$ resonances which are not part of the experimental "cocktail" and thus, are not subtracted from the measured yield, however, the relative contribution of $a_1$ and $\Delta$ decay is subdominant compared to the $\rho$ decay - cf. Figs. \ref{mass_spectra_RHIC}, \ref{mass_spectra_RHIC_200}.
The black squares represent theoretical calculations from the PHSD model with the shaded gray area indicating the uncertainty in the 80\% centrality determination. These results are compared with experimental data from the STAR Collaboration -- blue stars \cite{STAR:2023wta} and red stars \cite{STAR:2024bpc} -- and the theoretical calculations from Rapp et al. (orange circles) \cite{Rapp:2014hha} and Endres et al. (olive triangles) \cite{Endres:2016tkg}. The figure shows a consistent trend where the dielectron excess yield increases with the collision energy $\sqrt{s_{NN}}$, highlighting the growing contributions from in-medium hadronic decays and the quark-gluon plasma at higher energies. 
We note that the STAR data point for $ \sqrt{s_{NN}} = 39 \, \text{GeV} $ is underestimated in spite that the total dilepton yield is in a good agreement with the STAR data as follows from the Fig. \ref{mass_spectra_RHIC}. The PHSD results are slightly lower than those of the other models, especially at higher energies, but they are still consistent with the general trend observed in the experimental data and in line with the early PHSD results \cite{Linnyk:2015rco}.

Fig. \ref{excess_yield_bmax}  presents the centrality dependence of the dielectron excess yield from the PHSD model, normalized by $dN_{ch}/dy$ and integrated in the mass range $0.4 < M_{ee} < 0.75 \, \text{GeV}/c^2$ in the impact parameter interval $[0, b_{max}]$.
The top panel shows the dielectron excess yield as a function of the maximal  impact parameter $b_{max}$ (in fm) for Au+Au collisions at three different center-of-mass energies per nucleon: $\sqrt{s_{NN}}=$ 7.7, 27 and 200 GeV. We observe that the dielectron excess yield decreases monotonically with increasing $b_{max}$, indicating a reduction in yield for more peripheral collisions. The excess is more significant at higher collision energies ($  200 \, \text{GeV} $) and decreases for lower energies ($ 7.7 \, \text{GeV} $).
The bottom panel focuses on Au+Au collisions at $\sqrt{s_{NN}} = 200 \, \text{GeV}$ and decomposes the dielectron excess yield into its individual components as a function of $b_{max}$: total contribution from the QGP and hadronic sources ($ \rho $, $ a_1 $, and $ \Delta $ resonances).
The total yield (solid blue) is a combination of both QGP and hadronic sources, reflecting the interplay between the two components across different centrality.
The QGP contribution remains subdominant for all centrality classes, while hadronic contributions are dominant. Since the relative contributions from $a_1$ and $ \Delta $ is very low (cf. Fig. \ref{mass_spectra_RHIC_200}), it means that the measured excess provides an information of the $\rho$ meson in-medium properties.

Fig. \ref{excess_AuAu_pp} presents the PHSD results for the dielectron excess yield integrated for $ 0.4 < M_{ee} < 0.75 \, \text{GeV}/c^2 $ as a function of the center-of-mass energy $\sqrt{s_{NN}} $ for proton-proton (p+p, red) and Au+Au ($ 0-80\% $, blue) collisions.
For Au+Au collisions, the total dielectron yield, represented by the solid blue line, increases steadily with energy. This yield includes contributions from hadronic decays, such as those from $ \rho $, $ a_1 $, and $ \Delta $ decays (dashed lines), as well as from the QGP (dotted line). At lower energies, the hadronic contribution dominates, but as $\sqrt{s_{NN}} $ increases, the QGP contribution grows significantly, becoming more relevant at higher energies.
In contrast, for p+p collisions, the total dielectron yield follows a similar energy dependence but remains considerably lower compared to heavy-ion collisions. The hadronic contribution is also the main source of dielectrons in p+p collisions, while the QGP contribution is very small.

The comparison between p+p and Au+Au collisions highlights the critical role of the QGP in heavy-ion systems, especially at high energies. The substantial increase in dielectron production in Au+Au collisions reflects the formation of a hot and dense medium, which is absent in proton-proton collisions.

\section{Excitation Function of Dielectron Yields: Contributions from QGP, $ D\bar{D} $, and DY Processes}\label{sec6}

\begin{figure}[t!]
    \centering
\includegraphics[width=0.98\linewidth]{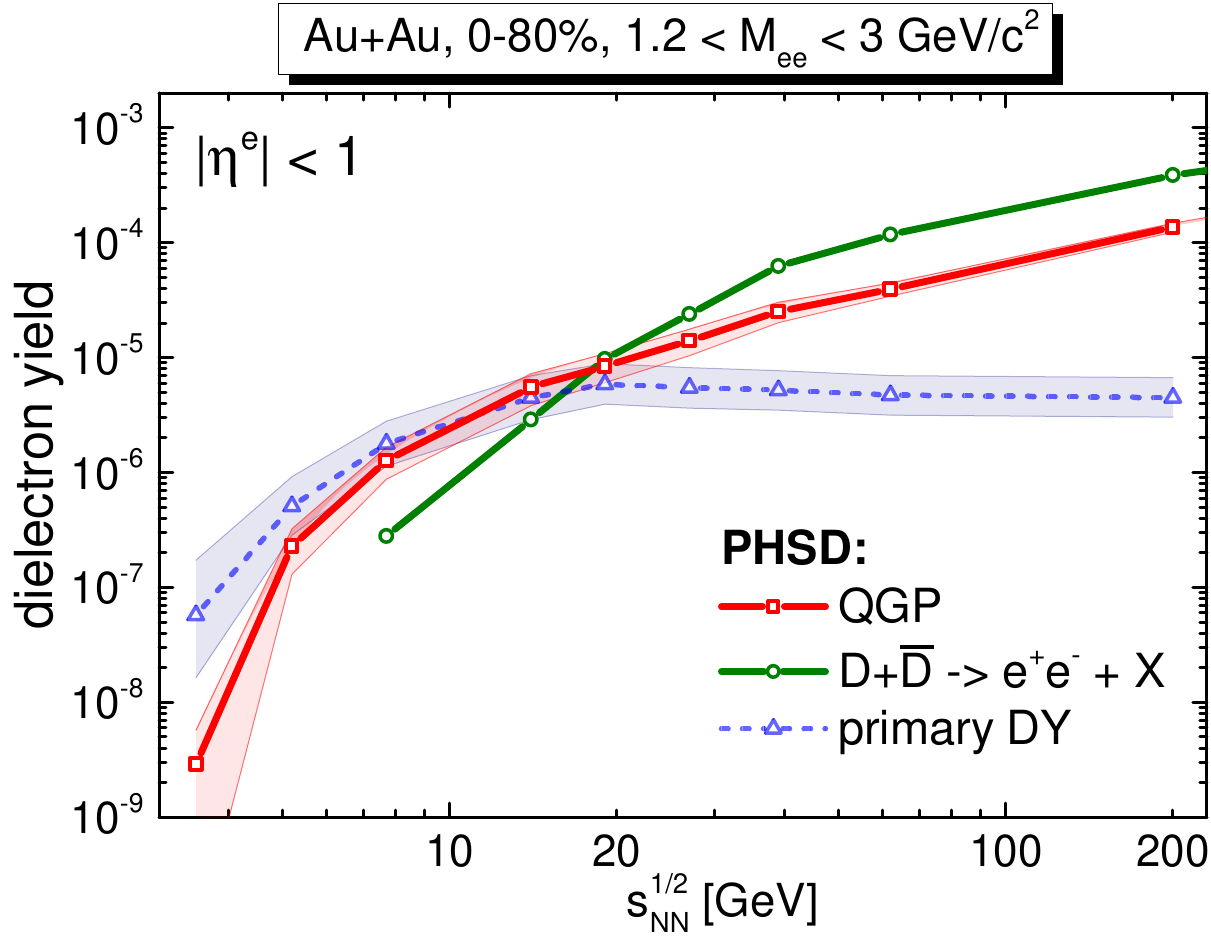}
\includegraphics[width=0.98\linewidth]{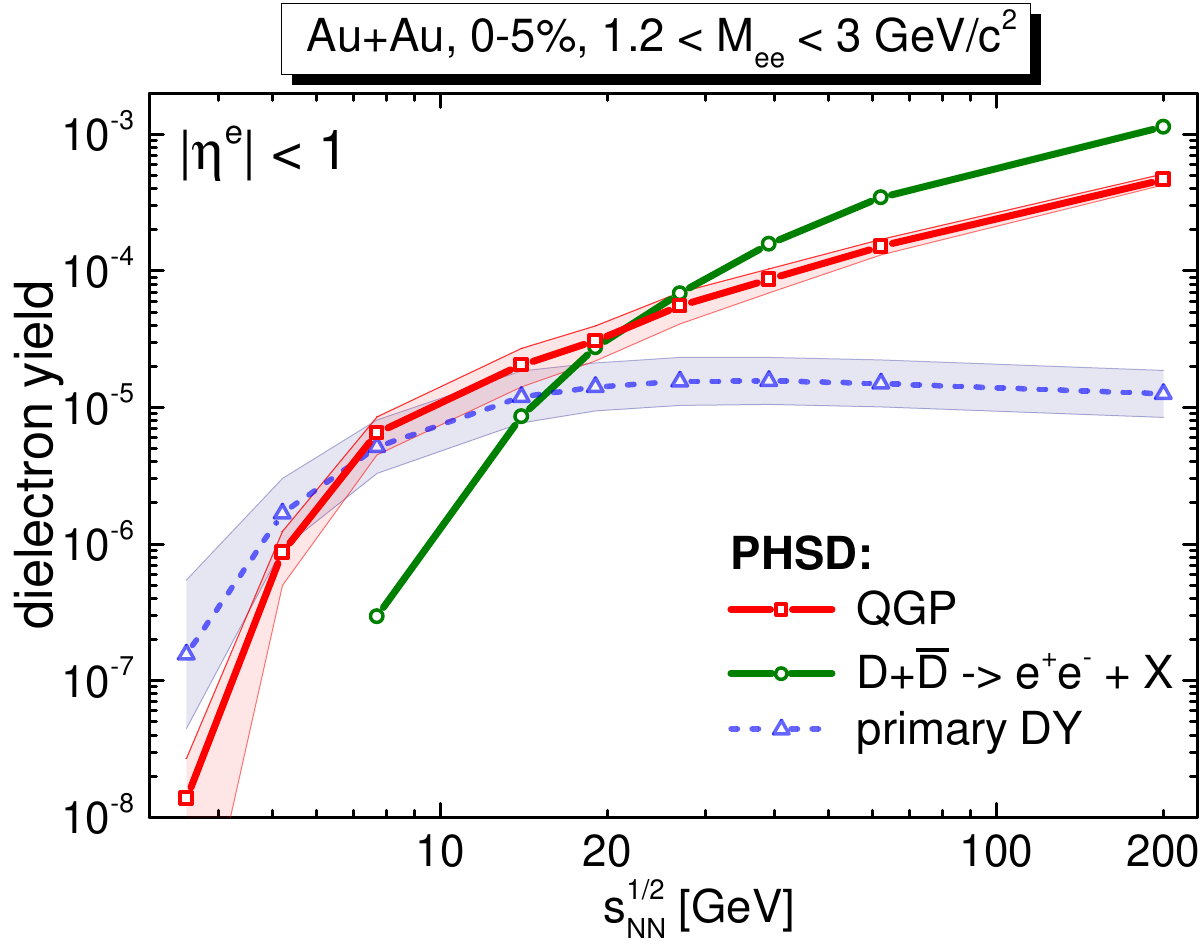}
   \caption{Comparison of the dielectron yield from the $D\bar{D}$, primary Drell-Yan (DY) and  QGP  contributions integrated in the interval $1.2 < M_{ee} < 3$ GeV/c$^2$ as a function of the center of mass energy $\sqrt{s_{NN}}$ from 3.5 GeV to 200 GeV, for  Au + Au collision 0-80\% (upper) and 0-5\% (lower). We display the total sum of the quark-gluon plasma (red), DY (dashed blue) and $D\bar{D}$ (green) yield, respectively.   }
    \label{excitation_function}
\end{figure}

In this section, we explore the excitation function of dielectron yields for intermediate masses ($ 1.2 < M_{ee} < 3 \, \text{GeV}/c^2 $) as a function of the center-of-mass energy $\sqrt{s_{NN}} $ in Au+Au collisions from 3.5 to 200 GeV. We focus on two centrality classes: min-bias collisions (0-80\%, left panel) and central collisions (0-5\%, right panel), here we applied a pseudorapidity cut $|\eta^e|<1$.  This study extends our earlier findings presented in Ref. \cite{Song:2018xca} to cover a broader range of collision energies and also incorporating the Drell-Yan contribution.

Fig. ~\ref{excitation_function} shows the contributions to the dielectron yield from three primary sources: the quark-gluon plasma (QGP, red lines), correlated charm $ D\bar{D} $ decays (green lines), and the primary Drell-Yan process (DY, dark-blue dashed lines). For lower energies ($ \sqrt{s_{NN}} < 10 \, \text{GeV} $), the dielectron yield is dominated by QGP and DY contributions, while the $ D\bar{D} $ channel becomes increasingly significant at higher collision energies ($ \sqrt{s_{NN}} > 20 \, \text{GeV} $).

The comparison between min-bias and central collisions reveals that the total dielectron yield is systematically higher in central collisions due to the increased particle production and medium density. However, the relative contributions from the QGP and $ D\bar{D} $ decays remain consistent across both centrality classes. The Drell-Yan process, on the other hand, exhibits a weaker energy dependence and contributes marginally to the total yield at high energies, while becomes important at low energies.

The excitation function observed in Fig. ~\ref{excitation_function} highlights the transition in dielectron production mechanisms, where the dominance shifts from QGP at low energies to $ D\bar{D} $ decays at higher energies, this transition energy is $\sqrt{s_{NN}} $=27 GeV for central collisions and $\sqrt{s_{NN}} =19.6$ GeV for min-bias collisions (in line with Ref. \cite{Song:2018xca}).   Thus, the QGP dileptons could be accessible experimentally at even low energies when the contribution of Drell-Yan is subtracted by the corresponding measurement in p+p collisions.

\section{Transverse Mass Spectra of Intermediate-Mass Dileptons}\label{sec4}

\begin{figure*}[p]
    \centering
\includegraphics[width=0.49\linewidth]{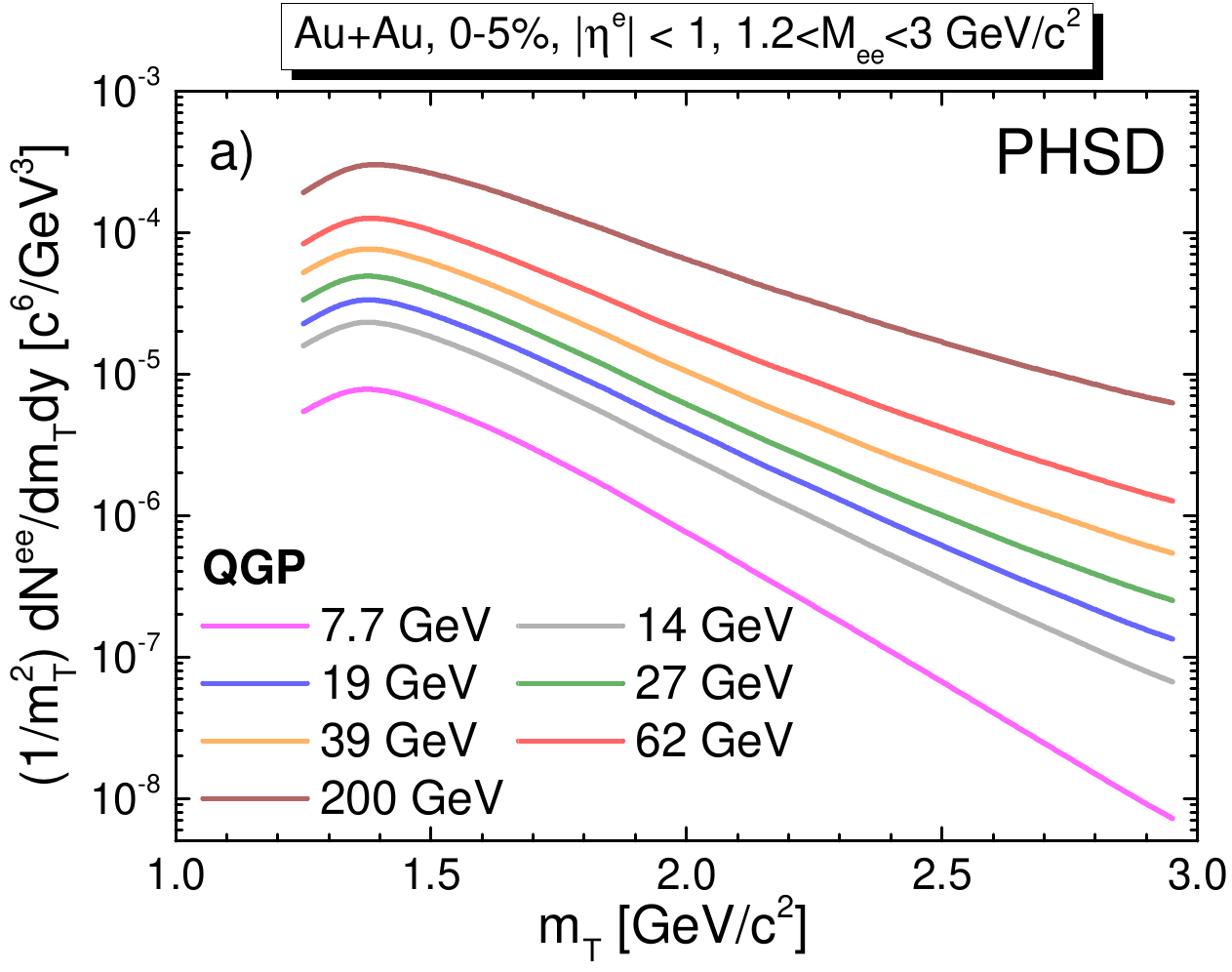}
\includegraphics[width=0.49\linewidth]{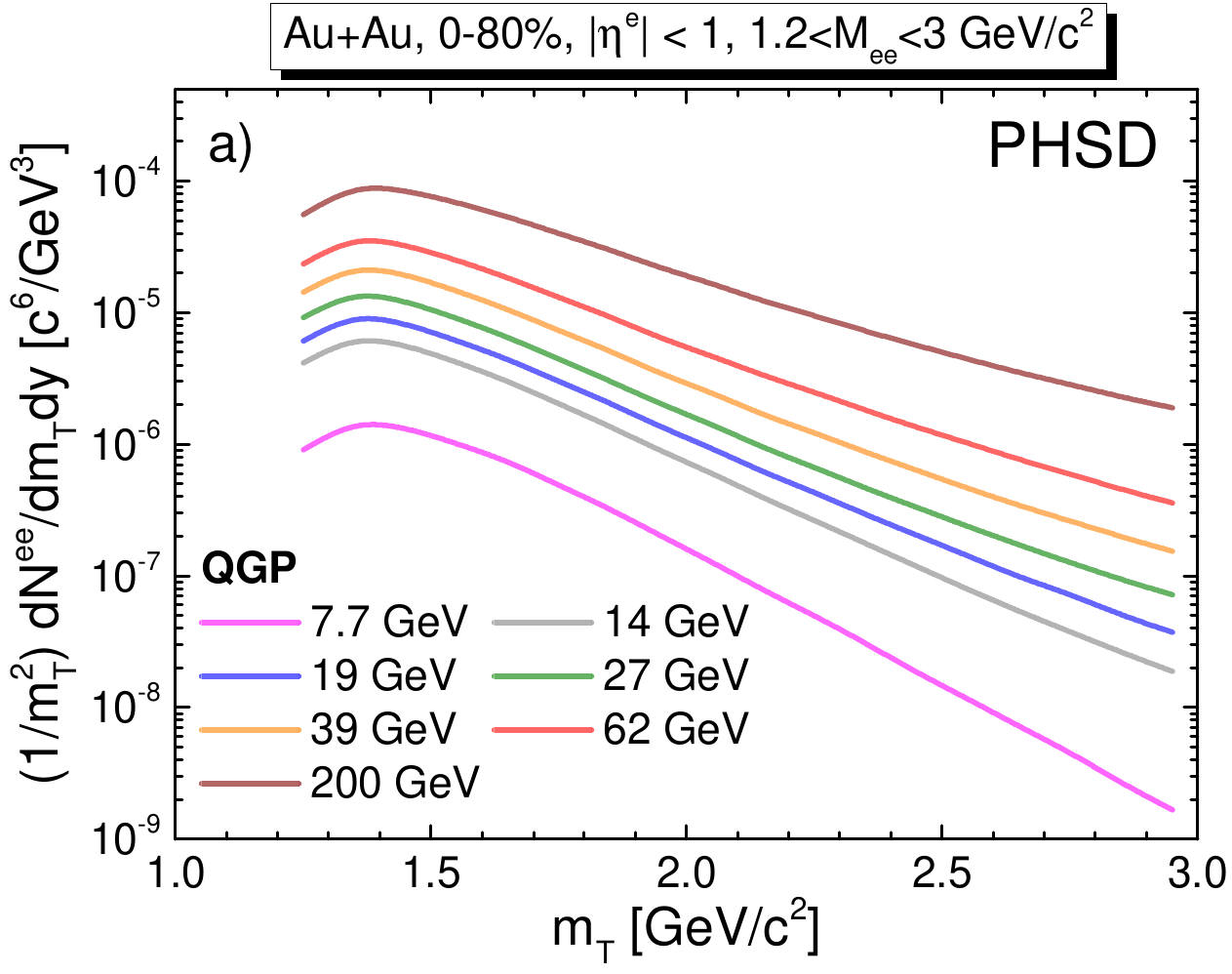}
\includegraphics[width=0.49\linewidth]{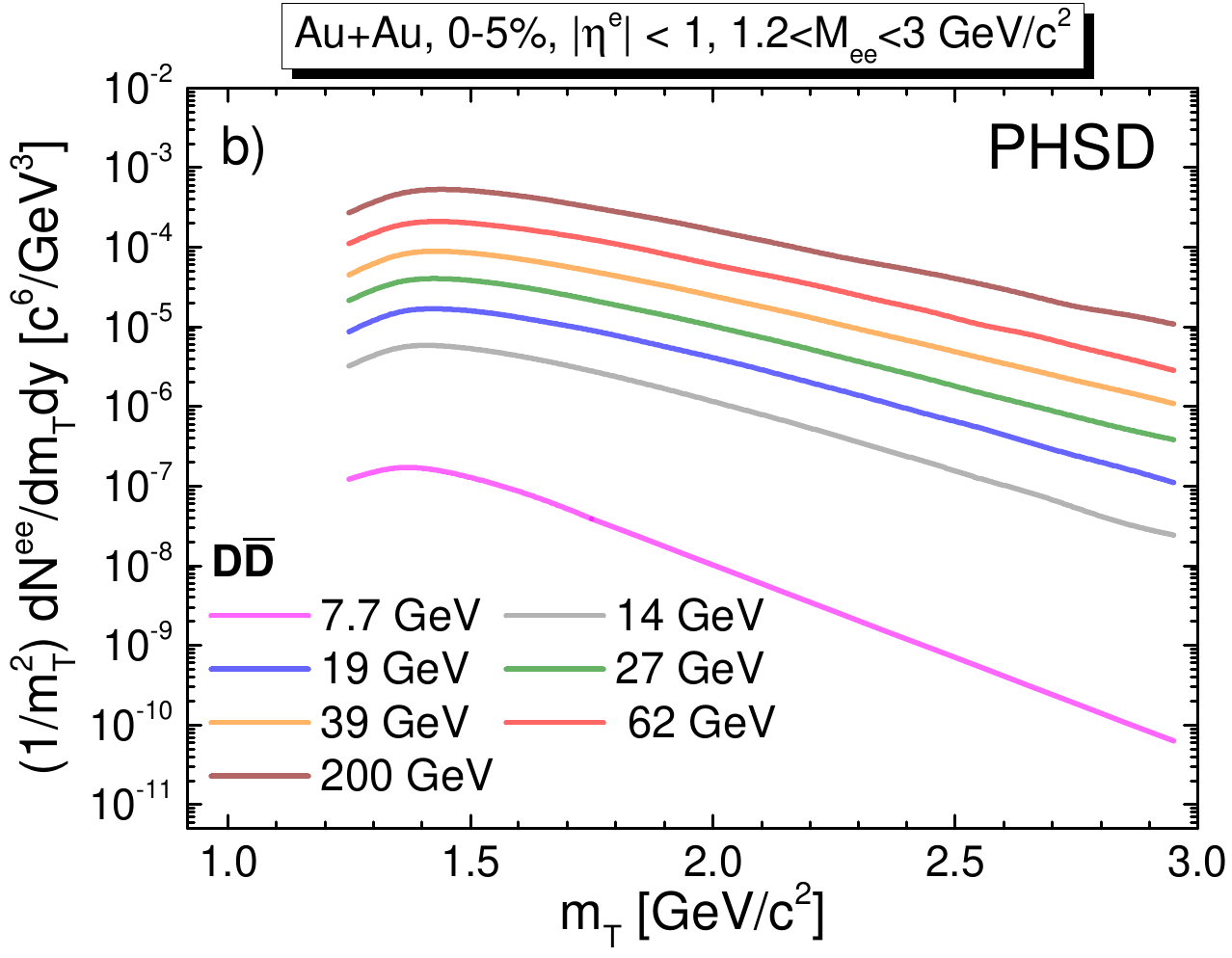}
\includegraphics[width=0.49\linewidth]{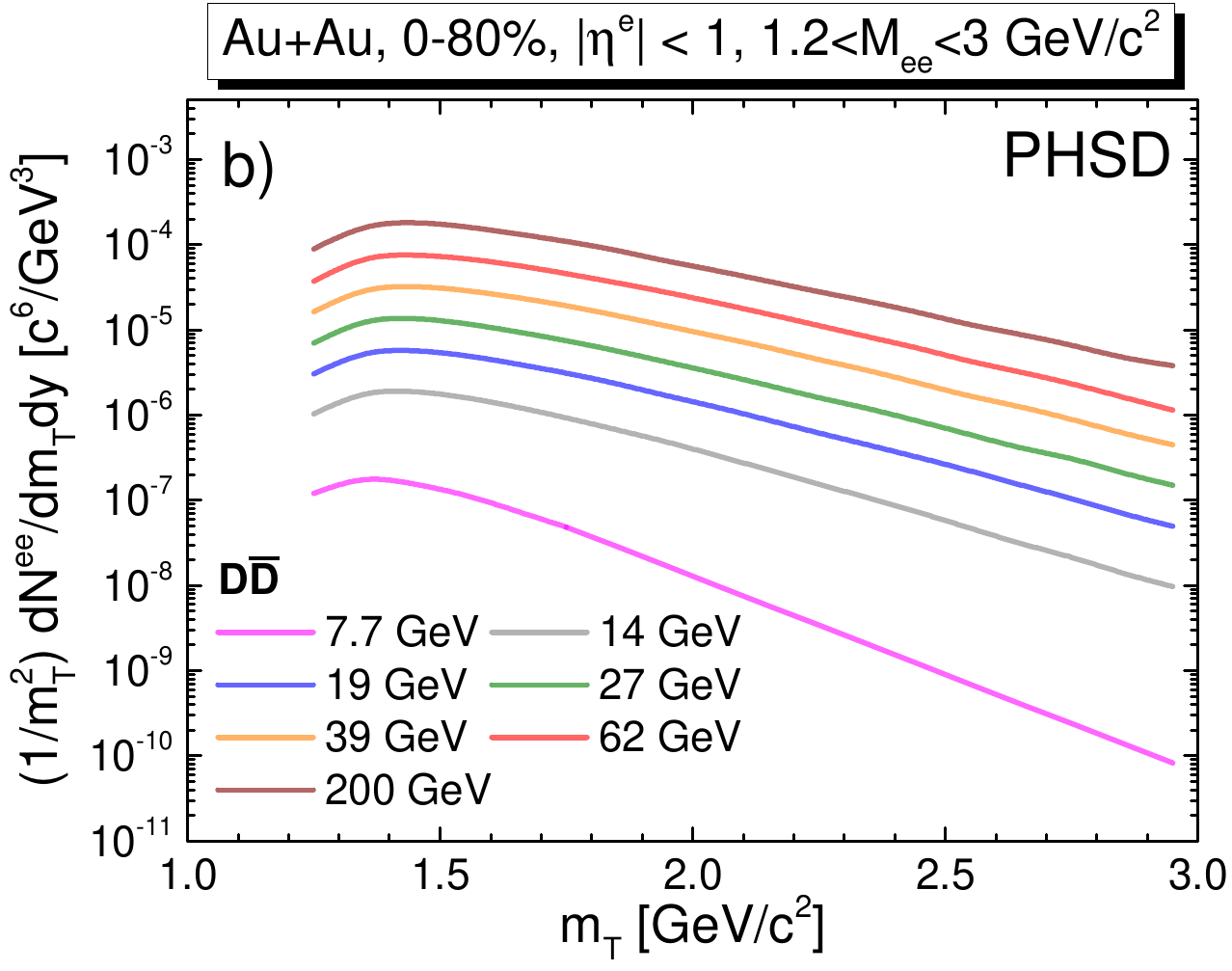}
\includegraphics[width=0.49\linewidth]{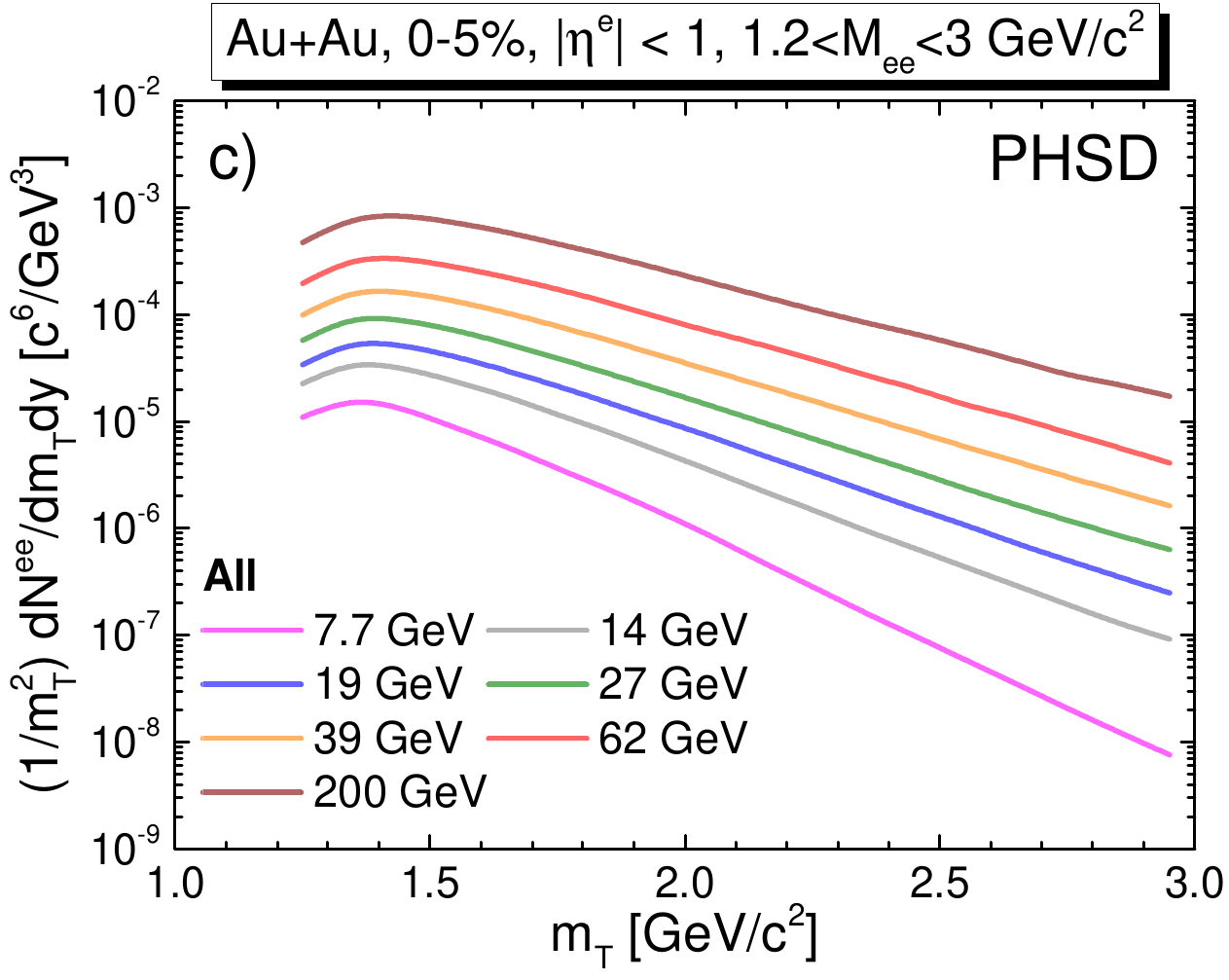}
\includegraphics[width=0.49\linewidth]{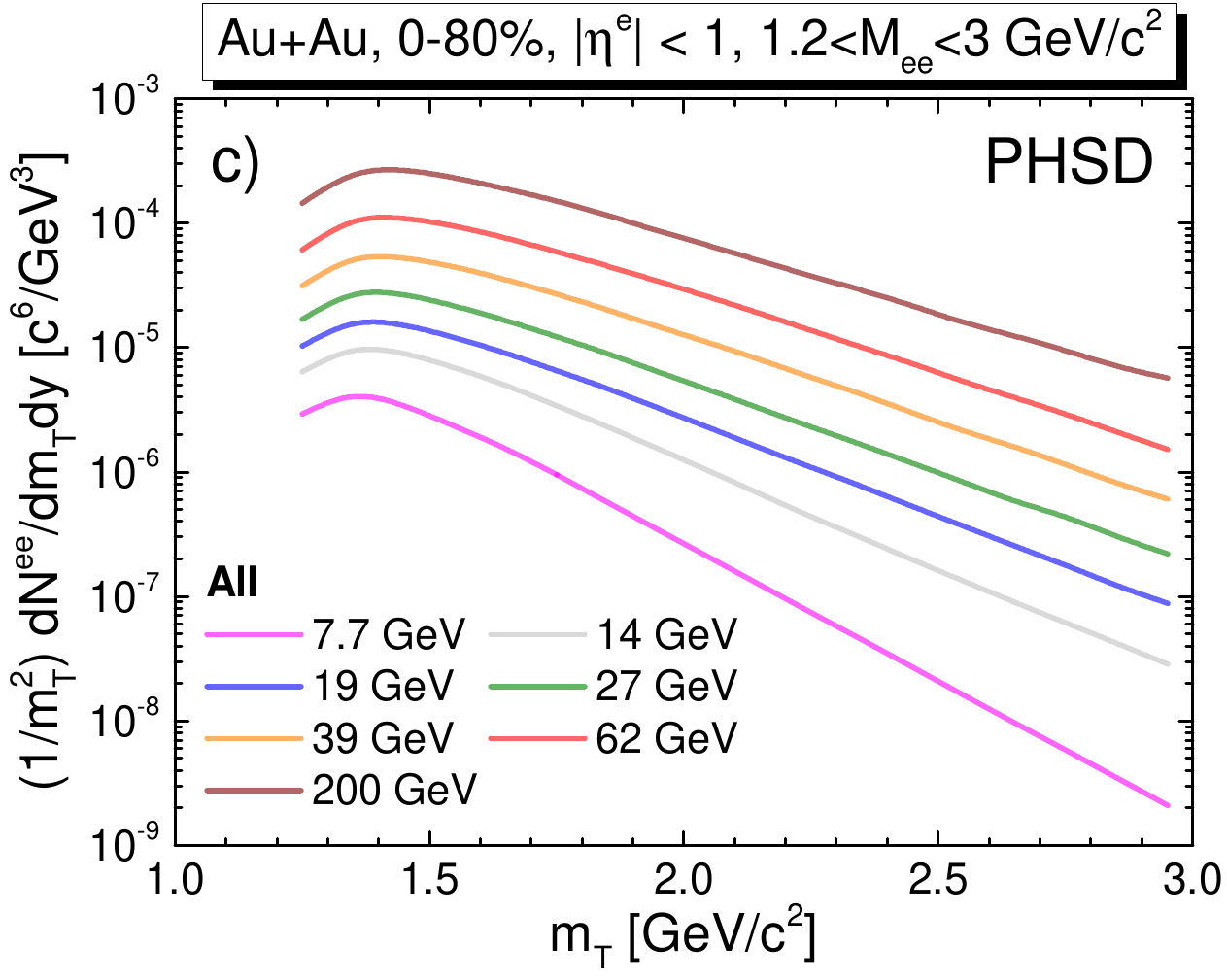}
     \caption{  The transverse mass spectra of dileptons for the invariant mass range $ 1.2 < M_{ee} < 3 \, \text{GeV}/c^2 $ from the QGP (a), $D\bar{D}$ pairs (b), and all sources (c) in  Au+Au collisionsat $\sqrt{s_{NN}}$ = 7.7, 14.5, 19.6, 27, 39, 62.4 and 200 GeV from the PHSD for 0-5\% (left column) and 0-80\% (right column) centralities.
The thick solid lines show exponential fits to the PHSD results in the transverse mass range [1.6, 2.9] GeV/c$^2$.}
    \label{Transversemass_dNdmt}
\end{figure*}

In this section we analyze the transverse mass spectra of intermediate-mass dileptons produced in Au+Au collisions at RHIC energies, focusing on central (0-5\%) and min-bias (0-80\%) collisions at midrapidity. The study emphasizes the contributions from the quark-gluon plasma (QGP) and decays of open charm pairs ($ D\bar{D} $) to the total dilepton yield in the invariant mass range $ 1.2 < M_{ee} < 3 \, \text{GeV}/c^2 $. By examining the Lorentz-invariant transverse mass distributions, we investigate the thermal properties of the medium, characterized by the inverse slope parameter $\beta = 1/T_{\text{eff}}$, where $T_{\text{eff}}$ represents an effective temperature. This parameter is obtained by fitting the transverse momentum spectra of the QGP and $D\bar{D}$ pairs with a thermal distribution function:
\begin{equation}
    \frac{1}{m_T^2} \frac{d \sigma}{d m_T dy} \sim e^{\beta m_T} \, .
    \label{expTeff}
\end{equation}
Furthermore, we analyze its dependence on collision centrality and energy.
 
Fig. \ref{Transversemass_dNdmt} presents the Lorentz-invariant transverse mass distributions for dielectrons in Au+Au collisions for the centrality of 0-5\% (left panels) and 0-80\% (right panels) at $\sqrt{s_{NN}}$ = 7.7, 14.5, 19.6, 27, 39, 62.4 and 200 GeV. The spectra correspond to dielectrons with invariant masses in the range $1.2 <M_{ee}<$ 3 GeV/c$^2$  and are categorized as originating from three primary sources: the QGP (panel a), decays of correlated $ D\bar{D} $ mesons (panel b), and the total dilepton yield, which is dominated by contributions from $D \bar{D}$ decays and QGP radiations (panel c). Across all energies the transverse mass distributions exhibit an approximately exponential decrease, fitted by Eq.~(\ref{expTeff}) in the range  $1.6 <m_{T}< 2.9$ GeV/c$^2$ to obtain  $T_{eff}$.
We note that the $m_T$ spectra in Fig. \ref{Transversemass_dNdmt}  correspond to the $(T,\mu_B)$-dependent EoS ("DQPM($T,\mu_B$)"). However, our explicit calculations for the "DQPM($T,\mu_B=0$)" case show rather similar slopes for the transverse mass distributions. Thus, we do not observe a valuable $\mu_B$-dependence of the $m_T$ dilepton spectra.

\begin{figure}[h!]
    \centering
\includegraphics[width=0.98\linewidth]{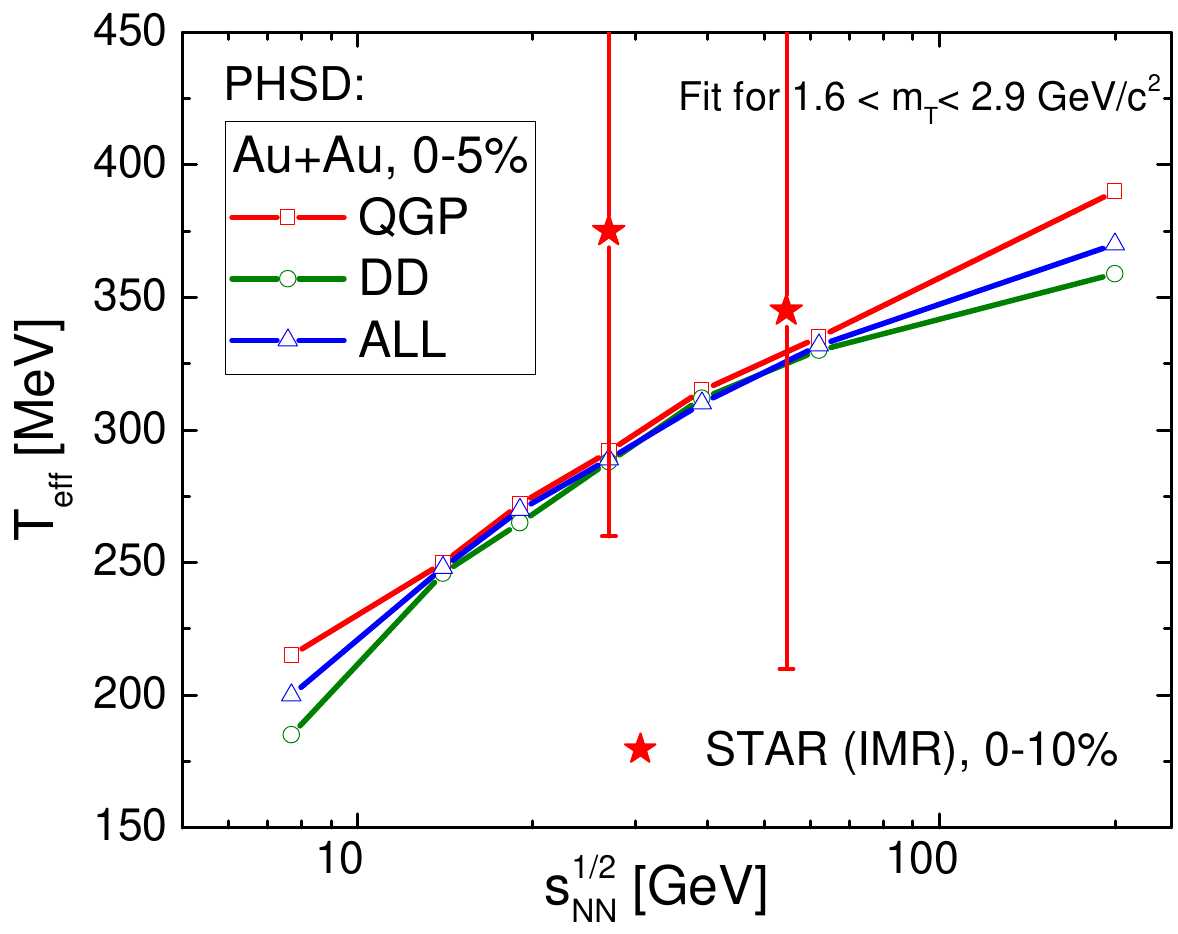}
\includegraphics[width=0.98\linewidth]{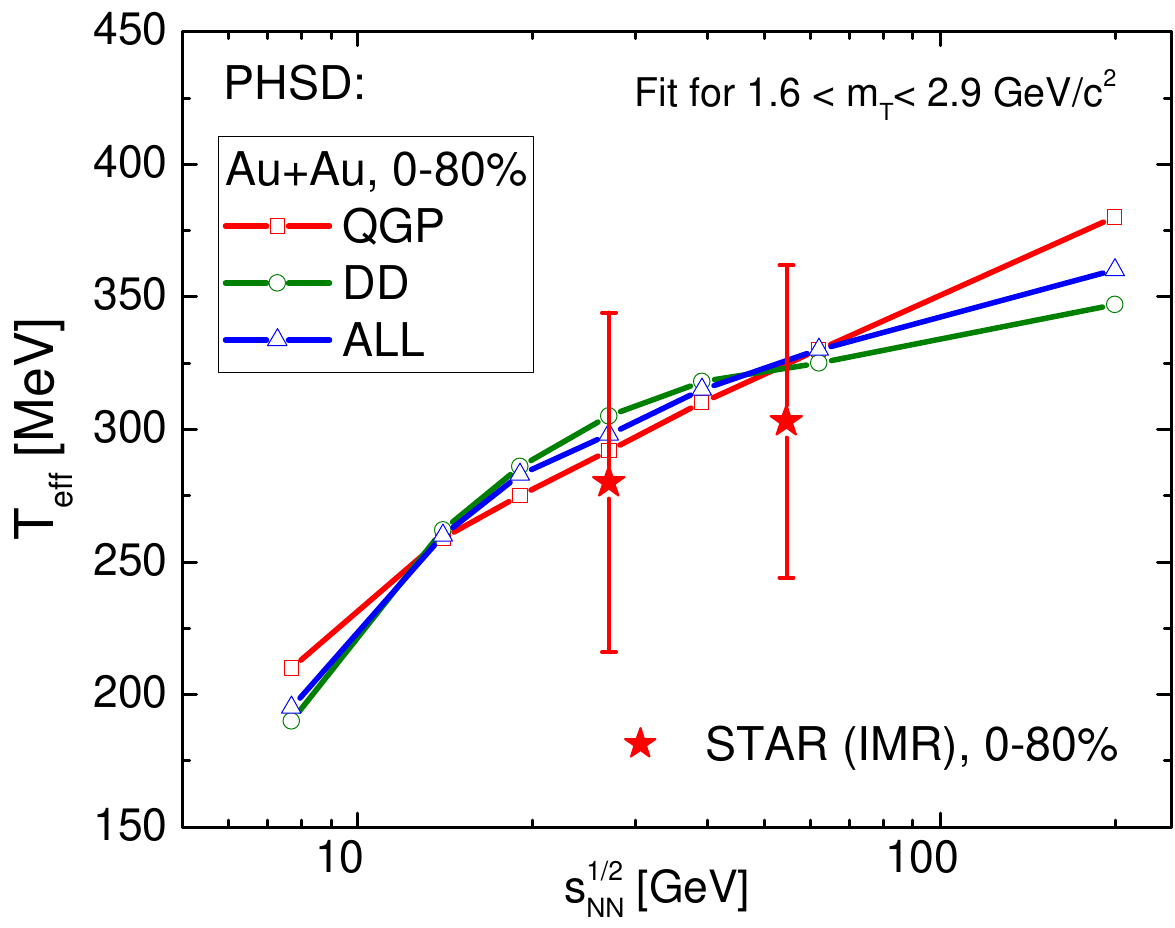}
   \caption{ PHSD results for the effective temperature $T_{eff}$ of intermediate-mass ($1.2 < M_{ee} < 2.9$ GeV/$c^2$) dielectrons fitted in the range $1.6 <m_{T}< 2.9$ GeV/c$^2$ (cf. Fig. \ref{Transversemass_dNdmt})
   from the QGP (red line with squares), $D\bar{D}$ pairs (green line with circles), and all sources (blue line with triangles) from Au+Au collisions for 0-5\% (top panel) and 0-80\% (bottom panel) centralities at $\sqrt{s_{NN}}$ = 7.7, 14.5, 19.6, 27, 39, 62.4 and 200 GeV. 
   The red stars (with error bars) show the effective temperature extracted by the STAR Collaboration \cite{STAR:2024bpc} from the invariant mass spectra of the dilepton excess in the intermediate mass region (IMR)  $1.0 < M_{ee} < 2.9$ GeV/$c^2$. }
    \label{slope_eff_temperature}
\end{figure}

The excitation function of the effective temperature $T_{eff}$ - fitted in the range $1.6 <m_{T}< 2.9$ GeV/c$^2$ - is presented in Fig. \ref{slope_eff_temperature} for the three cases discussed in Fig. \ref{Transversemass_dNdmt}: dileptons with an invariant mass between 1.2 and 3 GeV/c$^2$ originating from the QGP (red line with squares), $D\overline{D}$ pairs (green line with circles), and the overall dilepton sources (blue line with triangles) in  Au+Au collisions for centralities of 0-5\% (top panel) and 0-80\% (bottom panel) at  $\sqrt{s_{NN}}$ = 7.7, 14.5, 19.6, 27, 39, 62.4 and 200 GeV. 
The extracted parameter $T_{eff}$ for the QGP is roughly comparable to that of the $D \bar{D}$ contribution at intermediate energies for central collisions and becomes larger at top RHIC energies at well as low energies, where the charm production is reduced. A similar trend is observed for the min-bias Au+Au collisions for 0-80\% centrality.
The $T_{eff}$ for the QGP grows approximately logarithmically with bombarding energy in line with the increasing size of the QGP fireball and its temperature as discussed in Section IV. 

Moreover, we find that our result for $T_{eff}$ is in line with recent experimental data from the STAR Collaboration: the red stars in Fig. \ref{slope_eff_temperature} show the effective temperature $T_{eff}$ extracted by the STAR Collaboration \cite{STAR:2024bpc} from the invariant mass spectra of dilepton excess data $(dN/dM_{ee}/dy)/(dN_{ch}/dy$ for Au+Au, 0-10\% (upper plot) 0-80\% (lower plot),  at 27 and 54.4 GeV by fitting the Bolzmann function $f(M_{ee}) \sim M_{ee}^{3/2} \exp(M_{ee}/T_{eff})$ in the intermediate mass region (IMR) $1.0 < M_{ee} < 2.9$ GeV/$c^2$. Since  the correlated charm contribution has been subtracted from the excess spectra, the $T_{eff}$ from the STAR Collaboration could be mainly associated with the QGP contribution and agrees well with the $T_{eff}$ from the QGP extracted from the PHSD $m_T$ spectra integrated in a slightly different mass interval (to avoid the influence of the tails of vector meson distributions) as well as centrality (for upper plot). We also note that the $T_{eff} (QGP)$  fits to the PHSD invariant mass distribution $dN/dM_{ee}$ agree quite well in the intermediate-mass region $1.2 < M_{ee} < 2.9$ GeV/$c^2$. 
Thus, the measurement of $T_{eff}$ allows to penetrated inside the hot and dense matter and to probe its thermal properties.

\section{Conclusions}\label{conclusions}

A systematic study of dilepton production over a wide range of bombarding energies — from SIS energies of a few GeV per nucleon up to the multi-TeV regime at the LHC — has been carried out employing the Parton-Hadron-String Dynamics transport approach. By combining hadronic and partonic degrees of freedom within an off-shell transport framework, PHSD provides a comprehensive description of heavy-ion collisions, covering the entire evolution from the initial nucleon-nucleon interactions to the final-state hadronic freeze-out. Comparisons of the PHSD results with experimental data from HADES, STAR, and ALICE collaborations show a good description of dilepton spectra at different collision energies and system sizes.

Our findings can be summarized as follows: \\
--  We show (confirming our early studies \cite{Bratkovskaya:2007jk, Bratkovskaya:2013vx}) that at SIS energies (1–2\,$A$GeV), hadronic processes such as bremsstrahlung and $\pi^0$, $\eta$ and $\Delta$ Dalitz decays dominate the dilepton yield in the low-mass region, whereas the $\rho$ meson broadening plays a decisive role in reproducing the observed spectral shape.  The enhancement of the dilepton yield in A+A compared to p+p collisions (which grow with the system size $A$) - as observed by the HADES collaboration - is attributed to  in-medium effects such as multi-particle reactions - as $\Delta$ regeneration which leads to enhanced dilepton radiation as well as the issue of partial chiral symmetry restoration manifested in the collisional broadening of the vector meson spectral functions.  

-- The collisional broadening of vector meson ($\rho, \omega, \phi$) spectral functions provides a good description of the dilepton enhancement (relative to the hadrionic "cocktail") observed at higher energies (BES RHIC) as well as the excitation function of the integrated dilepton excess yields in the invariant mass range $0.4 < M_{ee} < 0.75$ GeV/c$^2$ as measured by the STAR collaboration.

-- In the intermediate-mass region (1.2--3\,GeV/$c^2$) the dominant contribution comes from partonic processes as well as correlated charm and bottom decays. The calculated excitation function of the integrated dilepton yield in this mass window shows that the QGP partons overshine the correlated charm at  $\sqrt{s_{NN}} \sim $ 30 GeV for central collisions and $\sqrt{s_{NN}} \sim 20$ GeV for min-bias collisions (in line with Ref. \cite{Song:2018xca}). Although the correlated charm is a dominant source of dileptons at LHC energies, the QGP contribution is very large, too. When subtracting the charm contribution from the total spectra, one gets direct access to a very hot QGP matter for $\mu_B \to 0$.

-- We have found that the primary Drell-Yan process plays an important role at intermediate energies, too.  There are rather large uncertainties in the calculation of the primary Drell-Yan process, especially at intermediate energies due to the limited phase space. Moreover, the DY yield is very sensitive to the PDF employed;  thus our calculations can be considered as upper estimates.
A precise dilepton measurement in p+p collisions is needed to subtract the DY contribution from A+A data in order to get an access to the QGP sources of dileptons. 

-- We have studied the properties of the partonic medium in terms of the averaged energy density $\langle \varepsilon \rangle$, temperature $\langle T \rangle$ and baryon chemical potential $\langle \mu_B\rangle$ at the emission point of the dileptons and found that even at rather low energies of  $\sqrt{s_{NN}} \sim $ 3.5 GeV (which is in the energy range of future FAIR experiments), there are  "hot spots" of partonic matter whose fraction rapidly grows with increasing bombarding energy.
Thus, the dileptons from partonic processes allow to penetrate inside the hot and dense phase of heavy-ion collisions  - including an early non-equilibrium phase of QGP formation -  and can be used as a "barometer" of the reaction by measurements of the dilepton invariant mass spectra and the inverse slope parameter of the $m_T$ spectra at intermediate masses. 

--  We have investigated the influence of the QGP EoS with an explicit $\mu_B$ dependence - interpreted in terms of properties of quasiparticles and their interactions within the effective DQPM model - on dilepton production. This study has been done for the first time within the PHSD. In spite that the QGP dilepton spectra show a visible modification due to the $\mu_B$ dependence of the EoS (additionally to its $T$-dependence) in the low mass region, the total dilepton spectra are practically not affected since the QGP contribution is small relative to the hadronic sources at intermediate energies where $\mu_B$ is large. 

-- We have provided predictions for the dilepton spectra for future FAIR experiments and shown that  at FAIR SIS100 energies the in-medium hadronic contributions will continue to be dominant in the low-mass dilepton region, but a sizable partonic component may emerge in the intermediate-mass range. These findings highlight the potential of SIS100 to explore the parton-hadron transition at high net-baryon density.

-- Our calculations of the dilepton production in p+p collisions has shown that at top RHIC and especially at LHC energies there is a subdominant contribution from partonic processes covered by the correlated charm/bottom decays and primary DY dileptons.
When subtracting these contributions one get access to the partonic dileptons from the "hot spots"  in high multiplicity p+p collisions.

Overall, our results underscore the multiple aspects of dilepton production in relativistic nuclear collisions and reaffirm the power of electromagnetic probes in mapping out both hadronic and partonic stages of the evolving medium.
New measurements at FAIR, NICA, and BES RHIC as well as the high-luminosity runs at the LHC will deepen our understanding of QCD matter under extreme conditions and solidify the role of dileptons as penetrating probes of the strongly interacting medium.

\begin{acknowledgments}
The authors thanks to J. Aichelin, M. Bleicher, T. Galatyuk,  I. Grishmanovskii, P. Salabura, O. Soloveva for useful discussions.
We thank  F. Geurts for the guiding us with the STAR measurements.  
We are grateful to W. Cassing for a careful reading of the manuscript and valuable suggestions.
A.R.J. expresses gratitude for the financial support from the Stiftung Giersch. 
We also acknowledge the support by the Deutsche Forschungsgemeinschaft (DFG) through the grant CRC-TR 211 "Strong-interaction matter under extreme conditions" (Project number 315477589 - TRR 211). 
The computational resources utilized for this work were provided by the Center for Scientific Computing (CSC) at Goethe University Frankfurt.

\end{acknowledgments}


\appendix
\section{Drell-Yan Cross Section for Dilepton Production in Proton-Proton Collisions}
\label{appendix:DY}

The cross-section for dilepton production in pp collisions through Drell-Yan process \cite{Drell:1970wh} is given by 
\begin{eqnarray}
\sigma^{pp\rightarrow l^+l^-}(s)=\sum_{q=u,d,s}\int dx_1 dx_2 \nonumber\\
\times D_q(x_1,Q) D_{\bar{q}}(x_2,Q) \sigma^{q\bar{q}\rightarrow l^+l^-}(M),
\label{factorization}
\end{eqnarray}
where $D_{q(\bar{q})}$ is the parton distribution function (PDF) of the proton with $x_1$ and $x_2$ being longitudinal  momentum fractions of quark and antiquark, respectively, and $M=\sqrt{x_1 x_2 s}$ is the invariant  mass of the dilepton.
Since the process is a s-channel process, one can take the scale $Q=M$.

From the energy-momentum coordinates of the initial quark and antiquark,
\begin{eqnarray}
p_1&=&\bigg(x_1\frac{\sqrt{s}}{2}, \vec{0}, x_1\frac{\sqrt{s}}{2}\bigg),\nonumber\\
p_2&=&\bigg(x_2\frac{\sqrt{s}}{2}, \vec{0}, -x_2\frac{\sqrt{s}}{2}\bigg),
\label{coordinate}
\end{eqnarray}
one can find 
\begin{eqnarray}
M^2&=&(x_1+x_2)^2\frac{s}{4}-(x_1-x_2)^2\frac{s}{4}=x_1x_2s,\nonumber\\
y&=&\frac{1}{2}\ln\frac{x_1}{x_2},
\label{rel1}
\end{eqnarray}
or 
\begin{eqnarray}
x_1&=&\frac{M}{\sqrt{s}}e^y,\nonumber\\
x_2&=&\frac{M}{\sqrt{s}}e^{-y}.
\label{rel2}
\end{eqnarray}

The scattering amplitude squared of $q\bar{q}\rightarrow l^+l^-$ is given by~\cite{Song:2018xca} 
\begin{eqnarray}
\overline{|\mathcal{M}|}^2=e_q^2\frac{(4\pi\alpha)^2}{N_c}\bigg(1-\frac{4m_l^2}{s}\bigg)(1+\cos^2\theta),
\end{eqnarray}
where $e_q=2/3$ for an up quark and $1/3$ for down and strange quarks, $\alpha=1/137$, $m_l$ is the lepton mass and $\theta$ the scattering angle.

Integrating over the phase space,
\begin{eqnarray}
\sigma^{q\bar{q}\rightarrow l^+l^-}(M)=\frac{1}{64\pi^2 M^2}\int d\Omega \overline{|\mathcal{M}|}^2\nonumber\\
=e_q^2\frac{4\pi\alpha^2}{3N_c M^2}.
\label{cs}
\end{eqnarray}

Substituting Eq.~(\ref{cs}) into Eq.~(\ref{factorization}) and using the Jacobian,
\begin{eqnarray}
dMdy=\frac{\sqrt{s}}{2\sqrt{x_1x_2}}dx_1dx_2,
\label{jacobian}
\end{eqnarray}
the differential cross section
\begin{eqnarray}
\frac{d\sigma^{pp\rightarrow l^+l^-}}{dMdy}(s)=2\sum_{q=u,d,s}2\sqrt{\frac{x_1x_2}{s}} D_q(x_1,Q)\nonumber\\
\times D_{\bar{q}}(x_2,Q) e_q^2\frac{4\pi\alpha^2}{3N_c M^2}.
\label{DY}
\end{eqnarray}
We note that the Drell-Yan cross section is multiplied by 2 since the quark is contributing either from the first proton or from the second proton in a p+p collision.

\begin{figure}[t!]
    \centering
   \includegraphics[width=0.98\linewidth]{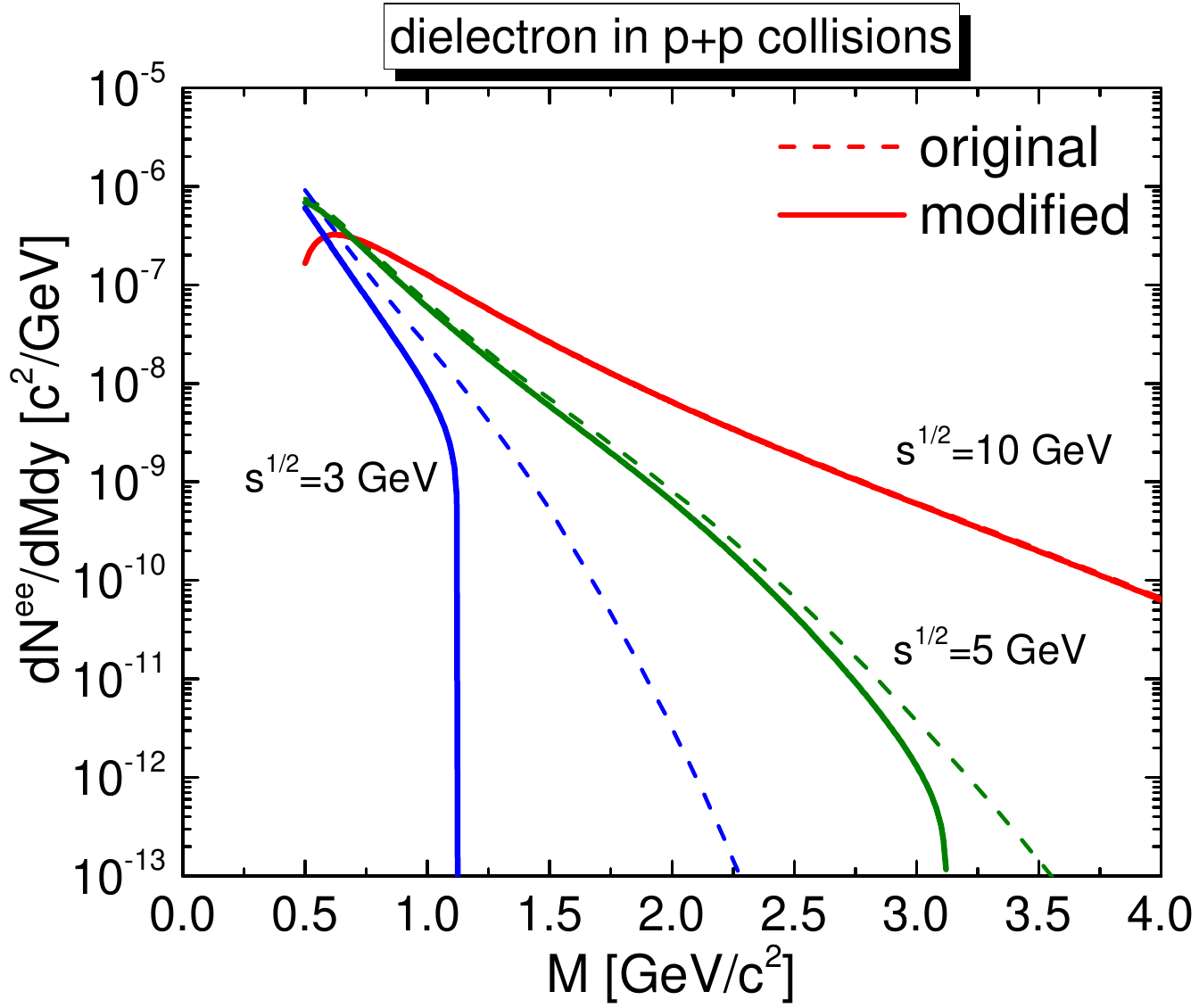}
   \caption{$d^2N/dMdy$ of dielectron at mid-rapidity ($y=0$) in pp collisions at $\sqrt{s}=$3, 5 and 10 GeV.}
    \label{dndm-beta}
\end{figure}

\begin{figure}[h!]
    \centering
\includegraphics[width=0.98\linewidth]{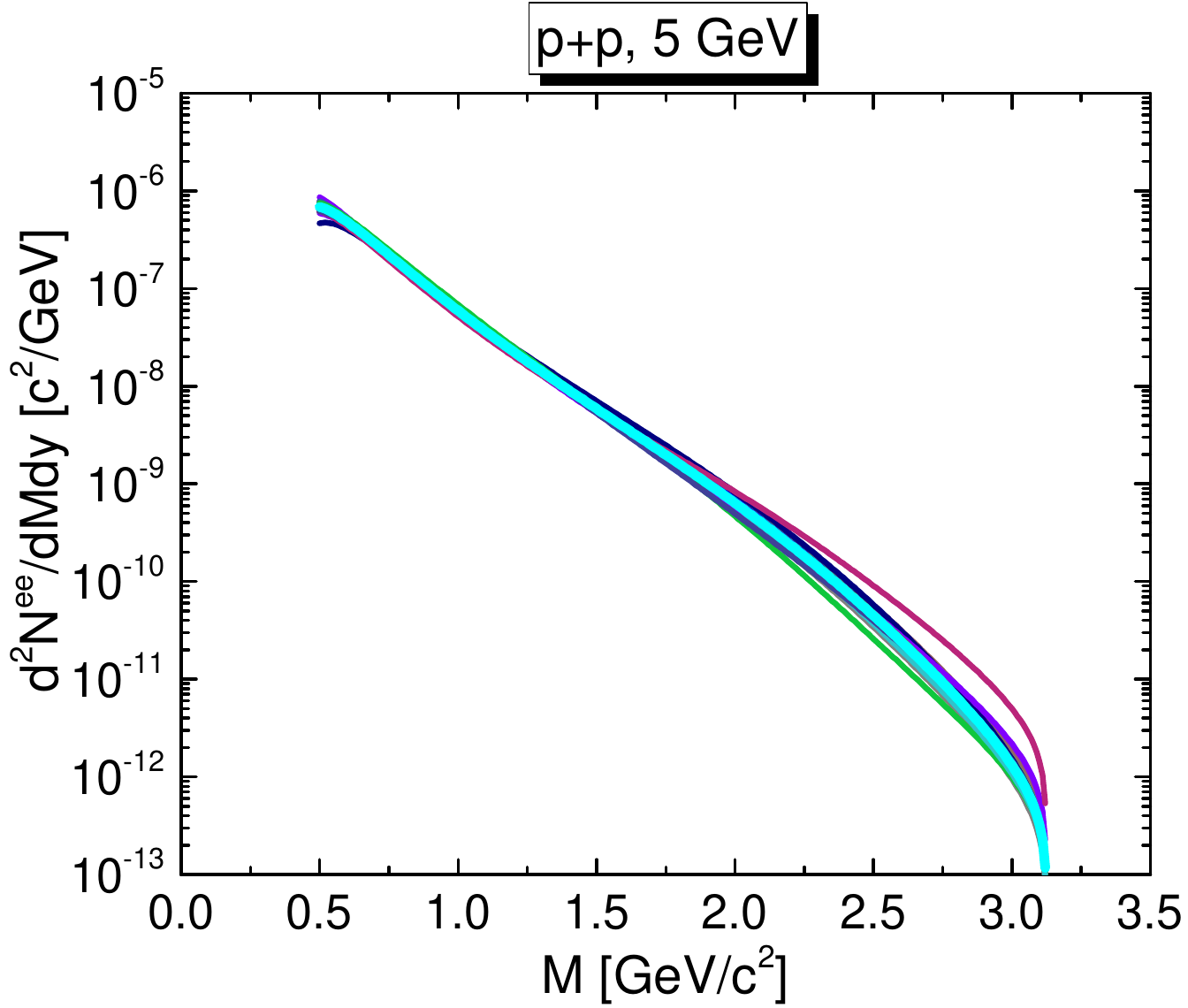}
\includegraphics[width=0.98\linewidth]{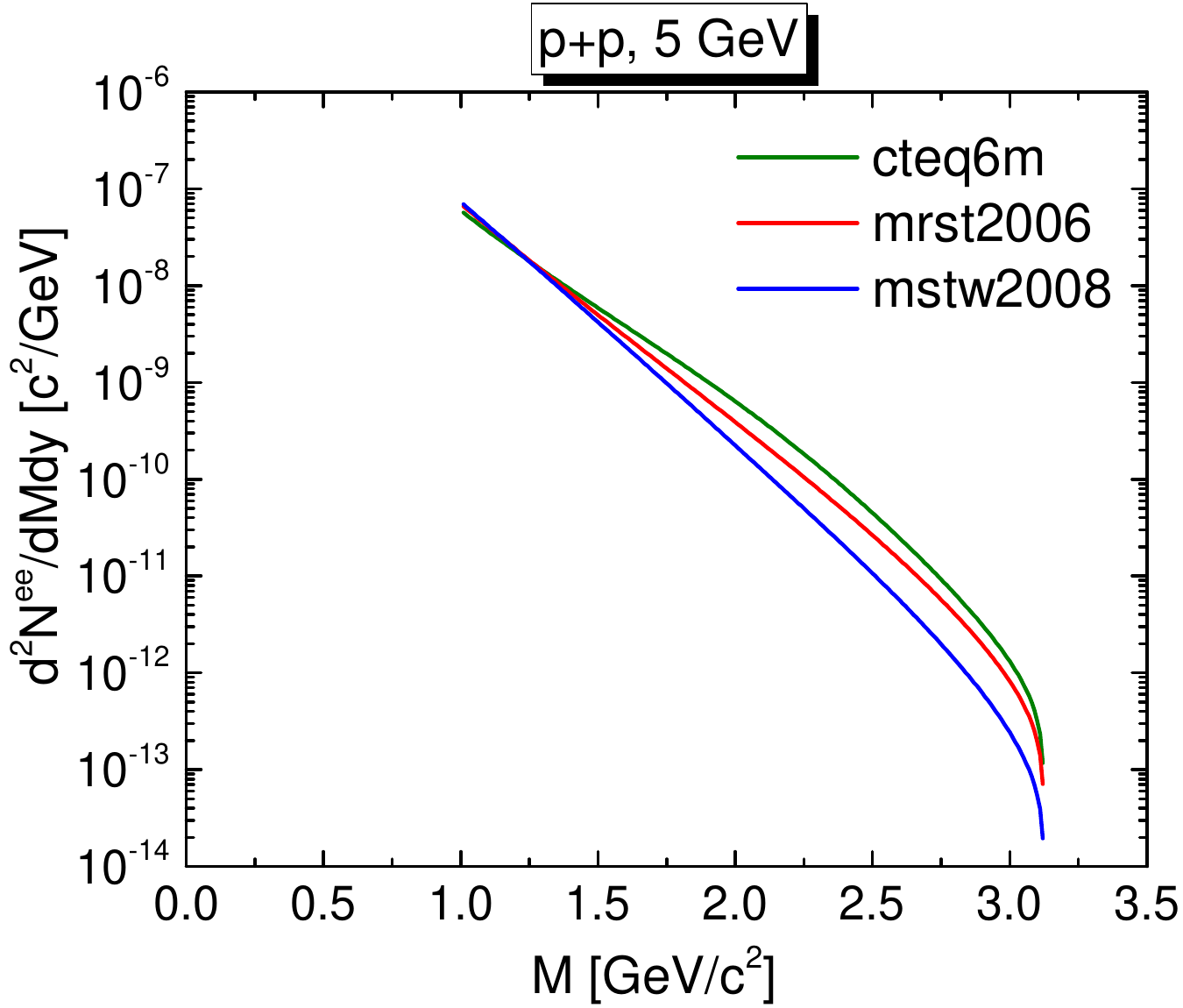}
   \caption{(Upper) Various $d^2N/dMdy$ at mid-rapidity from 41 different data sets in the "CTEQ6M" PDF with the thick skyblue line being the optimized one and (lower) three different $d^2N/dMdy$ at mid-rapidity from "CTEQ6M"~\cite{Nadolsky:2008zw}, "MRST2006"~\cite{Martin:2007bv} and "MSTW2008"~\cite{Martin:2009iq} PDFs in p+p collisions at 5 GeV.}
    \label{uncertainties}
\end{figure}

In low-energy collisions the energy conservation must be controlled.
For example, a p+p collision must have at least two baryons - apart from a dilepton - in the final state to conserve baryon number.
The most generous constraint  will be given by two nucleons in the final state whose phase space for the calculation of the cross section is expressed as
\begin{eqnarray}
\int\frac{d^3 p_a}{(2\pi)^3 2E_a}
\int\frac{d^3 p_b}{(2\pi)^3 2E_b}
\delta^2(\vec{p_T})\delta(p_z+M\sinh y)\nonumber\\
\times \delta(E_a+E_b+M\cosh y -\sqrt{s}),~~~~~~
\label{phase-space}
\end{eqnarray}
where $\vec{p_T}=\vec{p_a}_T+\vec{p_b}_T$ and $p_z=p_{az}+p_{bz}$ with $p_a$ and $p_b$ denoting momenta of the two nucleons.
Since Eq.~(\ref{phase-space}) is Lorentz-invariant, one can move to the rest frame of two nucleons for simplicity: 
\begin{eqnarray}
\int\frac{d^3 p_a}{(2\pi)^3 2E_a}
\int\frac{d^3 p_b}{(2\pi)^3 2E_b}
\delta^2(\vec{p_T})\delta(p_z)\nonumber\\
\times \delta(E_a+E_b+M\cosh (2y) -\sqrt{s})\nonumber\\
=\frac{p_a}{2(2\pi)^5(E_a+E_b)}~~~~~~~~~~~~~~~~~~\nonumber\\
=\frac{\sqrt{\{\sqrt{s}-M\cosh(2y)\}^2-4m_N^2}}{4(2\pi)^5\{{\sqrt{s}-M\cosh(2y)\}}},
\label{phase-space2}
\end{eqnarray}
where $m_N$ is the nucleon mass.

Considering the physical condition, $\sqrt{s}-M\cosh (2y)\ge 2 m_N$, one can find the constraint,
\begin{eqnarray}
x_1+x_2\le \sqrt{\frac{2M}{s}(\sqrt{s}+M-2m_N)}.
\end{eqnarray}

The phase space correction of Eq.~(\ref{phase-space2}) should be 1 for large $\sqrt{s}$  as in the factorization formula.
Therefore it should be normalized into 
\begin{eqnarray}
\frac{\sqrt{\{\sqrt{s}-M\cosh(2y)\}^2-4m_N^2}}{\{{\sqrt{s}-M\cosh(2y)\}}}.
\label{normal}
\end{eqnarray}
We note that Eq.~(\ref{normal}) is the minimum suppression of phase space, because the final state in general is more complex.

Fig.~\ref{dndm-beta} displays  the invariant mass distribution of dielectron produced through the Drell-Yan process in p+p collisions at $\sqrt{s}=$3, 5 and 10 GeV, where the "CTEQ6M" PDF is used for the calculations~\cite{Nadolsky:2008zw}.
The dashed lines are the results without the suppression factor of Eq.~(\ref{normal}) and the solid lines include it. 
One can see that the dielectron production is strongly suppressed for large invariant mass ($M\sim \sqrt{s}-2M_N$) to conserve the total energy.

We point out that the results depend on the PDF.
The upper panel of Fig.~\ref{uncertainties} shows the results for 41 different data sets in the CTEQ6M PDF with the thick skyblue line being the optimized one.
The uncertainties increase with decreasing $M$ because the parton scattering in pQCD is approaching the nonperturbative region.
On the other hand, the uncertainties also increase with increasing $M$, where the antiquark with a large energy fraction contributes but it has  large uncertainties in the PDF.

In the lower panel we show the results from three different PDFs.
Since the minimum value of $Q$ in the "MRST2006"~\cite{Martin:2007bv} and "MSTW2008"~\cite{Martin:2009iq} is 1 GeV, the results are compared above $M=$1 GeV.
We again see large uncertainties close to the maximum mass ($M\sim \sqrt{s}-2M_N$) for the same reason as in the upper panel.

The PDF is modified in a nuclear environment.
For example, the parton distribution is suppressed at small $x$ which is called  'shadowing effect'.
This is e.g. realized by the help of EPS09 in Ref.~\cite{Eskola:2009uj}.
There is another nuclear matter effect on the DY process due to the parton energy loss~\cite{Garvey:2002sn,Giri:2025bfq}.
However, it is still challenging to separate the latter effect from the former one and we do not include these modifications in the present study. 

\nocite{jorge2025arxiv2503}
\bibliography{Dileptons_paper_2025}
\end{document}